\documentclass[11pt]{article}
\pdfoutput=1

\usepackage{color}
\usepackage{mhchem}
\usepackage{xcolor}
\usepackage{colortbl}
\usepackage{hhline}
\usepackage{cite}
\usepackage{hyperref}
\hypersetup{colorlinks=true,linkcolor=red,anchorcolor=black,citecolor=green}
\usepackage[toc,page]{appendix}
\usepackage{mathtools}
\usepackage{amsfonts}
\usepackage{bbold}
\usepackage{textcomp}
\usepackage[DIV13]{typearea}
\usepackage{amsmath, amsthm, amssymb, mathtools,empheq,latexsym,dsfont}
\usepackage{bbm}
\usepackage{amsmath,amssymb,latexsym,dsfont}
\usepackage{slashed, simplewick}
\usepackage[utf8]{inputenc}
\usepackage{graphicx}
\usepackage{graphicx,placeins}
\usepackage{makeidx}
\usepackage[font=small,labelfont=bf]{caption}
\usepackage{nicefrac}
\usepackage{subfigure}
\usepackage{array, bigdelim,multirow,multicol}
\usepackage{mathrsfs}
\usepackage{adjustbox}
\usepackage{booktabs}
\usepackage{indentfirst}
\usepackage{booktabs}
\usepackage{tabulary}
\usepackage{longtable}
\usepackage{fancybox}
\usepackage{graphicx}
\usepackage{bm}
\usepackage{bbm}
\usepackage{tikz}
\usepackage{diagbox}
\usepackage{enumitem}
\usepackage[integrals]{wasysym}
\usepackage{pdflscape}

\graphicspath{{immagini/}}

\definecolor{Gray}{gray}{0.92}
\definecolor{highlight}{RGB}{255,255,0}

\newcommand{\ignore}[1]{}

\newcommand{\be}{\begin{equation}}
\newcommand{\ee}{\end{equation}}
\newcommand{\bea}{\begin{eqnarray}}
\newcommand{\eea}{\end{eqnarray}}

\DeclareMathOperator{\diag}{diag}

\definecolor{lightred}{rgb}{1,0.4,0.4}
\definecolor{lightgreen}{rgb}{0.4,1,0.4}
\definecolor{LightCyan}{rgb}{0.88,1,1}

\newcounter{Thm}[section]
\renewcommand{\theThm}{\arabic{section}.\arabic{Thm}}

\newcounter{nodecount}

\tikzstyle{every picture}+=[remember picture,baseline]
\tikzstyle{every node}+=[inner sep=0pt,anchor=base,
text depth=.25ex,outer sep=1.5pt]
\tikzstyle{every path}+=[thick, rounded corners]
\tikzset{
        plabel/.style={inner sep=2pt}
}

\makeatletter
\@addtoreset{equation}{section}
\makeatother

\begin{document}
\title{\begin{center}
{\Large\bf Neutrino mixing parameters and masses from $\Delta(96)\rtimes H_{CP}$ in the tri-direct CP approach}
\end{center}}
\date{}
\author{Li-Na Yan$^{a,b,c}$\footnote{E-mail: {\tt
			202321634@stumail.nwu.edu.cn}},  \
Xiang-Yan Gao$^{a,b,c}$\footnote{E-mail: {\tt
			202421359@stumail.nwu.edu.cn}},  \
Gao-Da Liu$^{a,b,c}$\footnote{E-mail: {\tt
			2022112109@stumail.nwu.edu.cn}},  \
	Cai-Chang Li$^{a,b,c}$\footnote{E-mail: {\tt ccli@nwu.edu.cn}} \ \\*[20pt]
	\centerline{\begin{minipage}{\linewidth}
			\begin{center}
				$^a${\it\small School of Physics, Northwest University, Xi'an 710127, China}\\[2mm]
				$^b${\it\small Shaanxi Key Laboratory for Theoretical Physics Frontiers, Xi'an 710127, China}\\[2mm]
				$^c${\it\small NSFC-SPTP Peng Huanwu Center for Fundamental Theory, Xi'an 710127, China}\\[2mm]
			\end{center}
	\end{minipage}} \\[10mm]}
\maketitle

\thispagestyle{empty}

\centerline{\large\bf Abstract}
\begin{quote}
\indent
We present a comprehensive model independent analysis of all breaking patterns resulting from $\Delta(96)\rtimes H_{CP}$ in the tri-direct CP approach of the minimal seesaw model with two right-handed neutrinos. The three generations of left-handed lepton doublets are assumed  to transform as the irreducible triplet $\bm{3_{0}}$ of $\Delta(96)$, and the two right-handed neutrinos are assigned to singlets. In the case that  both flavon fields $\phi_{\text{atm}}$ and $\phi_{\text{sol}}$ transform as triplet $\bm{\bar{3}_{0}}$, only one phenomenologically viable lepton mixing pattern is obtained for normal ordering neutrino masses.  The lepton mixing matrix is predicted to be the TM1 pattern, with neutrino masses, mixing angles, and CP violation phases depending on only three real input parameters.  When $\phi_{\text{sol}}$ is assigned to the $\bm{\bar{3}_{1}}$ representation, an additional real parameter $x$ must be included. Then we find 42 (12) independent phenomenologically interesting mixing patterns for normal (inverted) ordering neutrino masses, and the corresponding predictions for  lepton mixing parameters and neutrino masses are obtained. Furthermore,  we perform a detailed numerical analysis for five (one) example breaking patterns with some benchmark values of $x$ for normal (inverted) ordering. For the five normal examples, the absolute values of the  first columns of the PMNS matrix are fixed to be $\left(\sqrt{\frac{2}{3}},\frac{1}{\sqrt{6}},\frac{1}{\sqrt{6}}\right)^{T}$, $\frac{1}{5}\left(\sqrt{17},2,2\right)^{T}$, $\frac{1}{\sqrt{38}}\left(5,2,3\right)^{T}$, $\frac{1}{\sqrt{57}}\left(\sqrt{37},\sqrt{10},\sqrt{10}\right)^{T}$ and $\frac{1}{3}\left(\sqrt{6},1,\sqrt{2}\right)^{T}$, respectively. For the inverted example, the absolute value of the third column of the PMNS matrix is $\frac{1}{2\sqrt{11}}\left(1,5,3\sqrt{2}\right)^{T}$.

\end{quote}

\newpage

\section{Introduction}

The confirmation of neutrino oscillations demonstrates that neutrinos possess tiny masses and that flavor mixing occurs within the lepton sector. The Standard Model (SM) does not provide any insight into the origin of fermion mass hierarchy, flavor mixing and CP violation. This remains a fundamental and unsolved flavor puzzle in physics. A definitive theoretical framework capable of resolving this puzzle has yet to emerge~\cite{King:2014nza,King:2015aea,King:2017guk,Feruglio:2019ybq,Xing:2020ijf,Almumin:2022rml}. The observation of non-zero neutrino masses, and large  solar and atmospheric neutrino mixing angles open up a window to the new physics beyond SM. The discrete flavor symmetry approach, as a promising framework for the lepton flavor mixing angles, is now widely used to account for large neutrino mixing angles~\cite{Altarelli:2010gt,Ishimori:2010au,King:2013eh,King:2014nza,King:2015aea,King:2017guk,Petcov:2017ggy,Xing:2020ijf,Feruglio:2019ybq,Almumin:2022rml,Ding:2024ozt}. Non-Abelian discrete flavor symmetry combined with  CP (gCP) symmetry provides a powerful framework to explain the lepton mixing angles and predict lepton CP violation phases~\cite{Feruglio:2012cw,Holthausen:2012dk,Ding:2013hpa,Ding:2013bpa,Li:2013jya,Ding:2013nsa,
Ding:2014ssa,Ding:2014hva,Li:2014eia,Ding:2014ora,Chen:2014wxa,Everett:2015oka,Branco:2015hea,Li:2015jxa,DiIura:2015kfa,Ballett:2015wia,Branco:2015gna,Chen:2015nha,Ding:2015rwa,
Chen:2015siy,Li:2016ppt,Chen:2016ptr,Yao:2016zev,Li:2016nap,Lu:2016jit,Everett:2016jsk,Li:2017zmk,Li:2017abz,Lu:2018oxc,Lu:2019gqp,Chen:2018lsv,Hagedorn:2016lva,Delgadillo:2018tza}.  Furthermore, a simultaneous description of quark and lepton flavor mixing and CP violation can be achieved through spontaneous breaking of a discrete family symmetry and gCP symmetry~\cite{Li:2017abz,Lu:2018oxc,Lu:2019gqp}. The type I seesaw mechanism, which introduces heavy right-handed (RH) Majorana neutrinos, appears as the most elegant solution for generating tiny neutrino masses~\cite{Minkowski:1977sc,Mohapatra:1979ia,Schechter:1980gr}.

Although the type I seesaw mechanism offers a qualitative explanation for the small neutrino mass scale via heavy right-handed neutrinos (RHNs), without additional assumptions, the three RHNs seesaw model contains too many parameters to yield specific predictions for neutrino mass and mixing. Furthermore, there is a mass hierarchy between the two mass squared differences $\Delta m^2_{21}\equiv m^2_2-m^2_1$ and $|\Delta m^2_{3l}|\equiv |m^2_3-m^2_{l}|$($\Delta m^2_{3l}\equiv m^2_3-m^2_{1}>0$ for normal ordering (NO) and $\Delta m^2_{3l}\equiv m^2_3-m^2_{2}<0$ for inverted ordering (IO)). To address these limitations, the sequential dominance (SD) framework~\cite{King:1998jw,King:1999cm} provides an effective approach. It naturally results in the observed hierarchy between solar and atmospheric neutrino mass splittings $\Delta m^2_{21}$ and $\Delta m^2_{3l}$~\cite{Esteban:2024eli} by imposing a strong mass hierarchy among the heavy Majorana neutrinos: $M_\text{atm}\ll M_\text{sol}\ll M_\text{dec}$. This originates from the proposal that a dominant heavy RHN generates the atmospheric neutrino mass, a heavier subdominant RHN generates the solar neutrino mass, and a potential third, largely decoupled RHN generates the lightest neutrino mass. This results in an effective two right-handed neutrino (2RHN) model with a massless lightest neutrino~\cite{King:1999mb,Frampton:2002qc}, yielding simple and falsifiable predictions. To further increase predictive power  of the minimal seesaw framework, models with one~\cite{King:2002nf} or two~\cite{Frampton:2002qc} texture zeros in the Yukawa coupling were proposed. However, current data excludes two-zero models for normal ordering neutrino masses~\cite{Guo:2006qa,Harigaya:2012bw,Zhang:2015tea}.

The so-called CSD($n$) model~\cite{King:2005bj,Antusch:2011ic,King:2013iva,King:2015dvf,King:2016yvg,Ballett:2016yod,King:2018fqh,King:2013xba,King:2013hoa,Bjorkeroth:2014vha} represents a particularly predictive minimal seesaw model with one texture zero. Its Dirac neutrino mass matrix has two columns aligned with the directions $(0,1, -1)$ and $(1, n, 2-n)$ respectively in the RHN diagonal basis, where the parameter $n$ was usually assumed to be a positive integer, but can generally be any real number. The CSD($n$) framework, including models like the CSD($3$) (also called Littlest Seesaw model)~\cite{King:2013iva,King:2015dvf,King:2016yvg,Ballett:2016yod,King:2018fqh}, CSD($4$) models~\cite{King:2013xba,King:2013hoa} and CSD($-1/2$)~\cite{Chen:2019oey} can give rise to phenomenologically viable predictions for lepton mixing parameters and  neutrino masses  as special cases of TM1 mixing in which the first column of the tri-bimaximal mixing matrix is preserved.

In recent years, the idea of modular invariance~\cite{Feruglio:2017spp} has provided a powerful framework to explain the lepton masses and mixing parameters, see Refs.~\cite{Kobayashi:2023zzc,Ding:2023htn} for recent review.  As was observed, modular symmetry indicates CSD($1-\sqrt{6}$) $\approx$ CSD($-1.45$) which requires multiple moduli~\cite{Ding:2019gof,Ding:2021zbg}. The first complete Littlest Modular Seesaw model based on CSD($1-\sqrt{6}$) was recently developed, where the three required moduli have been incorporated into complete models of leptons at the field theory level~\cite{deMedeirosVarzielas:2022fbw},  or in 10-dimensional orbifolds~\cite{deAnda:2023udh}. A new related possibility based on CSD($1+\sqrt{6}$) $\approx$ CSD($3.45$) has been  proposed~\cite{deMedeirosVarzielas:2022fbw}. Furthermore, the Littlest Modular Seesaw model has been extended to a Grand Unified scenario based on $SU(5)$ endowed with three modular $S_{4}$ symmetries~\cite{deMedeirosVarzielas:2023ujt}.

Based on the  non-Abelian discrete flavor symmetry $G_{f}$ and gCP symmetry $H_{CP}$ compatible with it, we propose a new method called the tri-direct CP approach~\cite{Ding:2018fyz} for predicting neutrino masses and lepton mixing parameters in the minimal seesaw model with two right-handed neutrinos. In the tri-direct CP approach,  the high energy symmetry $G_{f}\rtimes H_{CP}$ is spontaneously broken down to $G_{l}$, $G_{\text{atm}}\rtimes H^{\text{atm}}_{CP}$ and $G_{\text{sol}}\rtimes H^{\text{sol}}_{CP}$ in the charged lepton, ``atmospheric'' and ``solar'' right-handed neutrino sectors, respectively~\cite{Ding:2018fyz}. Here $G_{l}$ is assumed to be an Abelian subgroup of $G_f$ and it is capable of distinguishing the three generations. The residual subgroups $G_{\text{atm}}\rtimes H^{\text{atm}}_{CP}$ and $G_{\text{sol}}\rtimes H^{\text{sol}}_{CP}$ fix  the two columns of Dirac neutrino mass matrix. In the framework of the tri-direct CP approach, all phenomenologically interesting breaking patterns which can arise from $S_{4}\rtimes H_{CP}$ have been discussed  for both NO and IO neutrino mass spectrum~\cite{Ding:2018tuj}. The model construction along the tri-direct CP approach was also illustrated~\cite{Ding:2018fyz,Ding:2018tuj,Chen:2019oey}. Furthermore, the phenomenologically viable golden Littlest Seesaw model with GR1 mixing which preserves the first column of the golden ratio mixing matrix has been  obtained from the breaking of  $A_5$~\cite{Ding:2017hdv}.

In the present work, we perform a comprehensive  model independent analysis of lepton mixing patterns resulting from the breaking of $\Delta(96)\rtimes H_{CP}$  via the tri-direct CP approach, considering both NO and IO neutrino mass spectrums. The three generations of the left-handed (LH) lepton doublets are assumed  to transform as the irreducible triplet $\bm{3_{0}}$ of $\Delta(96)$, and the 2RHN are assigned to the singlets $\bm{1_{m}}$ ($m=0,1$). Then flavon fields $\phi_{\text{atm}}$ and $\phi_{\text{sol}}$  are necessary to break the high energy symmetry $\Delta(96)\rtimes H_{CP}$ down to $G_{\text{atm}}\rtimes H^{\text{atm}}_{CP}$ and $G_{\text{sol}}\rtimes H^{\text{sol}}_{CP}$ in the atmospheric and solar neutrino sectors, respectively. Furthermore, they must transform as $\bm{\bar{3}_{m}}$  to preserve $\Delta(96)$ and gCP symmetry. For the NO case, if both  $\phi_{\text{atm}}$ and $\phi_{\text{sol}}$ are assigned to the triplet $\bm{\bar{3}_{0}}$, we find that only one independent lepton breaking pattern can accommodate the experimental data in lepton sector, in which lepton mixing parameters and neutrino masses depend on three real input parameters $|m_{a}|$, $r$ and $\eta$. As a consequence, the corresponding lepton mixing matrix is predicted to be TM1 pattern. When $\phi_{\text{sol}}$ transforms as $\bm{\bar{3}_{1}}$, one more real free parameter $x$ would be imposed.  Our analysis reveals that 42 independent breaking patterns are consistent with experimental data on neutrino masses and mixing parameters. We subsequently conduct a detailed analysis of five phenomenologically interesting mixing patterns with benchmark values for parameters $x$ and $\eta$. We identify four distinct mixing matrices that differ from the TM1 mixing scheme. Similarly, for the IO case, we find 12 phenomenologically viable breaking patterns when $\phi_{\text{atm}}$ and $\phi_{\text{sol}}$ are assigned to the representations $\bm{\bar{3}_{0}}$ and $\bm{\bar{3}_{1}}$, respectively. We then perform a detailed analysis of one particularly interesting example among them.

We emphasise that there are good motivations to study $\Delta(96)$ flavor symmetry. As a flavor symmetry, $\Delta(96)$ is well established and widely discussed~\cite{deAdelhartToorop:2011nfg,deAdelhartToorop:2011re,Ding:2014ssa,Ding:2012xx}, and breaking it can produce well-known mixing matrices such as tri-bimaximal, bimaximal, and Toorop-Feruglio-Hagedorn (TFH). $\Delta(96)$ is a group with complex triplet and sextet representations, thus offering new model building possibilities absent in $A_4$, $S_4$ and $A_5$. Such models readily extend to grand unified schemes like $\Delta(96)\times SU(5)$~\cite{King:2009mk}. Moreover, complex triplets forbid the trivial unit mass matrix which would destroy the mass hierarchies (unlike in  $A_4$, $S_4$ and  $A_5$ where it would be allowed). While analysis of these grand unified models lies beyond our scope, the formal results established here enable future exploration.

The remaining parts of this paper are organized as follows.  The framework of the tri-direct CP approach for 2RHN models, based on the discrete flavor group $G_{f}$  and gCP symmetry, is reviewed in  section~\ref{sec:framework}. In section~\ref{sec:lepton_pred}, we perform a model independent analysis of all possible lepton mixing patterns achievable from $\Delta(96)\rtimes H_{CP}$, and the phenomenological predictions for the lepton mixing angles, CP violation phases and neutrino masses are obtained. Six example breaking patterns are presented and a thorough numerical analysis is performed to discuss their phenomenology in section~\ref{sec:viable_model_x}. We summarize our results and give the conclusion in section~\ref{sec:conclusion}. The group theory of $\Delta(96)$ is presented in appendix~\ref{sec:Delta96_group_theory},  including the conjugacy classes, the irreducible representations, the Kronecker products, the Clebsch-Gordan (CG) coefficients,  the Abelian subgroups and corresponding invariant  vacuum expectation values (VEVs). The appendix~\ref{sec:Diag_mnup} describes the diagonalization of a general block-diagonal neutrino mass matrix.

\section{\label{sec:framework}Framework}

It is highly non-trivial to combine a family symmetry $G_{f}$ with the gCP symmetry. In order to consistently combine flavor symmetry $G_{f}$ with gCP symmetry $H_{CP}$, the so-called consistency conditions have to be fulfilled~\cite{Feruglio:2012cw,Holthausen:2012dk}:
\begin{equation}\label{eq:consistency_cond}
X_{\bm{r}}\rho_{\bm{r}}^{*}(g)X_{\bm{r}}^{\dagger}=\rho_{\bm{r}}(g^{\prime}), \qquad g,g^{\prime}\in G_{f}, \quad  X_{\bm{r}}\in H_{CP} ,
\end{equation}	
where $\rho_{\bm{r}}(g)$ represents the matrix of group element $g$ in the irreducible representation $\bm{r}$ of $G_{f}$, while $X_{\bm{r}}$ is the gCP transformation matrix. Crucially, the relation between $g$ and $g^{\prime}$ must hold identically across all irreducible representations $\bm{r}$. This equation reveals that $X_{\bm{r}}$ induces a mapping from group element $g$ to $g^{\prime}$, thereby establishing a connection between gCP transformations and automorphisms $u:G_{f}\rightarrow G_{f}$. Moreover, it was recently shown that  the physically well-defined CP transformations have to be given by a class-inverting automorphism of $G_{f}$~\cite{Chen:2014tpa}. Notably, the composite transformation $\rho_{\bm{r}}(h)X_{\bm{r}}$ ($\forall h\in G_{f}$) corresponds to an automorphism formed by combining $u$ with an inner automorphism $\mu_{h}:g\rightarrow hgh^{-1}$~\cite{Girardi:2013sza,Li:2014eia}. Hence the mathematical structure of the full symmetry group comprising family symmetry $G_{f}$ and gCP symmetry is in general a semi-direct product. Consequently, the complete symmetry group is $G_{f}\rtimes H_{CP}$.

In the present work, we shall concentrate on the flavor symmetry $G_{f}=\Delta(96)$ which can be generated by four generators $a$, $b$, $c$ and $d$. The group theory of $\Delta(96)$, the representation matrices of generators, all Abelian subgroups and all the CG coefficients in our basis are reported in appendix~\ref{sec:Delta96_group_theory}. Now we determine the explicit form of these gCP transformation matrices in our working basis.  After performing a comprehensive study of the automorphism group of $\Delta(96)$, we find a generic class-inverting automorphism $u$ of the $\Delta(96)$ group, and its actions on the generators $a$, $b$, $c$ and $d$ are as follows
\begin{equation}
\label{eq:physical_aut}a\stackrel{u}{\longmapsto}a,\qquad
b\stackrel{u}{\longmapsto}b,\qquad c\stackrel{u}{\longmapsto}c^{3},\qquad
d\stackrel{u}{\longmapsto}d^{3}\,.
\end{equation}
By examining the conjugacy classes of $\Delta(96)$ listed in Eq.~\eqref{eq:Delta96_CC}, one can readily verify that the automorphism $u$ maps every group element to the conjugacy class of its inverse. This establishes the constraints for determining the gCP transformation  $X_{\bm{r}}(u)$ through the following consistency equations:
\begin{eqnarray}
\nonumber&&X_{\bm{r}}(u)\rho^{*}_{\bm{r}}(a)X^{\dagger}_{\bm{r}}(u)=\rho_{\bm{r}}(a)\,,\qquad
X_{\bm{r}}(u)\rho^{*}_{\bm{r}}(b)X^{\dagger}_{\bm{r}}(u)=\rho_{\bm{r}}\left(b\right)\,,\\
\label{eq:conmsistency_equations}&&X_{\bm{r}}(u)\rho^{*}_{\bm{r}}(c)X^{\dagger}_{\bm{r}}(u)=\rho_{\bm{r}}\left(c^{3}\right)\,,\qquad 
X_{\bm{r}}(u)\rho^{*}_{\bm{r}}(d)X^{\dagger}_{\bm{r}}(u)=\rho_{\bm{r}}\left(d^{3}\right)\,.
\end{eqnarray}
As evident from the representation basis specified in table~\ref{tab:Delta96_Reps}, the complex conjugated representations satisfy 
\begin{equation}
\rho^{*}_{\bm{r}}(a)=\rho_{\bm{r}}(a), \qquad \rho^{*}_{\bm{r}}(b)=\rho_{\bm{r}}(b), \qquad \rho^{*}_{\bm{r}}(c)=\rho_{\bm{r}}(c^{3}), \qquad \rho^{*}_{\bm{r}}(d)=\rho_{\bm{r}}(d^{3}),
\end{equation}
which implies that the gCP transformation matrix  $X_{\bm{r}}(u)$ is fixed to be unity matrix up to an overall phase, i.e. 
\begin{equation}
X_{\bm{r}}(u)=\mathbb{1}_{\bm{r}}\,.
\end{equation}
When incorporating additional inner automorphisms as outlined below Eq.~\eqref{eq:consistency_cond}, the complete set of gCP transformations compatible with $\Delta(96)$ symmetry can be expressed as
\begin{equation}
\label{eq:Delta96_GCP_full}
X_{\bm{r}}=\rho_{\bm{r}}(g),\qquad g\in \Delta(96)\,.
\end{equation}
This indicates that the gCP transformation compatible with $\Delta(96)$ shares an identical structure with the family group transformation in our chosen basis, although they act on a field multiplet in different ways: under group elements $\varphi(x)\stackrel{g}{\longmapsto}\rho_{\bf{r}}(g)\varphi(x)$, whereas under CP transformation, $\varphi(x)\stackrel{CP}{\longmapsto}X_{\bf r}\varphi^{*}(x_P)=\rho_{\bf r}(g)\varphi^{*}(x_P)$, where $x_{P}=(t,-\vec{x})$. This leads to the consequence that all coupling constants become real numbers in a $\Delta(96)\rtimes H_{CP}$ model, owing to the reality of the CG coefficients demonstrated in table~\ref{tab:Delta96_CG}.

\subsection{\label{sec:Tridirect_approach}The tri-direct approach}

In the present work, neutrinos are treated as Majorana particles with masses generated through the Type I minimal seesaw model~\cite{King:1999mb}, which only involves 2RHN $N^c_{\text{atm}}$ (``atmospheric'' neutrino) and $N^c_{\text{sol}}$ (``solar'' neutrino). The third right-handed neutrino is assumed to be almost decoupled, and 2RHN model provides a viable leading order approximation. In the RHN  diagonal basis, the Lagrangian for the charged lepton and neutrino masses can be written as
\begin{eqnarray}
 \nonumber \mathcal{L}&=&-y_{l}L\phi_{l}E^{c}-y_{\text{atm}}L\phi_{\text{atm}}N_{\text{atm}}^{c}-y_{\text{sol}}L\phi_{\text{sol}}N_{\text{sol}}^{c}\\
 \label{eq:Lagrangian}&&-\frac{1}{2}x_{\text{atm}}\xi_{\text{atm}}N_{\text{atm}}^{c}N_{\text{atm}}^{c}
 -\frac{1}{2}x_{\text{sol}}\xi_{\text{sol}}N_{\text{sol}}^{c}N_{\text{sol}}^{c}+\text{h.c.} \,,
\end{eqnarray}
where the fermion fields above are adopted to two-component fermion notation. In Eq.~\eqref{eq:Lagrangian}, $L$ and $E^{c}$ represent left-handed lepton doublets and right-handed charged leptons, respectively. The flavons $\xi_{\text{atm}}$ and $\xi_{\text{sol}}$ are the Standard Model and flavor symmetry singlets, whereas the flavons $\phi_{l}$, $\phi_{\text{sol}}$ and $\phi_{\text{atm}}$ can be either Higgs fields or combinations of the electroweak Higgs doublet together with flavons. Note that, in order to avoid cross term $N_{\text{sol}}^{c}N_{\text{atm}}^{c}$ in Eq.~\eqref{eq:Lagrangian}, the right-handed neutrinos $N_{\text{sol}}^{c}$ and $N_{\text{atm}}^{c}$ must be assigned to different singlets under $\Delta(96)$. Otherwise, the model would require an additional $Z_{n}$ ($n\geq2$) symmetry, under which the two fields carry distinct charges.

In the model construction, the three generations of left-handed lepton doublets $L$ are assigned to the faithful three-dimensional representations $\bm{3_{m}}$ or $\bm{\bar{3}_{m}}$, while  $N^c_{\text{atm}}$ and $N^c_{\text{sol}}$ transform as singlets under the $\Delta(96)$ group. Then the flavon fields $\phi_{\text{atm}}$  and $\phi_{\text{sol}}$  are required to be  triplets $\bm{\bar{3}_{m}}$ or $\bm{3_{m}}$, whereas $\xi_{\text{atm}}$ and $\xi_{\text{sol}}$ remain invariant under $\Delta(96)$. Additionally, the combination of the flavon $\phi_{l}$ and the right-handed charged leptons  $E^{c}$ must reside within the faithful three-dimensional representation of $\Delta(96)$. Notably, when the flavor symmetry $\Delta(96)$ is combined with gCP symmetry, all coupling constants $y_{l}$, $y_{\text{atm}}$, $y_{\text{sol}}$, $x_{\text{atm}}$ and $x_{\text{sol}}$ in the Lagrangian of Eq.~\eqref{eq:Lagrangian}  become constrained to real values.  Following the breaking of electroweak and flavor symmetries, the Dirac and Majorana neutrino mass matrices can be systematically constructed using the CG coefficients provided in table~\ref{tab:Delta96_CG}, leading to the following explicit expressions:
\begin{equation}
m_{D}=\begin{pmatrix}y_{\text{atm}}\langle\phi_{\text{atm}}\rangle,&y_{\text{sol}}\langle\phi_{\text{sol}}\rangle\end{pmatrix}, \qquad 
m_{N}=\begin{pmatrix}x_{\text{atm}}\langle\xi_{\text{atm}}\rangle&0\\0&x_{\text{sol}}\langle\xi_{\text{sol}}\rangle\end{pmatrix} \,,
\end{equation}
where $\langle\phi_{\text{atm}}\rangle$, $\langle\phi_{\text{sol}}\rangle$, $\langle\xi_{\text{atm}}\rangle$ and $\langle\xi_{\text{sol}}\rangle$ denote the vacuum alignments of flavons $\phi_{\text{atm}}$, $\phi_{\text{sol}}$, $\xi_{\text{atm}}$ and $\xi_{\text{sol}}$, respectively. The Dirac neutrino mass matrix $m_{D}$ is defined by the convention $(\nu_{L})^{T}m_{D}N^{c}$. The light neutrino mass matrix derived via the seesaw mechanism can be expressed as
\begin{equation}\label{eq:mnu}
m_{\nu}=-m_{D}m_{N}m^T_{D}=m_{a}\left(\bm{v}_{\text{atm}}\bm{v}^T_{\text{atm}}+re^{i\eta}\bm{v}_{\text{sol}}\bm{v}^T_{\text{sol}}\right)\,,
\end{equation}
where $\bm{v}_{\text{atm}}\equiv\langle\phi_{\text{atm}}\rangle/v_{\phi_a}$ and $\bm{v}_{\text{sol}}\equiv\langle\phi_{\text{sol}}\rangle/v_{\phi_s}$ represent three-dimensional dimensionless column vectors aligned with the vacuum directions of $\phi_{\text{atm}}$ and $\phi_{\text{sol}}$, respectively. Here, $v_{\phi_a}$ and $v_{\phi_s}$ denote the corresponding respective flavon field VEV magnitudes, and they are taken to be real in the following. The other parameters in Eq.~\eqref{eq:mnu} are defined as
\begin{eqnarray}
\nonumber && m_{a}=-y^2_{\text{atm}}v^2_{\phi_a}/(x_{\text{atm}}\langle\xi_{\text{atm}}\rangle), \quad r=|y^2_{\text{sol}}v^2_{\phi_s}x_{\text{atm}}\langle\xi_{\text{atm}}\rangle/(y^2_{\text{atm}}v^2_{\phi_a}x_{\text{sol}}\langle\xi_{\text{sol}}\rangle)|, \\
\label{eq:mareta_def}&& \eta=\text{arg}\left[x_{\text{atm}}\langle\xi_{\text{atm}}\rangle/(x_{\text{sol}}\langle\xi_{\text{sol}}\rangle)\right]\,,
\end{eqnarray}
where the phase of $m_{a}$ corresponds to an unphysical global phase that can be eliminated through lepton field redefinitions. Consequently, $m_{a}$ can consistently be treated as a positive real parameter in subsequent analyses.

In the setup, the lepton mixing and neutrino masses can be predicted from $\Delta(96)\rtimes H_{CP}$ breaking into different remnant symmetries $G_{l}$, $G_{\text{atm}}\rtimes H^{\text{atm}}_{CP}$ and $G_{\text{sol}}\rtimes H^{\text{sol}}_{CP}$ in the charged lepton sector, the atmospheric sector and solar neutrino sector respectively, where the Abelian subgroup $G_{l}$ is required to be capable of distinguishing the three generations, $G_{\text{atm}}$ and $G_{\text{sol}}$ are Abelian subgroups of $\Delta(96)$.  Since both the remnant family symmetry and remnant CP symmetries remain preserved following symmetry breaking in both the atmospheric and solar neutrino sectors, their mutual compatibility becomes necessary. This consequently imposes the following constrained consistency conditions that must be satisfied:
\begin{subequations}
\begin{eqnarray}
\label{eq:nu_atm_consis}&&X^{\text{atm}}_{\bm{r}}\rho^{*}_{\bm{r}}(g^{\text{atm}}_{i})(X^{\text{atm}}_{\bm{r}})^{-1}
=\rho_{\bm{r}}(g^{\text{atm}}_{j}),\qquad g^{\text{atm}}_{i},g^{\text{atm}}_{j}\in G_{\text{atm}},\quad  X^{\text{atm}}_{\bm{r}}\in H^{\text{atm}}_{CP}\,,\\
\label{eq:nu_sol_consis}&&X^{\text{sol}}_{\bm{r}}\rho^{*}_{\bm{r}}(g^{\text{sol}}_{i})(X^{\text{sol}}_{\bm{r}})^{-1}
=\rho_{\bm{r}}(g^{\text{sol}}_{j}),\qquad g^{\text{sol}}_{i},g^{\text{sol}}_{j}\in G_{\text{sol}},\quad  X^{\text{sol}}_{\bm{r}}\in H^{\text{sol}}_{CP}\,.
\end{eqnarray}
\end{subequations}
Then the residual CP symmetries $H^{\text{atm}}_{CP}$ and $H^{\text{sol}}_{CP}$ can be systematically derived through resolution of the constraints outlined in Eqs. \eqref{eq:nu_atm_consis} and \eqref{eq:nu_sol_consis}, respectively. Notably, the predictions for the lepton flavor mixing and neutrino masses only depend on the assumed symmetry breaking patterns and are independent of the details of a specific implementation scheme, such as the possible additional symmetries of the model,  the involved flavon fields, or their assignments. In the following, we shall firstly review how the tri-direct CP approach allows us to predict the lepton mixing and neutrino masses in terms of few parameters.

Now let us outline the detailed analytical procedures for systematically investigating the phenomenological predictions of the tri-direct CP approach in a model independent framework. Specifically, we derive generic formulations for both the lepton mixing matrix and neutrino mass spectrum based on the residual symmetry combination $\{G_{l},G_{\text{atm}}\rtimes H^{\text{atm}}_{CP},G_{\text{sol}}\rtimes H^{\text{sol}}_{CP}\}$.  The condition that the charged lepton mass term maintains the symmetry $G_{l}$ necessitates that the hermitian form $m^{\dagger}_{l}m_{l}$ must remain invariant under the residual symmetry group $G_{l}$. Formally, for every element $g_{l}\in G_{l}$, the following equality must hold:
\begin{equation}\label{eq:ch_dia}
\rho^{\dagger}_{\bm{r}}(g_{l})m^{\dagger}_{l}m_{l}\rho_{\bm{r}}(g_{l})=m^{\dagger}_{l}m_{l}\,,
\end{equation}
where $m_{l}$, the charged lepton mass matrix, is defined using the convention $(E^{c})^{T}m_{l}E_{L}$. The representation $\bm{r}$ corresponds to faithful, irreducible three-dimensional representation of the finite group $\Delta(96)$. With $G_{l}$ defined, the general structure of $m^{\dagger}_{l}m_{l}$ is directly formulated using Eq.~\eqref{eq:ch_dia}. The unitary matrix $U_{l}$, satisfying $U^{\dagger}_{l}m^{\dagger}_{l}m_{l}U_{l}=\text{diag}(m^2_{e},m^2_{\mu},m^2_{\tau})$, governs the diagonalization of this hermitian product. As revealed by Eq.~\eqref{eq:ch_dia}, $U_{l}$ is determined through the eigenvalue decomposition relation:
\begin{equation}
\label{eq:Ul}U^{\dagger}_{l}\rho_{\bm{r}}(g_{l})U_{l}=\rho^{\text{diag}}_{\bm{r}}(g_{l})\,,
\end{equation}
where $\rho^{\text{diag}}_{\bm{r}}(g_{l})$ denotes the diagonal matrix containing the eigenvalues of $\rho_{\bm{r}}(g_{l})$ as its entries.

 In the atmospheric and solar neutrino sectors, the vacuum alignment directions $\bm{v}_{\text{atm}}$ and $\bm{v}_{\text{sol}}$ of flavons $\phi_{\text{atm}}$ and $\phi_{\text{sol}}$ must remain invariant under their respective residual symmetry groups $G_{\text{atm}}\rtimes H^{\text{atm}}_{CP}$ and $G_{\text{sol}}\rtimes H^{\text{sol}}_{CP}$, respectively. This requirement translates to the constraints:
\begin{equation}\label{eq:nu_consis_vev}
\rho_{\bm{r}}(g^{\text{atm}})\bm{v}_{\text{atm}}=\bm{v}_{\text{atm}}, \quad X^{\text{atm}}_{\bm{r}}\bm{v}_{\text{atm}}^*=\bm{v}_{\text{atm}}\,, \qquad \rho_{\bm{r}}(g^{\text{sol}})\bm{v}_{\text{sol}}=\bm{v}_{\text{sol}}, \quad X^{\text{sol}}_{\bm{r}}\bm{v}_{\text{sol}}^*=\bm{v}_{\text{sol}}\,,
\end{equation}
which determine the structures of $\bm{v}_{\text{atm}}$ and $\bm{v}_{\text{sol}}$. Notably, the cross product vector
\begin{equation}
\bm{v}_{\text{fix}}\equiv \bm{v}_{\text{atm}}\times \bm{v}_{\text{sol}}
\end{equation}
corresponds to the zero eigenvalue $m_{1}$=0 of the light neutrino mass matrix $m_{\nu}$ in the NO case, and to $m_{3}$=0 in the IO case.
 Hence, the normalized version $\bm{\hat{v}}_{\text{fix}}\equiv \bm{v}_{\text{fix}}/\sqrt{\bm{v}^{\dagger}_{\text{fix}} \bm{v}_{\text{fix}}}$ corresponds to the first column of the diagonalizing matrix $U_{\nu}$ satisfying $U^{T}_{\nu}m_{\nu}U_{\nu}=\text{diag}(0,m_{2},m_{3})$ for the NO case and the third column of the diagonalizing matrix $U_{\nu}$ satisfying $U^{T}_{\nu}m_{\nu}U_{\nu}=\text{diag}(m_{1},m_{2},0)$ for the IO case. As a consequence, the first (third) column of the PMNS matrix is fixed to be $U^{\dagger}_{l}\bm{\hat{v}}_{\text{fix}}$ for the NO (IO) case.

To diagonalize the neutrino mass matrix $m_{\nu}$  in Eq.~\eqref{eq:mnu} and find the general form of  $U_{\nu}$, the remaining two columns  of $U_{\nu}$ must be determined. These columns are orthogonal to  $\bm{\hat{v}}_{\text{fix}}$. The light neutrino mass matrix $m_{\nu}$ can be block diagonalized by performing the following unitary transformation
\begin{equation}\label{eq:Unu1}
	U_{\nu1}=\left\{\begin{array}{lll}{(\boldsymbol{\hat{v}}_{\mathrm{fix}},\boldsymbol{\hat{v}}_{\mathrm{atm}}^{*},\boldsymbol{\hat{v}}_{\mathrm{sol}}^{\prime})}&{\mathrm{for}}&{\mathrm{NO}\,,}\\{(\boldsymbol{\hat{v}}_{\mathrm{atm}}^{*},\boldsymbol{\hat{v}}_{\mathrm{sol}}^{\prime},\boldsymbol{\hat{v}}_{\mathrm{fix}})}&{\mathrm{for}}&{\mathrm{IO}\,.}\end{array}\right.
\end{equation}
with
\begin{equation}
\bm{\hat{v}}_{\text{atm}}=\frac{ \bm{v}_{\text{atm}}}{\sqrt{\bm{v}^{\dagger}_{\text{atm}} \bm{v}_{\text{atm}}}},\qquad
\bm{\hat{v}}^\prime_{\text{sol}}=\bm{\hat{v}}^*_{\text{fix}}\times \bm{\hat{v}}_{\text{atm}}\,.
\end{equation}
 Note that $U_{\nu1}$  in Eq.~\eqref{eq:Unu1} is non-unique: when $\bm{v}^\dagger_{\text{sol}}\bm{v}_{\text{atm}}\neq0$, for the NO case, it may include a (23) rotation, while for the IO case, it may include a (12) rotation. Consequently, the neutrino mass matrix $m_{\nu}$ can be reduced to a simpler structure:
\begin{equation}\label{eq:mnup}
	m_\nu^{\prime}=U_{\nu1}^Tm_\nu U_{\nu1}=\begin{cases}\begin{pmatrix}0&0&0\\0&y&z\\0&z&w\end{pmatrix}&\mathrm{for}\quad\mathrm{NO\,,}\\\begin{pmatrix}y&z&0\\z&w&0\\0&0&0\end{pmatrix}&\mathrm{for}\quad\mathrm{IO\,.}\end{cases}
\end{equation}	
where the expressions of the parameters $y$, $z$ and $w$ are
\begin{eqnarray}
\nonumber y&=& m_{a}\left[\bm{v}^{\dagger}_{\text{atm}} \bm{v}_{\text{atm}}+re^{i\eta}\left(\bm{\hat{v}}^{\dagger}_{\text{atm}} \bm{v}_{\text{sol}}\right)^2\right]\,, \\
\nonumber  z&=& m_{a}re^{i\eta}\sqrt{\left(\bm{\hat{v}}_{\text{atm}}\times \bm{v}_{\text{sol}}\right)^\dagger \left(\bm{\hat{v}}_{\text{atm}}\times \bm{v}_{\text{sol}}\right)}\left(\bm{\hat{v}}^{\dagger}_{\text{atm}} \bm{v}_{\text{sol}}\right)\,, \\
\label{eq:yzw}  w&=& m_{a}re^{i\eta}\left(\bm{\hat{v}}_{\text{atm}}\times \bm{v}_{\text{sol}}\right)^\dagger \left(\bm{\hat{v}}_{\text{atm}}\times \bm{v}_{\text{sol}}\right)\,.
\end{eqnarray}
The light neutrino mass matrix $m_\nu^{\prime}$ can be diagonalized by the unitary matrix $U_{\nu2}$ following standard methods detailed in Refs.~\cite{Ding:2013bpa,Ding:2018fyz,Ding:2018tuj}. The general parametrization of $U_{\nu2}$ is expressed as:
\begin{equation}\label{eq:Unu2}
	U_{\nu2}=\left\{\begin{array}{ccc}\begin{pmatrix}1&0&0\\0&\cos\theta e^{i(\psi+\rho)/2}&\sin\theta e^{i(\psi+\sigma)/2}\\0&-\sin\theta e^{i(-\psi+\rho)/2}&\cos\theta e^{i(-\psi+\sigma)/2}\end{pmatrix}&\mathrm{for}&\mathrm{NO\,,}\\\\\begin{pmatrix}\cos\theta e^{i(\psi+\rho)/2}&\sin\theta e^{i(\psi+\sigma)/2}&0\\-\sin\theta e^{i(-\psi+\rho)/2}&\cos\theta e^{i(-\psi+\sigma)/2}&0\\0&0&1\end{pmatrix}&\mathrm{for}&\mathrm{IO\,.}\end{array}\right.
\end{equation}
For the NO (IO) case, the nonzero neutrino masses $m_{2}$ and $m_{3}$ ($m_{1}$ and $m_{2}$), along with the parameters $\theta$, $\psi$, $\rho$ and $\sigma$ are analytically determined by the parameters $y$, $z$ and $w$ in the simplified mass matrix $m_\nu^{\prime}$, as established in Refs.~\cite{Ding:2013bpa,Ding:2018fyz,Ding:2018tuj}, and the explicit expressions of them can be found in appendix~\ref{sec:Diag_mnup}.

In short, within the residual mixing pattern $\{G_{l},G_{\text{atm}}\rtimes H^{\text{atm}}_{CP},G_{\text{sol}}\rtimes H^{\text{sol}}_{CP}\}$, the explicit forms of $U_{l}$, $\bm{v}_{\text{atm}}$ and $\bm{v}_{\text{sol}}$ can be straightforwardly determined.  However, due to the undetermined charged lepton mass hierarchy in the tri-direct CP approach, the PMNS matrix remains defined only up to independent row permutations. Consequently, the lepton mixing matrix for this residual symmetry pattern is resolved modulo row reordering.  Thus the general expression for the mixing matrix is derived as
\begin{equation} \label{eq:genral_UPMNS}
U_{PMNS}=Q_{l}P_{l}U_{l}^{\dagger}U_{\nu1}U_{\nu2} ,
\end{equation}
where the unphysical phase matrix $Q_{l}$ can be absorbed by the charged lepton fields, and $P_{l}$ denotes a permutation matrix with six distinct forms:
\begin{equation}\label{eq:permutation_matrices}
\begin{array}{lll}
P_{123}=\begin{pmatrix}
1  &~ 0  ~&  0 \\
0  &~ 1  ~&  0\\
0  & ~0~  &  1
\end{pmatrix},~~&~~ P_{132}=\begin{pmatrix}
1  &  ~0~ &  0 \\
0  &  ~0~ &  1 \\
0  &  ~1~ &  0
\end{pmatrix},~~&~~ P_{213}=\begin{pmatrix}
0  &  ~1~  &  0 \\
1  &  ~0~  &  0 \\
0  &  ~0~  &  1
\end{pmatrix},\\
& & \\[-10pt]
P_{231}=\begin{pmatrix}
0   &  ~1~   &  0 \\
0   &  ~0~   &  1  \\
1   &  ~0~   &  0
\end{pmatrix},~~&~~ P_{312}=\begin{pmatrix}
0   &  ~0~  &   1  \\
1   &  ~0~  &   0 \\
0   &  ~1~  &  0
\end{pmatrix},~~&~~ P_{321}=\begin{pmatrix}
0    &   ~0~    &   1  \\
0    &   ~1~    &   0  \\
1    &   ~0~    &   0
\end{pmatrix}\,.
\end{array}
\end{equation}
For subsequent analysis, only the permutation yielding mixing parameters consistent with experimental data is retained. In the present work, the standard parametrization of the PMNS matrix follows the convention in Ref.~\cite{ParticleDataGroup:2024cfk},
\begin{equation}\label{eq:PMNS_def}
U_{PMNS}=\begin{pmatrix}
c_{12}c_{13}  &   s_{12}c_{13}   &   s_{13}e^{-i\delta_{CP}}  \\
-s_{12}c_{23}-c_{12}s_{13}s_{23}e^{i\delta_{CP}}   &  c_{12}c_{23}-s_{12}s_{13}s_{23}e^{i\delta_{CP}}  &  c_{13}s_{23}  \\
s_{12}s_{23}-c_{12}s_{13}c_{23}e^{i\delta_{CP}}   & -c_{12}s_{23}-s_{12}s_{13}c_{23}e^{i\delta_{CP}}  &  c_{13}c_{23}
\end{pmatrix}\text{diag}(1,e^{i\frac{\beta}{2}},1)\,,
\end{equation}
where $c_{ij}\equiv \cos\theta_{ij}$, $s_{ij}\equiv \sin\theta_{ij}$, $\delta_{CP}$ is the Dirac CP violation phase and $\beta$ is the Majorana CP phase. For quantitative characterization of CP violation, two basis independent quantities are introduced: $J_{CP}$~\cite{Jarlskog:1985ht} corresponding to the Dirac phase and $I_1$~\cite{Branco:1986gr,Nieves:1987pp,Nieves:2001fc,Jenkins:2007ip,Branco:2011zb} associated with the Majorana phase, defined respectively by:
\begin{eqnarray}
		\nonumber  J_{CP}&=&\Im{(U_{11}U_{33}U^{*}_{13}U^{*}_{31})}=\frac{1}{8}\sin2\theta_{12}\sin2\theta_{13}\sin2\theta_{23}\cos\theta_{13}\sin\delta_{CP}\,, \\
		\label{eq:CP_invariants} I_{1}&=&\left\{\begin{array}{lll}\Im{(U^{2}_{12}U^{*\,2}_{13})}=\frac{1}{4}\sin^2\theta_{12}\sin^22\theta_{13}\sin(\beta+2\delta_{CP}) & \text{for } & \text{NO} \,,\\
		\Im{(U^{2}_{12}U^{*\,2}_{11})}=\frac{1}{4}\cos^4\theta_{13}\sin^22\theta_{12}\sin\beta & \text{for} & \text{ IO} \,.\end{array}\right.
\end{eqnarray}

\section{\label{sec:lepton_pred}Lepton mixing predictions from $\Delta(96)\rtimes H_{CP}$}	

This section conducts a comprehensive model independent analysis of viable lepton mixing patterns within the framework of $\Delta(96)\rtimes H_{CP}$ by using the tri-direct CP approach. Through systematic scanning of all possible remnant symmetry combinations $\{G_{l},G_{\text{atm}}\rtimes H^{\text{atm}}_{CP},G_{\text{sol}}\rtimes H^{\text{sol}}_{CP}\}$,  the corresponding resulting predictions for lepton mixing parameters and neutrino masses are obtained. Unlike model dependent approaches requiring symmetry breaking mechanisms via charged scalar fields, our method focuses solely on residual symmetry structures. The PMNS matrix emerges from misalignment among the three residual symmetries, bypassing explicit vacuum alignment dynamics.

We shall restrict ourselves to working with the three left-handed lepton doublet generations $L$ assigned to a three-dimensional faithful irreducible representation of $\Delta(96)$. Among the four distinct three-dimensional faithful irreducible representations ($\bm{3_{m}}$ and their complex conjugates $\bm{\bar{3}_{m}}$ with $m=0,1$, as shown in appendix~\ref{sec:Delta96_group_theory}), we specifically adopt the $\bm{3_{0}}$ representation for lepton doublet unification. Notably, equivalent physical predictions would emerge when using the representation $\bm{3_{1}}$ instead, as these representations differ solely through an overall sign of the generator $b$. For conjugate representations $\bm{\bar{3}_{m}}$, the resulting lepton mass matrices and mixing matrices would manifest as complex conjugates of those derived from $\bm{3_{m}}$. The theoretical framework assumes that right-handed neutrinos $N^{c}_{\text{atm}}$ and $N^{c}_{\text{sol}}$ transform as singlet representations $\bm{1_{m}}$ and $\bm{1_{n}}$ with $m,n=0,1$, respectively. To construct $\Delta(96)$ invariant Yukawa couplings through the contractions  $\left(L\phi_{\text{atm}}\right)_{\bm{1_{m}}}$ and $\left(L\phi_{\text{sol}}\right)_{\bm{1_{n}}}$, the flavon fields $\phi_{\text{atm}}$ and $\phi_{\text{sol}}$ must transform under conjugate triplet representations $\bm{\bar{3}_{m}}$ and $\bm{\bar{3}_{n}}$, respectively. This representation pairing ensures the formation of $\Delta(96)$ contractions through tensor product combinations $\bm{3_{m}}\otimes \bm{\bar{3}_{n}}= \bm{1_{[m+n]}}\oplus\bm{2}\oplus\bm{6}$.

Let us systematically analyze the distinct breaking patterns arising from the breaking of the flavor symmetry group $\Delta(96)\rtimes H_{CP}$ within the tri-direct CP framework. In order to  distinguish the three generations in the charged lepton sector, the order of the residual flavor symmetry $G_{l}$ in the charged lepton sector can not be smaller than 3. This constraint allows $G_{l}$ to correspond to any of the 41 subgroups $Z_{3}$, $K_{4}$, $Z_{4}$ and $Z_{8}$ of $\Delta(96)$. In the neutrino sector, the vacuum alignments of $\phi_{\text{atm}}$ and $\phi_{\text{sol}}$ must preserve distinct residual symmetries $G_{\text{atm}}\rtimes H^{\text{atm}}_{CP}$ and $G_{\text{sol}}\rtimes H^{\text{sol}}_{CP}$ for atmospheric and solar neutrino mass terms, respectively. Here, $G_{\text{atm}}$ and $G_{\text{sol}}$ represent those Abelian subgroups whose generator matrices must contain eigenvalue 1 in either representations $\bm{\bar{3}_{0}}$ or $\bm{\bar{3}_{1}}$. Notably, exchanging the residual symmetry pairs $G_{\text{atm}}\rtimes H^{\text{atm}}_{CP}$ and $G_{\text{sol}}\rtimes H^{\text{sol}}_{CP}$ induces column permutation between the first and second generations in the Dirac mass matrix $M_{D}$.  However, this exchange leaves the overall neutrino mass matrix $m_{\nu}$ invariant, rendering the resulting mixing patterns physically indistinguishable\footnote{This equivalence arises because exchanging Yukawa couplings between $N^{c}_{\text{atm}}$ and $N^{c}_{\text{sol}}$ can be compensated by swapping their Majorana masses $M_{1}$ and $M_{2}$, maintaining identical low-energy phenomenology.}. Furthermore, if a pair of residual flavor symmetries $\{G^{\prime}_{l},G^{\prime}_{\text{atm}}, G^{\prime}_{\text{sol}}\}$  is conjugated to the pair of groups $\{G_{l},G_{\text{atm}}, G_{\text{sol}}\}$ under an element of $\Delta(96)$, i.e.,
\begin{equation}\label{eq:conjugate}
G^{\prime}_{l}=hG_{l}h^{-1},\qquad G^{\prime}_{\text{atm}}=hG_{\text{atm}}h^{-1}, \qquad G^{\prime}_{\text{sol}}=hG_{\text{sol}}h^{-1}, \qquad h\in \Delta(96)\,,
\end{equation}
then these two breaking patterns will yield identical phenomenological predictions for mixing parameters~\cite{Ding:2013bpa,Li:2014eia,Ding:2018tuj}. Therefore, it is sufficient  to analyze only a select few representative remnant symmetries that produce distinct outcomes for lepton mixing parameters and neutrino masses.

As outlined in appendix~\ref{sec:Delta96_group_theory}, all the $Z_{3}$ and $Z_{8}$ subgroups of $\Delta(96)$ are conjugate to each other. In contrast, the twelve $Z_{4}$ subgroups separate into three distinct conjugacy classes under similarity transformations within $\Delta(96)$. Significantly, the representation matrices of the generators of the three $Z_{4}$ subgroups $Z_{4}^{cd^2}$, $Z_{4}^{cd^3}$ and $Z_{4}^{c^2d^3}$ manifest eigenvalue degeneracy, making them incompatible with residual symmetry assignments in the charged lepton sector. The Klein subgroup $K^ {(c^2,d^2)} _{4} $ is a normal subgroup, whereas all other $K_{4}$ subgroups belong to a single conjugacy class. Consequently, only specific residual symmetry combinations need to be considered:
\begin{equation}
 G_{l}\in\{Z_{3}^{ac},~Z_{4}^{c},~Z_{4}^{abd^2},~Z_{8}^{abd},~K_{4}^{(c^2,d^2)},~K_{4}^{(abc,c^2)}\}\,,
\end{equation}
while $G_{\text{atm}} $ and $G_{\text{sol}}$ remain unrestricted and may adopt any of the 56 Abelian subgroups of $\Delta(96)$ under exchange equivalence. Importantly, the charged lepton diagonalization matrices arising from the Klein subgroups $K_{4}^{(c^2,d^2)}$ and $K_{4}^{(abc,c^2)}$ exhibit identical structures to those derived from $Z_{4}^{c}$ and $Z_{4}^{abd^2}$, respectively. This equivalence ultimately reduces the independent mixing patterns to those generated by the subset:
\begin{equation}
 G_{l}\in\{Z_{3}^{ac},~Z_{4}^{c},~Z_{4}^{abd^2},~Z_{8}^{abd}\}\,.
\end{equation}
When $G_{l}$  is assumed to be any of these four subgroup types, the diagonalization matrix $U_{l}$ of the hermitian combination $m^{\dagger}_{l}m_{l}$  can be uniquely determined (up to column vector permutations and phase redefinitions) as:
\begin{eqnarray}
\nonumber
U^{(1)}_{l}&=&\frac{1}{\sqrt{3}}\begin{pmatrix}
	-i \omega  & -i \omega ^2 & -i \\
	\omega ^2 & \omega  & 1 \\
	1 & 1 & 1 \end{pmatrix} \qquad \qquad
\text{for} \qquad G_{l}=Z_{3}^{ac} \,,\\
\nonumber
U^{(2)}_{l}&=&
\begin{pmatrix}
	1 & 0 & 0 \\
	0 & 1 & 0 \\
	0 & 0 & 1 \\
\end{pmatrix}
\qquad \qquad \qquad \qquad \quad ~\text{for}\qquad  G_{l}=Z_{4}^{c}\,, \\
\nonumber
U^{(3)}_{l}&=&\frac{1}{\sqrt{2}}\begin{pmatrix}
	i & -i & 0 \\
	1 & 1 & 0 \\
	0 & 0 & \sqrt{2} \\
\end{pmatrix}\qquad \qquad \qquad 
\text{for}\qquad G_{l}=Z_{4}^{abd^2}\,, \\
\label{eq:4Ul}U^{(4)}_l&=&\begin{pmatrix}
	\frac{e ^{\frac{\pi i}{4}}}{\sqrt{2}} & \frac{e ^{-\frac{3 \pi  i}{4}}}{\sqrt{2}} & 0 \\
	\frac{1}{\sqrt{2}} & \frac{1}{\sqrt{2}} & 0 \\
	0 & 0 & 1 \\
\end{pmatrix}\qquad \qquad \qquad ~
\text{for}\qquad G_{l}=Z_{8}^{abd}\,,
\end{eqnarray}
where the parameter $\omega$ is the cube root of unit with $\omega=e^{\frac{2\pi i}{3}}$.

In this framework, we examine scenarios where the flavon field $\phi_{\text{atm}}$  is assigned to the triplet representation $\bm{\bar{3}_{0}}$, while $\phi_{\text{sol}}$ may transform as either $\bm{\bar{3}_{0}}$ or $\bm{\bar{3}_{1}}$.  The most general VEVs for these flavons, consistent with residual symmetry preservation, are cataloged in table~\ref{tab:invariant_VEVs}. Notably, certain compatible residual CP transformations are omitted in the table. This simplification arises because invariant vacuum alignments for unlisted CP phases differ only by an overall factor of $e^{\pm i\frac{\pi}{4}}$ or $i$.  Such phase contributions are rendered physically inconsequential through parameter redefinitions $\eta\to\eta\pm\frac{\pi}{2}$ or $\eta+\pi$. For both $\phi_{\text{atm}}$ and $\phi_{\text{sol}}$ assigned to the $\bm{\bar{3}_{0}}$ representation, their VEVs remain independent of the real parameter $x$. Consequently, lepton mixing parameters and neutrino masses rely solely on three input parameters $|m_{a}|$, $r$ and $\eta$. When $\phi_{\text{sol}}$ transforms as $\bm{\bar{3}_{1}}$, however, its VEV introduces an additional real parameter $x$, expanding the dependence to four variables $|m_{a}|$, $r$, $\eta$ and $x$. Then we shall systematically analyze viable symmetry breaking patterns from these two assignments, subsequently evaluating their predictions for neutrino masses and mixing parameters.

Note that if both $\phi_{\text{atm}}$ and $\phi_{\text{sol}}$ are assigned to the $\bm{\bar{3}_{0}}$ representation, then the right-handed neutrinos $N_{\text{atm}}^{c}$ and $N_{\text{sol}}^{c}$ must both transform as $\bm{1_{0}}$ under $\Delta(96)$. This assignment, however, necessitates introducing an additional $Z_{n}$ ($n \geq 2$) symmetry to forbid the unwanted cross term $N_{\text{atm}}^{c}N_{\text{sol}}^{c}$. The compensating advantage of this setup is that it requires only three input parameters, eliminating the need for the additional parameter $x$. When $\phi_{\text{atm}}$ and $\phi_{\text{sol}}$ are assigned to $\bm{\bar{3}_{0}}$ and $\bm{\bar{3}_{1}}$ respectively, the right-handed neutrinos $N_{\text{atm}}^{c}$ and $N_{\text{sol}}^{c}$ must correspondingly transform as $\bm{1_{0}}$ and $\bm{1_{1}}$. In this case, the cross term vanishes naturally, eliminating the need for an additional symmetry to distinguish the two fields. The main drawback, however, is the requirement of an additional real parameter $x$.

\subsection{Lepton mixing patterns for   $\phi_{\text{atm}},\phi_{\text{sol}}\sim\mathbf{\bar{3}_{0}}$}

\begin{table}[t!]
\centering
\renewcommand{\arraystretch}{1.2}
\begin{tabular}{|c|c|c||c|c|c|c|c|}
\hline \hline
\multirow{2}{*}{Observables}  &  	\multicolumn{2}{c||}{NO}   &      \multicolumn{2}{c|}{IO}       \\ \cline{2-3} \cline{4-5}
	
& $\text{bf}\pm1\sigma$  & $3\sigma$ region & $\text{bf}\pm1\sigma$ & $3\sigma$ region   \\ \hline

&   &  &  &    \\[-0.150in]
	
$\sin^2\theta_{13}$ & $0.02215^{+0.00056}_{-0.00058}$ & $[0.02030,0.02388]$ & $0.02231^{+0.00056}_{-0.00056}$ &  $[0.02060,0.02409]$  \\ [0.050in]
	
$\sin^2\theta_{12}$ & $0.308^{+0.012}_{-0.011}$ & $[0.275,0.345]$ & $0.308^{+0.012}_{-0.011}$ & $[0.275,0.345]$  \\ [0.050in]
	
$\sin^2\theta_{23}$  & $0.470^{+0.017}_{-0.013}$  & $[0.435,0.585]$ & $0.550^{+0.012}_{-0.015}$  & $[0.440,0.584]$   \\ [0.050in]
	
$\delta_{CP}/\pi$  & $1.178^{+0.144}_{-0.228}$ & $[0.689,2.022]$  & $1.522^{+0.122}_{-0.139}$ & $[1.117,1.861]$  \\ [0.050in]

$\frac{\Delta m^2_{21}}{10^{-5}\text{eV}^2}$ & $7.49^{+0.19}_{-0.19}$ & $[6.92,8.05]$ & $7.49^{+0.19}_{-0.19}$ & $[6.92,8.05]$  \\ [0.050in]

$\frac{\Delta m^2_{3\ell}}{10^{-3}\text{eV}^2}$ & $2.513^{+0.021}_{-0.019}$ & $[2.451,2.578]$ & $-2.484^{+0.020}_{-0.020}$ & $[-2.547,-2.421]$ \\ [0.050in]

$\Delta m^2_{21}/\Delta m^2_{3\ell}$  &  $0.0298^{+0.00079}_{-0.00079}$  & $[0.0268,0.0328]$ & $-0.0302^{+0.00080}_{-0.00080}$ & $[-0.0333,-0.0272]$  \\ \hline \hline
\end{tabular}
\caption{\label{tab:bf_13sigma_data} The global best fit values, $1\sigma$ ranges and $3\sigma$ ranges of mixing parameters and lepton mass ratios. Here the experimental data and errors of the lepton mixing parameters and neutrino masses  are obtained from NuFIT 6.0 with Super-Kamiokande atmospheric data~\cite{Esteban:2024eli}. Note that the notation $\Delta m^2_{3\ell}$ denotes $\Delta m^2_{31} > 0$ for the NO and $\Delta m^2_{32} < 0$ for the IO.} 
\end{table}

To systematically identify unique viable symmetry breaking patterns within the tri-direct CP approach, we first establish all independent combinations of the residual symmetry structure $\{G_{l},G_{\text{atm}}\rtimes H^{\text{atm}}_{CP},G_{\text{sol}}\rtimes H^{\text{sol}}_{CP}\}$. These combinations must exclude redundancies arising from group conjugation (as defined in Eq.~\eqref{eq:conjugate}) and permutations between the atmospheric and solar neutrino sector symmetries. To perform a quantitative assessment of how well each breaking pattern aligns with neutrino oscillation data and mass parameters~\cite{Esteban:2024eli}, we define a $\chi^2$ function that quantifies the discrepancy between theoretical predictions and experimental results:
\begin{equation}\label{eq:chisq}
\chi^2 = \sum_{i=1}^6 \left( \frac{P_i(r, \eta)-O_i}{\sigma_i}\right)^2\,,
\end{equation}
where the observables $O_{i}\pm\sigma_{i}$ represent the global best fit values and $1\sigma$ uncertainties for the six observable quantities including the three mixing angles $\sin^2\theta_{12}$, $\sin^2\theta_{13}$ and $\sin^2\theta_{23}$, the Dirac CP phase $\delta_{CP}$ and the mass squared differences $\Delta m^2_{21}$, $\Delta m^2_{31}$ ($\Delta m^2_{32}$) for the NO (IO) case. In Eq.~\eqref{eq:chisq}, $P_{i}$ represents the theoretical predictions for those six observable quantities which are formulated as functions of the input parameters $m_{a}$, $r$ and $\eta$. For each $(m_{a}, r,\eta)$ parameter set, we compute both the predicted observables $P_{i}$ and the associated $\chi^{2}$ value. A comprehensive $\chi^{2}$ minimization analysis has been conducted across the parameter space.

\begin{figure}[t!]
\centering
\begin{tabular}{c}
\includegraphics[width=0.48\linewidth]{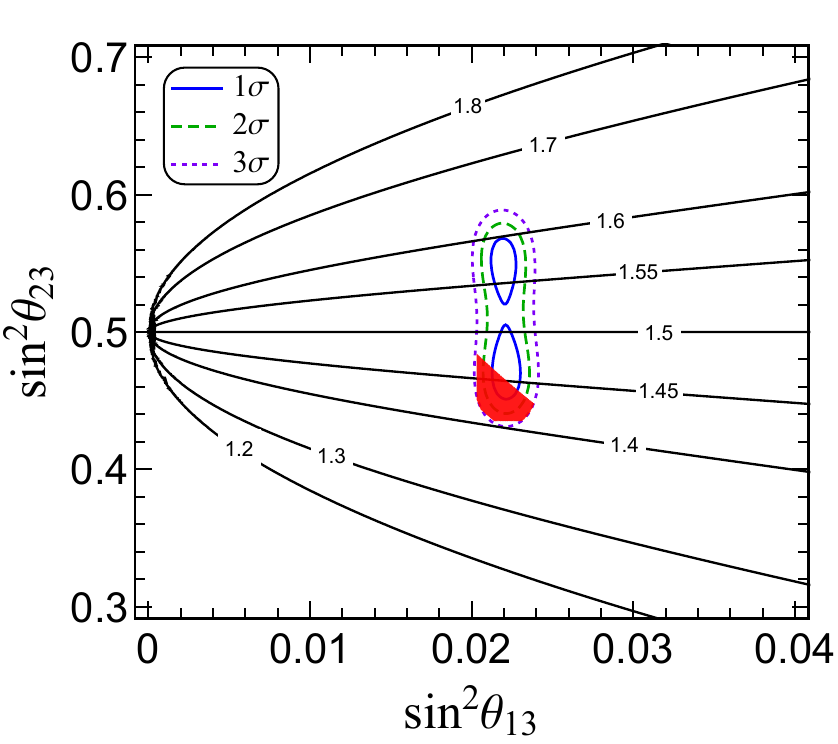}
\includegraphics[width=0.5\linewidth]{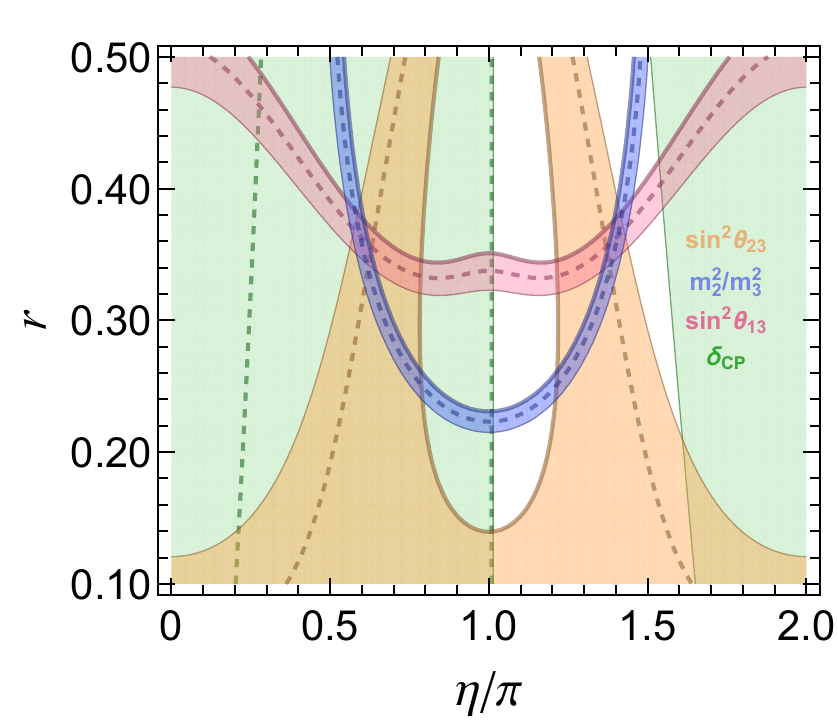}
\end{tabular}
\caption{\label{fig:contour_CP} Contour plots of $\delta_{CP}/\pi$ within $\sin^2\theta_{13}-\sin^2\theta_{23}$ plane, and the dimensionless observable quantities $\sin^{2}\theta_{23}$, $\sin^{2}\theta_{13}$, $\delta_{CP}$ and $m_{2}^2/m_{3}^2$ in the $\eta$/$\pi-r$ plane. On the left panel,  the results arise from applying the sum rule outlined in Eq.~\eqref{eq:delta_CP_sumrule1}. The red regions show allowed ranges of $\sin^2\theta_{13}$ and $\sin^2\theta_{23}$, calculated from random variations of input parameters $r$ and $\eta$ while ensuring agreement with experimental $3\sigma$ ranges for mixing angles, the Dirac CP phase, and  $\Delta m^2_{21}/\Delta m^2_{31}$ in table~\ref{tab:bf_13sigma_data}.  On the right panel, the orange, blue, claret and green regions represent the 3$\sigma$ ranges of $\sin^{2}\theta_{23}$, $m_{2}^2/m_{3}^2$, $\sin^{2}\theta_{13}$ and $\delta_{CP}$ respectively. The dashed lines denote the best fit values of them.}

\end{figure}

After performing the $\chi^2$ analysis for all possible independent breaking patterns, we find only one phenomenologically interesting mixing pattern. The representative combination of residual flavor symmetries is $\{G_{l},G_{\text{atm}},G_{\text{sol}}\}=(Z_{3}^{ac},Z_{2}^{a^2b},Z_{3}^{ad^2})$, and both the atmospheric and solar neutrino sectors exhibit trivial residual CP transformations represented by the identity matrix $\mathbb{1}_{3}$. In this mixing pattern, the diagonalization matrix of charged lepton mass matrix $U_{l}$, the VEVs $\langle\phi_{\text{atm}}\rangle$ and $\langle\phi_{\text{sol}}\rangle$ are taken to be 
\begin{equation}\label{eq:mix_vev}
  U_{l}=
  \frac{1}{\sqrt{3}}\begin{pmatrix}
	-i  & -i \omega ^2 & -i \omega \\
	1 & \omega  & \omega^2 \\
	1 & 1 & 1 \end{pmatrix}, \qquad
  \langle\phi_{\text{atm}}\rangle=\left(\begin{array}{c}0 \\1 \\ -1 \end{array}\right)v_{\phi_a}\,, \quad \langle\phi_{\text{sol}}\rangle=\left(\begin{array}{c}1\\ -1 \\ 1\end{array}\right)v_{\phi_s}\,,
\end{equation}
where both $v_{\phi_a}$ and $v_{\phi_s}$ are real. Employing the standard seesaw mechanism, the mass matrix for light neutrinos $m_{\nu}$ can be expressed as:
\begin{equation}
m_{\nu}= m_{a} \left[
\begin{pmatrix}
0 & 0 & 0 \\
0 & 1 & -1 \\
0 & -1 & 1 \\
\end{pmatrix}
+ re^{i \eta }
\begin{pmatrix}
1 & -1 & 1 \\
-1 & 1 & -1 \\
1 & -1 & 1 \\
\end{pmatrix}\right]\,.
\end{equation}
Following the methodology outlined in section~\ref{sec:Tridirect_approach}, we implement a unitary transformation $U_{\nu1}$ to the neutrino mass matrix with
\begin{equation} \label{eq:Uun1_model1}
U_{\nu1}=	\begin{pmatrix}
0  & 1 & 0  \\
-\frac{1}{\sqrt{2}} & 0 & \frac{1}{\sqrt{2}}   \\
-\frac{1}{\sqrt{2}} & 0 & -\frac{1}{\sqrt{2}}   \\
\end{pmatrix}\,.
\end{equation}
This transformation modifies the mass matrix to $m_{\nu}^{\prime}=U^{T}_{\nu1}m_{\nu}U_{\nu1}$, which is a block diagonal structure populated by nonzero parameters:
\begin{equation}
y= m_{a}re^{i \eta }, \qquad z=-\sqrt{2} m_{a} re^{i \eta } ,\qquad w= 2m_{a} (1+re^{i \eta })\,.
\end{equation}
As detailed in appendix~\ref{sec:Diag_mnup}, the transformed matrix $m_{\nu}^{\prime}$ subsequently undergoes diagonalization through another unitary matrix $U_{\nu2}$. The resulting neutrino masses $m_{2}$, $m_{3}$, along with the explicit form of $U_{\nu2}$, are directly derivable by substituting parameters $y$, $z$ and $w$ into the framework established by Eqs.~\eqref{eq:diag_2nd} and \eqref{eq:prs_U2}.  As a result, we can read out the lepton mixing matrix as
\begin{eqnarray} \label{eq:PMNS_model1}
U_{\text{PMNS}}=\begin{pmatrix}
\sqrt{\frac{2}{3}} & \frac{ \cos \theta }{\sqrt{3}} & \frac{\sin \theta e^{i\psi} }{\sqrt{3}} \\
-\frac{1}{\sqrt{6}} & \frac{ \cos\theta }{\sqrt{3}}-\frac{ e^{-i \psi } \sin\theta }{\sqrt{2}} & \frac{ e^{i \psi } \sin \theta }{\sqrt{3}}+\frac{ \cos \theta }{\sqrt{2}} \\
-\frac{1}{\sqrt{6}} & \frac{ \cos \theta }{\sqrt{3}}+\frac{ e^{-i \psi } \sin \theta }{\sqrt{2}} & \frac{ e^{i \psi } \sin \theta }{\sqrt{3}}-\frac{ \cos \theta }{\sqrt{2}} \\
\end{pmatrix}P_{\nu}\,,
\end{eqnarray}
where  an overall phase of each row has been absorbed by the charged lepton fields, and the diagonal phase matrix $P_\nu$ is defined as $P_{\nu}=\text{diag}(1, -ie^{i(\psi+\rho)/2}, -ie^{i(-\psi+\sigma)/2})$. We see that the first
column of the mixing matrix is in common with that of the tri-bimaximal mixing matrix, and the so-called TM1 mixing matrix is obtained. The three lepton mixing angles for this mixing matrix are
\begin{equation}
\sin^{2}\theta_{13}=\frac{\sin^{2}\theta}{3},\quad
\sin^{2}\theta_{12}=\frac{2\cos^{2}\theta}{5+\cos2\theta},
\quad\sin^{2}\theta_{23}=\frac{1}{2}+\frac{\sqrt{6}\cos\psi\sin2\theta}{5+\cos2\theta}\,.
\end{equation}
Obviously, the solar and the reactor mixing angles satisfy the following sum rule
\begin{equation}\label{eq:sum_rule_TM1_1}
\cos^2\theta_{12}\cos^2\theta_{13}=\frac{2}{3}\,,
\end{equation}
which holds true for all TM1 mixing matrices. The lepton mixing matrix in Eq.~\eqref{eq:PMNS_model1} yields explicit expressions for the CP  invariants:
\begin{equation}
J_{CP}=-\frac{\sin2\theta\sin\psi}{6\sqrt{6}},\qquad  I_{1 }=\frac{1}{36} \sin^{2}2 \theta\sin (\rho-\sigma) 
\end{equation}
Notably, the three mixing angles and the Jarlskog parameter $J_{CP}$ exhibit parametric dependence solely on $\theta$ and $\psi$, where these parameters are functions of the fundamental inputs $r$ and $\eta$. This relationship enables the Dirac CP phase $\delta_{CP}$ to be formulated explicitly through measurable mixing angles:
\begin{equation}\label{eq:delta_CP_sumrule1}
\cos\delta_{CP}=\frac{(3-5\cos2\theta_{13})\cot2\theta_{23}\csc\theta_{13}}{4 \sqrt{3\cos2\theta_{13}-1}}\,.
\end{equation}
Under maximal atmospheric mixing $\theta_{23}=\pi/4$, the vanishing cotangent term $\cot2\theta_{23}=0$ enforces $\cos\delta_{CP}=0$, thereby producing maximal CP violation with $\delta_{CP}=\pm\pi/2$. Figure~\ref{fig:contour_CP} displays the contour plot of $\delta_{CP}/\pi$ within the $\sin^2\theta_{13}-\sin^2\theta_{23}$ parameter plane. 
The red region corresponds to viable parameter space where input variables $r$ and $\eta$ are sampled within permissible ranges, subject to experimental constraints ($3\sigma$) on mixing angles,  Dirac CP phase and the ratio $\Delta m^2_{21}/\Delta m^2_{31}=m^{2}_{2}/m^{2}_{3}$. Remarkably, numerical simulations constrain $\delta_{CP}$ to a narrow interval near $1.45\pi$, demonstrating predictive power despite parametric freedom.

The absolute masses of neutrinos $m_2$ and $m_3$ demonstrate dependence on all three input parameters $m_{a}$, $\eta$ and $r$ in this breaking pattern. Through numerical optimization matching experimental constraints, the parameter configuration 
\begin{equation}
m_{a}=24.543\,\text{meV}, \qquad r=0.361, \qquad \eta=0.599\pi\,,
\end{equation}
emerges as the optimal solution, yielding the following theoretical predictions
\begin{eqnarray}
\nonumber &&\sin^2\theta_{13}=0.02195, \quad \sin^2\theta_{12}=0.318, \quad \sin^2\theta_{23}=0.450, \quad \delta_{CP}=1.429\pi,\quad \beta=1.631\pi\,, \\
\label{eq:best_fit_model1}&&m_1=0\,\text{meV}, \qquad m_2=8.685\,\text{meV}, \qquad m_3=50.114\,\text{meV}, \qquad m_{ee}=2.956\,\text{meV}\,,
\end{eqnarray}
where $m_{ee}$ refers to the neutrinoless double beta ($0\nu\beta\beta$) decay effective mass which is defined as
\begin{equation}
m_{ee}=|m_{1}U^2_{PMNS,11}+m_{2}U^2_{PMNS,12}+m_{3}U^2_{PMNS,13}|\,.
\end{equation}
The predictions for mixing parameters in Eq.~\eqref{eq:best_fit_model1} exhibit excellent consistency with current experimental measurements, with a minimum $\chi^2$ value of $\chi^2_{\text{min}}=5.545$.

Given the high flavor symmetry breaking scale, it is necessary in principle to account for renormalization group (RG) running down to the electroweak scale. Comprehensive treatments of the RG evolution of lepton mixing parameters and non-zero neutrino masses can be found in Refs.~\cite{Antusch:2005gp,King:2016yef,Geib:2017bsw}. These studies show that the corrections to the mixing parameters due to RG running are negligible, as they are significantly smaller than current experimental uncertainties. This behavior arises because, in the case of a massless lightest neutrino, the running lacks a strong enhancement mechanism. Therefore, our low-energy predictions remain robust against RG running effects.

Furthermore, we conduct an extensive numerical study by treating the input parameters $m_{a}$, $r$ and $\eta$ as randomly sampled real numbers within the intervals $[0,1]\text{eV}$, $[0,100]$ and $[0,2\pi]$, respectively. For each parameter set, we compute the lepton mixing parameters and neutrino masses. The computed values of the six observable quantities in table~\ref{tab:bf_13sigma_data}  are constrained to lie in their $3\sigma$ confidence regions derived from recent global analyses of neutrino oscillation data~\cite{Esteban:2024eli}. Our analysis reveals the following experimentally compatible parameter ranges $m_{a}/\text{meV}\in[23.600,25.840]$, $r\in[0.338,0.381]$ and $\eta/\pi\in[0.571,0.643]$. The model predicts tightly constrained neutrino mass and mixing parameters:
\begin{eqnarray}
\nonumber &&0.02030\leq \sin^2\theta_{13}\leq0.02388, \qquad 0.317\leq \sin^2\theta_{12}\leq0.320, \qquad  0.435\leq \sin^2\theta_{23}\leq0.484, \\
\nonumber &&1.406\leq\delta_{CP}/\pi\leq1.476, \qquad  1.578\leq\beta/\pi\leq1.654, \qquad  2.818\,\text{meV}\leq m_{ee}\leq3.098\,\text{meV}, \\
\nonumber &&8.319\,\text{meV}\leq m_2\leq8.972\,\text{meV}, \qquad 49.508\,\text{meV}\leq m_3\leq50.774\,\text{meV}, \\
\label{eq:model_prediction}&& 57.829\,\text{meV}\leq \sum_{i=1}^{3}m_{i}\leq59.745\,\text{meV}\,.
\end{eqnarray}
Moreover, we plot the contour regions for the $3\sigma$ intervals of mixing parameters $\theta_{13}$, $\theta_{23}$, $\delta_{CP}$ and  mass ratio $m^{2}_{2}/m^{2}_{3}$ in the plane $r$ and $\eta$ in figure~\ref{fig:contour_CP}. The result for $\theta_{12}$ is not displayed here, because it is related to the reactor angle $\theta_{13}$ by the TM1 mixing sum rule in Eq.~\eqref{eq:sum_rule_TM1_1}. The JUNO experiment~\cite{JUNO:2022mxj}, an upcoming medium-baseline reactor neutrino facility, is projected to constrain $\sin^2\theta_{12}$ within  $[0.3022, 0.3118]$ at $3\sigma$ after six years of operation. Complementary tests of $\sin^2\theta_{23}$ and $\delta_{CP}$ resolutions (anticipated ranges of $[0.462, 0.480]$ and $[1.099, 1.255]$ in radians after 15 years, respectively) will be enabled by long-baseline experiments DUNE~\cite{DUNE:2020ypp} and T2HK~\cite{Hyper-Kamiokande:2018ofw}. The predicted neutrino mass sum $\sum_{i=1}^{3} m_{i}$ aligns with cosmological constraints from Planck $+$ lensing $+$ BAO~\cite{Planck:2018vyg} and could be further probed by upcoming surveys like Euclid+CMB-S4+LiteBIRD~\cite{Euclid:2024imf}. The predicted effective Majorana mass $m_{ee}$ values fall below current KamLAND-Zen~\cite{KamLAND-Zen:2024eml}.Simultaneously, it also lies below the projected sensitivities of next-generation $0\nu\beta\beta$-decay experiments LEGEND-1000~\cite{LEGEND:2021bnm} and nEXO~\cite{nEXO:2021ujk}.  Furthermore, a very precise measurement of the lepton mixing parameters and neutrino masses might provide a good opportunity for probing this model.

\begin{figure}[t!]
\centering
\begin{tabular}{c}
\includegraphics[width=0.5\linewidth]{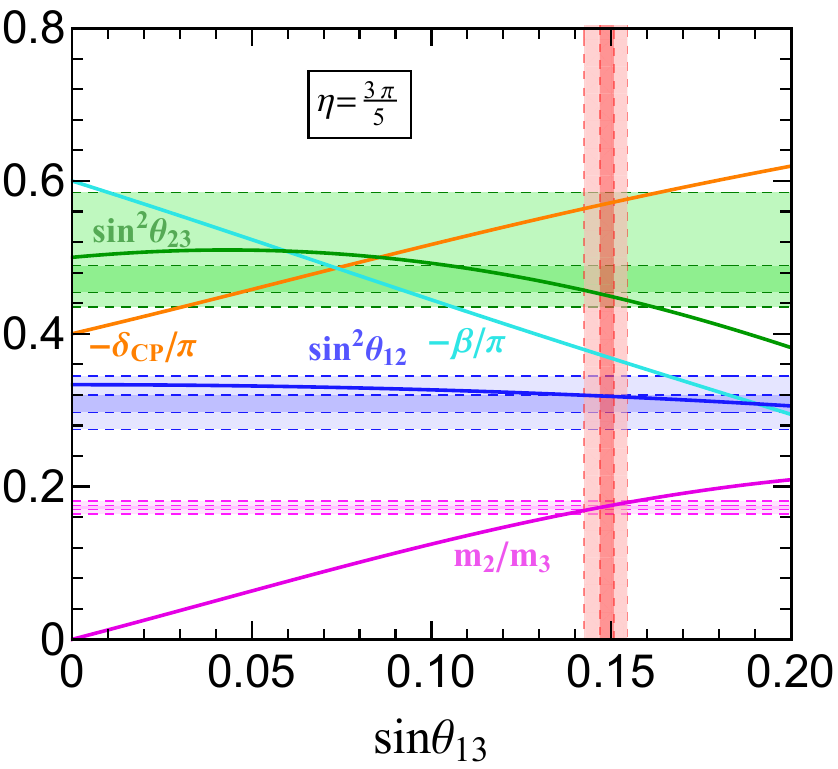}
\end{tabular}
\caption{\label{fig:Mix_par_M1} The predicted values of our model with $\eta=3\pi/5$ for the mixing angles $\sin^2\theta_{12}$, $\sin^2\theta_{23}$, the Dirac CP phase $\delta_{CP}$, the Majorana CP phase $\beta$ and mass ratio $m_2/m_3$ as a function of $\sin\theta_{13}$. Horizontal and vertical bands show the experimentally determined $1\sigma$ and $3\sigma$ ranges~\cite{Esteban:2024eli} for each parameter.
 }
\end{figure}

We stress that the numerical values of parameters $m_a$,  $r$ and $\eta$ remain undetermined by residual symmetry constraints. The phase parameter $\eta$ requires explicit model-building efforts for its determination. Upon specifying $\eta$, all neutrino mixing observables and mass eigenvalues become functions solely of $m_{a}$ and $r$ ultimately constrained through experimental measurements of $\Delta m^2_{21}$ and $\Delta m^2_{31}$. This framework consequently provides unique predictions for three lepton mixing angles, two CP violation phases and the absolute neutrino masses. Our analysis focuses on the minimal configuration  $\eta=\frac{3\pi}{5}$, a value chosen for mathematical simplicity and anticipated feasibility in dynamical model implementations. The subsequent numerical calculations systematically explore the correlations between input parameters  $m_a$,  $r$ and output observables. We find that the minimum value of $\chi^{2}$ function is $\chi^{2}_{\text{min}}=5.573$ when the free parameters take the following values
\begin{equation}
m_{a}=24.587\,\text{meV}, \qquad r=0.360\,.
\end{equation}
Accordingly the lepton masses and mixing parameters are determined to be
\begin{eqnarray}
\nonumber &&\sin^2\theta_{13}=0.02191, \quad \sin^2\theta_{12}=0.318, \quad \sin^2\theta_{23}=0.451, \quad \delta_{CP}=1.431\pi,\quad \beta=1.629\pi\,, \\
\label{eq:best_fit_model2}&&m_1=0\,\text{meV}, \qquad m_2=8.698\,\text{meV}, \qquad m_3=50.107\,\text{meV}, \qquad m_{ee}=2.954\,\text{meV}\,.
\end{eqnarray}
As outlined in preceding analysis, the full set of lepton mixing observables and the $m_2/m_3$ mass ratio demonstrate exclusive dependence on the fundamental parameter $r$. Following the removal of this free parameter, predictive relationships emerge between different measurable quantities. Furthermore, figure~\ref{fig:Mix_par_M1} systematically illustrates the interdependent patterns connecting the reactor angle $\theta_{13}$ with multiple physical quantities in this framework.

\subsection{Lepton mixing patterns for   $\phi_{\text{atm}}\sim\mathbf{\bar{3}_{0}}$, $\phi_{\text{sol}}\sim\mathbf{\bar{3}_{1}}$}

\begin{table}[t!]
	\begin{center}
		\renewcommand{\tabcolsep}{0.1mm}
\renewcommand{\arraystretch}{1.2}
		\begin{tabular}{|c|c|c|c||c|c|c|c|}\hline\hline
			\multicolumn{8}{|c|}{For the NO case}\\
			\hline
			&$(G_{l}$,$G_{\text{atm}}$,$G_{\text{sol}})$&$(X^{\text{atm}}_{\bm{\bar{3}_{0}}}$,$X^{\text{sol}}_{\bm{\bar{3}_{1}}})$&$P_{l}$& &$(G_{l}$,$G_{\text{atm}}$,$G_{\text{sol}})$&$(X^{\text{atm}}_{\bm{\bar{3}_{0}}}$,$X^{\text{sol}}_{\bm{\bar{3}_{1}}})$&$P_{l}$\\
			\hline
			$\mathcal{T}_{1}$&($Z^{ac}_{3},Z^{b}_{2},Z^{bcd}_{2}$)&($\mathbb{1}_{3},\rho_{\bm{\bar{3}_{1}}}(d^2)$)&$P_{132}$&$\mathcal{T}_{2}$&($Z^{ac}_{3},Z^{b}_{2},Z^{ab}_{2}$)&($\mathbb{1}_{3},\rho_{\bm{\bar{3}_{1}}}(c^2)$)&$P_{132}$\\ \hline 
			$\mathcal{T}_{3}$&($Z^{ac}_{3},Z^{b}_{2},Z^{abc}_{2}$)&($\mathbb{1}_{3},\rho_{\bm{\bar{3}_{1}}}(c^3)$)&$P_{132}$&	$\mathcal{T}_{4}$&($Z^{ac}_{3},Z^{c^2}_{2},Z^{abc^3}_{2}$)&($\mathbb{1}_{3},\rho_{\bm{\bar{3}_{1}}}(d^2)$)&$P_{321}$\\
			\hline 
			$\mathcal{T}_{5}$&($Z^{ac}_{3},Z^{d^2}_{2},Z^{abc}_{2}$)&($\mathbb{1}_{3},\rho_{\bm{\bar{3}_{1}}}(d^2)$)&$P_{312}$&$\mathcal{T}_{6}$&($Z^{ac}_{3},Z^{ab}_{2},Z^{a^2bd^2}_{2}$)&($\mathbb{1}_{3},\rho_{\bm{\bar{3}_{1}}}(c^2d)$)&$P_{231}$\\ \hline
			$\mathcal{T}_{7}$&($Z^{ac}_{3},Z^{ab}_{2},Z^{abc^3}_{2}$)&($\mathbb{1}_{3},\rho_{\bm{\bar{3}_{1}}}(c)$)&$P_{231}$	&$\mathcal{T}_{8}$&($Z^{ac}_{3},Z^{ac^3}_{3},Z^{abc^3}_{2}$)&($\rho_{\bm{\bar{3}_{0}}}(d^2),\rho_{\bm{\bar{3}_{1}}}(d^2)$)&$P_{312}$	\\
			\hline
			$\mathcal{T}_{9}$&($Z^{ac}_{3},Z^{abc^2}_{2},Z^{b}_{2}$)&($\mathbb{1}_{3},\rho_{\bm{\bar{3}_{1}}}(cd^3)$)&$P_{132}$	&\cellcolor{yellow!70}$\mathcal{T}_{10}$&\cellcolor{yellow!70}($Z^{ac}_{3},Z^{bcd}_{2},Z^{abc^2}_{2}$)&\cellcolor{yellow!70}($\rho_{\bm{\bar{3}_{0}}}(d^2),\mathbb{1}_{3}$)&\cellcolor{yellow!70}$P_{321}$\\
			\hline
			$\mathcal{T}_{11}$&($Z^{ac}_{3},Z^{abc^2}_{2},Z^{abc^3}_{2}$)&($\mathbb{1}_{3},\rho_{\bm{\bar{3}_{1}}}(c)$)&$P_{132}$&$\mathcal{T}_{12}$&($Z^{ac}_{3},Z^{acd^2}_{3},Z^{abc^3}_{2}$)&($\rho_{\bm{\bar{3}_{0}}}(d^2),\rho_{\bm{\bar{3}_{1}}}(c)$)&$P_{321}$\\
			\hline
			$\mathcal{T}_{13}$&($Z^{ac}_{3},Z^{acd^2}_{3},Z^{a^2bd^2}_{2}$)&($\rho_{\bm{\bar{3}_{0}}}(d^2),\rho_{\bm{\bar{3}_{1}}}(d^2)$)&$P_{321}$&$\mathcal{T}_{14}$&($Z^{ac}_{3},Z^{ac^2d^2}_{3},Z^{bcd}_{2}$)&($\mathbb{1}_{3},\rho_{\bm{\bar{3}_{1}}}(d^2)$)&$P_{321}$\\
			\hline
			$\mathcal{T}_{15}$&($Z^{ac}_{3},Z^{ac^2d^2}_{3},Z^{abc}_{2}$)&($\mathbb{1}_{3},\rho_{\bm{\bar{3}_{1}}}(d^2)$)&$P_{321}$&$\mathcal{T}_{16}$&($Z^{ac}_{3},Z^{ac^2d^2}_{3},Z^{a^2bd^2}_{2}$)&($\mathbb{1}_{3},\rho_{\bm{\bar{3}_{1}}}(c^2d^3)$)&$P_{312}$\\
			\hline
			$\mathcal{T}_{17}$&($Z^{ac}_{3},Z^{ac^2d^2}_{3},Z^{a^2bd^2}_{2}$)&($\mathbb{1}_{3},\rho_{\bm{\bar{3}_{1}}}(d^2)$)&$P_{321}$&	$\mathcal{T}_{18}$&($Z^{ac}_{3},Z^{bcd}_{2},Z^{ab}_{2}$)&($\rho_{\bm{\bar{3}_{0}}}(d^2),\mathbb{1}_{3}$)&$P_{321}$\\
			\hline
			$\mathcal{T}_{19}$&($Z^{ac}_{3},Z^{abc^2}_{2},Z^{abc}_{2}$)&($\mathbb{1}_{3},\rho_{\bm{\bar{3}_{1}}}(c)$)&$P_{123}$&$\mathcal{T}_{20}$&($Z^{ac}_{3},Z^{bcd}_{2},Z^{abc^3}_{2}$)&($\rho_{\bm{\bar{3}_{0}}}(d^2),\rho_{\bm{\bar{3}_{1}}}(c)$)&$P_{312}$\\
			\hline
			$\mathcal{T}_{21}$&($Z^{ac}_{3},Z^{abc^2}_{2},Z^{abc^3}_{2}$)&($\mathbb{1}_{3},\rho_{\bm{\bar{3}_{1}}}(c^3)$)&$P_{132}$&	$\mathcal{T}_{22}$&($Z^{ac}_{3},Z^{b}_{2},Z^{a^2bd^2}_{2}$)&($\mathbb{1}_{3},\mathbb{1}_{3}$)&$P_{123}$\\
			\hline
			$\mathcal{T}_{23}$&($Z^{ac}_{3},Z^{a^2bd^3}_{2},Z^{abc^3}_{2}$)&($\rho_{\bm{\bar{3}_{0}}}(c^2d^2),\rho_{\bm{\bar{3}_{1}}}(c^3)$)&$P_{231}$&$\mathcal{T}_{24}$&($Z^{ac}_{3},Z^{ac^3d}_{3},Z^{bc^3d^3}_{2}$)&($\rho_{\bm{\bar{3}_{0}}}(bcd^3),\rho_{\bm{\bar{3}_{1}}}(c^2)$)&$P_{321}$\\
			\hline
			$\mathcal{T}_{25}$&($Z^{ac}_{3},Z^{bc^2d^2}_{2},Z^{abc}_{2}$)&($\mathbb{1}_{3},\rho_{\bm{\bar{3}_{1}}}(c)$)&$P_{231}$&$\mathcal{T}_{26}$&($Z^{ac}_{3},Z^{bc^2d^2}_{2},Z^{a^2bd^2}_{2}$)&($\mathbb{1}_{3},\mathbb{1}_{3}$)&$P_{213}$\\
			\hline
			$\mathcal{T}_{27}$&($Z^{ac}_{3},Z^{bc^2d^2}_{2},Z^{bcd}_{2}$)&($\mathbb{1}_{3},\rho_{\bm{\bar{3}_{1}}}(d^2)$)&$P_{231}$&	$\mathcal{F}_{1}$&($Z^{c}_{4},Z^{a}_{3},Z^{abc^2}_{2}$)&($\mathbb{1}_{3},\rho_{\bm{\bar{3}_{1}}}(c^3d^2)$)&$P_{321}$\\
			\hline
			$\mathcal{F}_{2}$&($Z^{c}_{4},Z^{a}_{3},Z^{abc^2}_{2}$)&($\mathbb{1}_{3},\rho_{\bm{\bar{3}_{1}}}(cd^2)$)&$P_{321}$&$\mathcal{F}_{3}$&($Z^{c}_{4},Z^{a}_{3},Z^{abc^2}_{2}$)&($\mathbb{1}_{3},\rho_{\bm{\bar{3}_{1}}}(c^2)$)&$P_{321}$\\
			\hline
			\cellcolor{yellow!70}$\mathcal{F}_{4}$&\cellcolor{yellow!70}($Z^{c}_{4},Z^{ac}_{3},Z^{bc^3d^3}_{2}$)&\cellcolor{yellow!70}($\rho_{\bm{\bar{3}_{0}}}(d^2),\rho_{\bm{\bar{3}_{1}}}(d^2)$)&\cellcolor{yellow!70}$P_{213}$&	$\mathcal{F}_{5}$&($Z^{c}_{4},Z^{a^2bd}_{2},Z^{ab}_{2}$)&($\rho_{\bm{\bar{3}_{0}}}(c^2d^2),\rho_{\bm{\bar{3}_{1}}}(c^3d^2)$)&$P_{132}$\\
			\hline
			\cellcolor{yellow!70}$\mathcal{F}_{6}$&\cellcolor{yellow!70}($Z^{c}_{4},Z^{a^2bd^2}_{2},Z^{abc}_{2}$)&\cellcolor{yellow!70}($\mathbb{1}_{3},\rho_{\bm{\bar{3}_{1}}}(d^2)$)&\cellcolor{yellow!70}$P_{123}$&\cellcolor{yellow!70}$\mathcal{F}_{7}$&\cellcolor{yellow!70}($Z^{c}_{4},Z^{a^2bd^2}_{2},Z^{ab}_{2}$)&\cellcolor{yellow!70}($\mathbb{1}_{3},\mathbb{1}_{3}$)&\cellcolor{yellow!70}$P_{132}$\\
			\hline
			$\mathcal{S}_{1}$&($Z^{abd^2}_{4},Z^{ac}_{3},Z^{a^2bd}_{2}$)&($\rho_{\bm{\bar{3}_{0}}}(d^2),\rho_{\bm{\bar{3}_{1}}}(d)$)&$P_{123}$&\cellcolor{yellow!70}$\mathcal{S}_{2}$&\cellcolor{yellow!70}($Z^{abd^2}_{4},Z^{ac}_{3},Z^{ab}_{2}$)&\cellcolor{yellow!70}($\rho_{\bm{\bar{3}_{0}}}(d^2),\mathbb{1}_{3}$)&\cellcolor{yellow!70}$P_{123}$\\
			\hline
			$\mathcal{S}_{3}$&($Z^{abd^2}_{4},Z^{a^2b}_{2},Z^{a^2bd^3}_{2}$)&($\mathbb{1}_{3},\rho_{\bm{\bar{3}_{1}}}(c^2d^2)$)&$P_{132}$&$\mathcal{S}_{4}$&($Z^{abd^2}_{4},Z^{bcd}_{2},Z^{bc^2d^2}_{2}$)&($\rho_{\bm{\bar{3}_{0}}}(d^2),\rho_{\bm{\bar{3}_{1}}}(c^2d^2)$)&$P_{231}$\\
			\hline
			$\mathcal{S}_{5}$&($Z^{abd^2}_{4},Z^{bc^3d^3}_{2},Z^{bc^2d^2}_{2}$)&($\rho_{\bm{\bar{3}_{0}}}(d^2),\rho_{\bm{\bar{3}_{1}}}(c^2d^2)$)&$P_{213}$&$\mathcal{W}_{1}$&($Z^{abd}_{8},Z^{ab}_{2},Z^{a^2bd}_{2}$)&($\mathbb{1}_{3},\rho_{\bm{\bar{3}_{1}}}(d^3)$)&$P_{123}$\\
			\hline
			$\mathcal{W}_{2}$&($Z^{abd}_{8},Z^{ad^2}_{3},Z^{abc}_{2}$)&($\mathbb{1}_{3},\rho_{\bm{\bar{3}_{1}}}(c^3)$)&$P_{123}$&$\mathcal{W}_{3}$&($Z^{abd}_{8},Z^{acd^2}_{3},Z^{bc^2d^2}_{2}$)&($\rho_{\bm{\bar{3}_{0}}}(d^2),\rho_{\bm{\bar{3}_{1}}}(c^3d)$)&$P_{231}$\\
			\hline\hline 
			\multicolumn{8}{|c|}{For the IO case}\\  
			\hline
			$\mathcal{T}_{1}^{\prime}$&($Z^{ac}_{3},Z^{c^2}_{2},Z^{a^2b}_{2}$)&($\mathbb{1}_{3},\rho_{\bm{\bar{3}_{1}}}(c^2d^3)$)&$P_{231}$&$\mathcal{T}_{2}^{\prime}$&($Z^{ac}_{3},Z^{a}_{3},Z^{a^2bd^2}_{2}$)&($\mathbb{1}_{3},\rho_{\bm{\bar{3}_{1}}}(c^2d)$)&$P_{321}$\\ 
			\hline
			$\mathcal{T}_{3}^{\prime}$&($Z^{ac}_{3},Z^{ac^2}_{3},Z^{a^2bd^3}_{2}$)&($\mathbb{1}_{3},\rho_{\bm{\bar{3}_{1}}}(d)$)&$P_{312}$&$\mathcal{T}_{4}^{\prime}$&($Z^{ac}_{3},Z^{ac^2}_{3},Z^{a^2bd^3}_{2}$)&($\mathbb{1}_{3},\rho_{\bm{\bar{3}_{1}}}(c^2d^2)$)&$P_{321}$\\ 
			\hline
			$\mathcal{T}_{5}^{\prime}$&($Z^{ac}_{3},Z^{acd}_{3},Z^{abc^3}_{2}$)&($\rho_{\bm{\bar{3}_{0}}}(bc^3d),\rho_{\bm{\bar{3}_{1}}}(d^2)$)&$P_{312}$&$\mathcal{T}_{6}^{\prime}$&($Z^{ac}_{3},Z^{acd^2}_{3},Z^{b}_{2}$)&($\rho_{\bm{\bar{3}_{0}}}(d^2),\mathbb{1}_{3}$)&$P_{312}$\\ 
			\hline
		$\mathcal{F}_{1}^{\prime}$&($Z^{c}_{4},Z^{a}_{3},Z^{a^2bd}_{2}$)&($\mathbb{1}_{3},\rho_{\bm{\bar{3}_{1}}}(c^2d^2)$)&$P_{123}$	&\cellcolor{yellow!70}$\mathcal{F}_{2}^{\prime}$&\cellcolor{yellow!70}($Z^{c}_{4},Z^{ac^2}_{3},Z^{abc^3}_{2}$)&\cellcolor{yellow!70}($\mathbb{1}_{3},\rho_{\bm{\bar{3}_{1}}}(d^2)$)&\cellcolor{yellow!70}$P_{321}$\\ 
			\hline
		$\mathcal{S}_{1}^{\prime}$&($Z^{abd^2}_{4},Z^{a^2bd}_{2},Z^{b}_{2}$)&($\rho_{\bm{\bar{3}_{0}}}(c^2d^2),\mathbb{1}_{3}$)&$P_{321}$	&$\mathcal{S}_{2}^{\prime}$&($Z^{abd^2}_{4},Z^{a}_{3},Z^{bc^2d^2}_{2}$)&($\mathbb{1}_{3},\rho_{\bm{\bar{3}_{1}}}(c^3d)$)&$P_{231}$\\
			\hline
			$\mathcal{W}_{1}^{\prime}$&($Z^{abd}_{8},Z^{a^2b}_{2},Z^{b}_{2}$)&($\mathbb{1}_{3},\rho_{\bm{\bar{3}_{1}}}(cd^3)$)&$P_{321}$&$\mathcal{W}_{2}^{\prime}$&($Z^{abd}_{8},Z^{ac^2}_{3},Z^{abc^3}_{2}$)&($\mathbb{1}_{3},\rho_{\bm{\bar{3}_{1}}}(c^3)$)&$P_{132}$\\
			\hline\hline
		\end{tabular}
		\caption{\label{tab:viable_BPs}The 54 viable breaking patterns, where $\mathcal{T}_{i}$ ($\mathcal{T}_{i}^{\prime}$), $\mathcal{F}_{i}$ ($\mathcal{F}_{i}^{\prime}$), $\mathcal{S}_{i}$ ($\mathcal{S}_{i}^{\prime}$) and $\mathcal{W}_{i}$ ($\mathcal{W}_{i}^{\prime}$) denote those patterns in which the residual symmetries in the charged lepton sector correspond to the cyclic groups $Z^{ac}_{3}$, $Z^{c}_{4}$, $Z^{abd^2}_{4}$ and $Z^{abd}_{8}$, respectively. $P_{l}$ represents the undetermined permutation matrix of PMNS matrix in Eq.~\eqref{eq:genral_UPMNS}.  }
	\end{center}
\end{table}

For the representation assignments of the lepton fields and flavon fields in this framework, we identify four distinct residual symmetries $G_{l}$ for the charged lepton sector, with their corresponding diagonalization matrices  $U_{l}$ explicitly provided in Eq.~\eqref{eq:4Ul}. Additionally, we comprehensively classify 31 independent residual symmetries $G_{\text{atm}}\rtimes H^{\text{atm}}_{CP}$ for the atmospheric neutrino sector and 48 independent residual symmetries  $G_{\text{sol}}\rtimes H^{\text{sol}}_{CP}$  for the solar neutrino sector. The invariant VEVs associated with these residual symmetry structures are systematically summarized in table~\ref{tab:invariant_VEVs}. Note that the VEVs of the flavon field $\phi_{\text{sol}}$ contain a real free parameter denoted as $x$. Consequently, all lepton mixing parameters and neutrino masses are determined by four real input parameters: $m_{a}$, $r$, $x$ and $\eta$.

For each of the 5,952 possible residual symmetry combinations, it is necessary to evaluate their compatibility with experimental data by determining whether the corresponding symmetry breaking patterns can reproduce observed results within experimental uncertainties. To this end, we perform a comprehensive numerical analysis for all possible breaking patterns. In our analysis, the $\chi^2$ statistic depends on four input parameters which are randomly sampled from the ranges $m_{a}\in[0,1]\text{eV}$, $r\in[0,100]$, $x\in[-20,20]$ and $\eta\in[0,2\pi]$, with the goal of identifying the minimum $\chi^2$ values. Furthermore, we concentrate on predicting symmetry breaking patterns that yield the three mixing angles $\theta_{13}$, $\theta_{12}$, $\theta_{23}$, the Dirac CP phase $\delta_{CP}$ and the mass squared differences $\Delta m^{2}_{21}$, $\Delta m^{2}_{31}$ ($\Delta m^{2}_{32}$) for the NO (IO) case within their respective $3\sigma$ ranges of global data analysis summarized in table~\ref{tab:bf_13sigma_data}. Through a comprehensive $\chi^2$ analysis of all 5,952 potential symmetry breaking configurations, we identify 42 (12) distinct patterns whose theoretical predictions closely align with experimental measurements of neutrino mass parameters and mixing angles in NO (IO) case\footnote{We only kept breaking patterns with $\chi^{2}_{\text{min}}\leq10$, as we think that those with  $\chi^{2}_{\text{min}}>10$ do not exhibit particularly good agreement with experiments.}. The 54 viable  breaking patterns and the corresponding residual symmetry combinations are summarized in table~\ref{tab:viable_BPs}. Specifically, among these 54 breaking patterns:  33 feature the charged lepton residual symmetry $G_{l}$ as $Z^{ac}_{3}$ (labeled $\mathcal{T}_{i}$ and $\mathcal{T}_{i}^{\prime}$), 9 exhibit $Z^{c}_{4}$ (labeled  $\mathcal{F}_{i}$ and $\mathcal{F}_{i}^{\prime}$), 7 demonstrate $Z^{abd^2}_{4}$ (labeled $\mathcal{S}_{i}$ and $\mathcal{S}_{i}^{\prime}$), and 5 display $Z^{abd}_{8}$ (labeled $\mathcal{W}_{i}$ and $\mathcal{W}_{i}^{\prime}$). For the NO (IO) case, the best fit values of the input parameters, three mixing angles $\theta_{12}$, $\theta_{13}$, $\theta_{23}$, Dirac CP violation phase $\delta_{CP}$, Majorana CP violation phase $\beta$,  the light neutrino masses  $m_{2}$, $m_{3}$ ($m_{1}$, $m_{2}$) and the effective mass $m_{ee}$ in $0\nu\beta\beta$-decay  at the minimum values of $\chi^2$ function for 54 viable breaking patterns are summarized in table~\ref{tab:viable_best_fit}.  In the case of the NO spectrum, the predicted values of $m_{ee}$ across all viable symmetry breaking patterns fall within the range $[2\,\text{meV},4\,\text{meV}]$ which is too small to be detectable by both current and future $0\nu\beta\beta$-decay experiments. These predictions are significantly below the latest KamLAND-Zen bound of $m_{ee}<(28-122)\,\text{meV}$~\cite{KamLAND-Zen:2024eml}, and also fall below the projected sensitivity thresholds of next-generation experiments such as LEGEND-1000 (targeting $m_{ee} < (9 \sim 21)\,$meV)~\cite{LEGEND:2021bnm} and nEXO (projected to reach $m_{ee} < (4.7 \sim 20.3)\,$meV)~\cite{nEXO:2021ujk}. In contrast, for the IO spectrum, future $0\nu\beta\beta$-decay experiments are anticipated to be able to test all viable breaking patterns.  After scanning the parameter space of $m_{a}$, $r$, $x$ and $\eta$, we identified their viable ranges consistent with neutrino oscillation data at $3\sigma$  level, yielding predictions for lepton mixing parameters and neutrino masses. Please see table~\ref{tab:viable_regions}.

\begin{figure}[t!]
\centering
\begin{tabular}{c}
\includegraphics[width=0.92\linewidth]{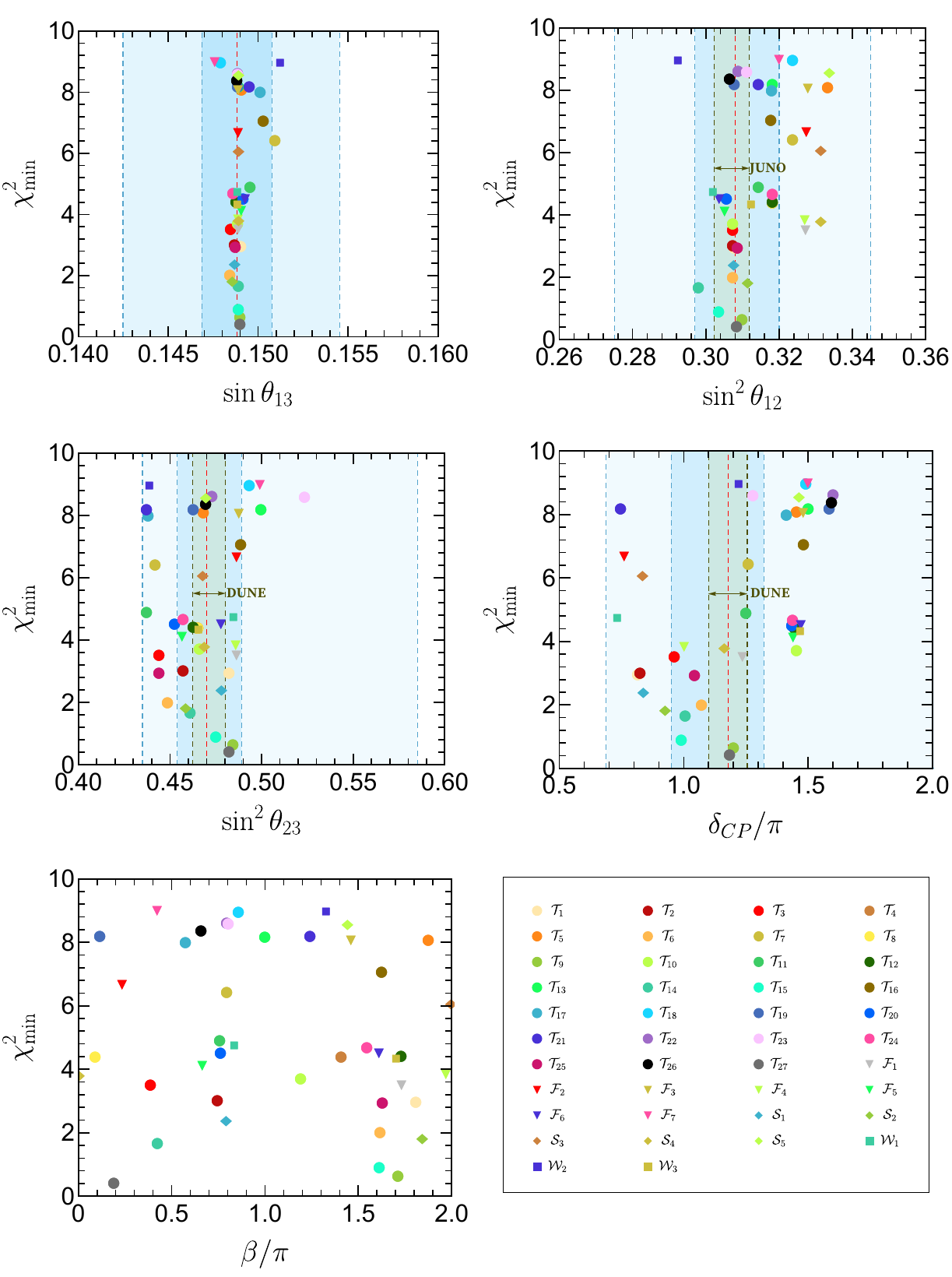}
\end{tabular}
\caption{\label{fig:bf_mixing} Best fit results for 42 viable breaking patterns showing $\chi^2$ minima, lepton mixing angles, and CP violation phases for the NO case. Red dashed lines indicate best fit values, light blue bands show the $1\sigma$ and $3\sigma$ ranges from NuFIT 6.0 with Super-Kamiokande atmospheric data~\cite{Esteban:2024eli}. The pale green band shows the predicted $3\sigma$ range for $\sin^{2}\theta_{12}$ after 6 years of JUNO data~\cite{JUNO:2022mxj}. The faint green areas indicate the resolution (in degrees) for $\sin^{2}\theta_{23}$ and $\delta_{CP}$ after 15 years of DUNE operation~\cite{DUNE:2020ypp}. }
\end{figure}

\begin{figure}[t!]
	\centering
	\begin{tabular}{c}
		\includegraphics[width=0.92\linewidth]{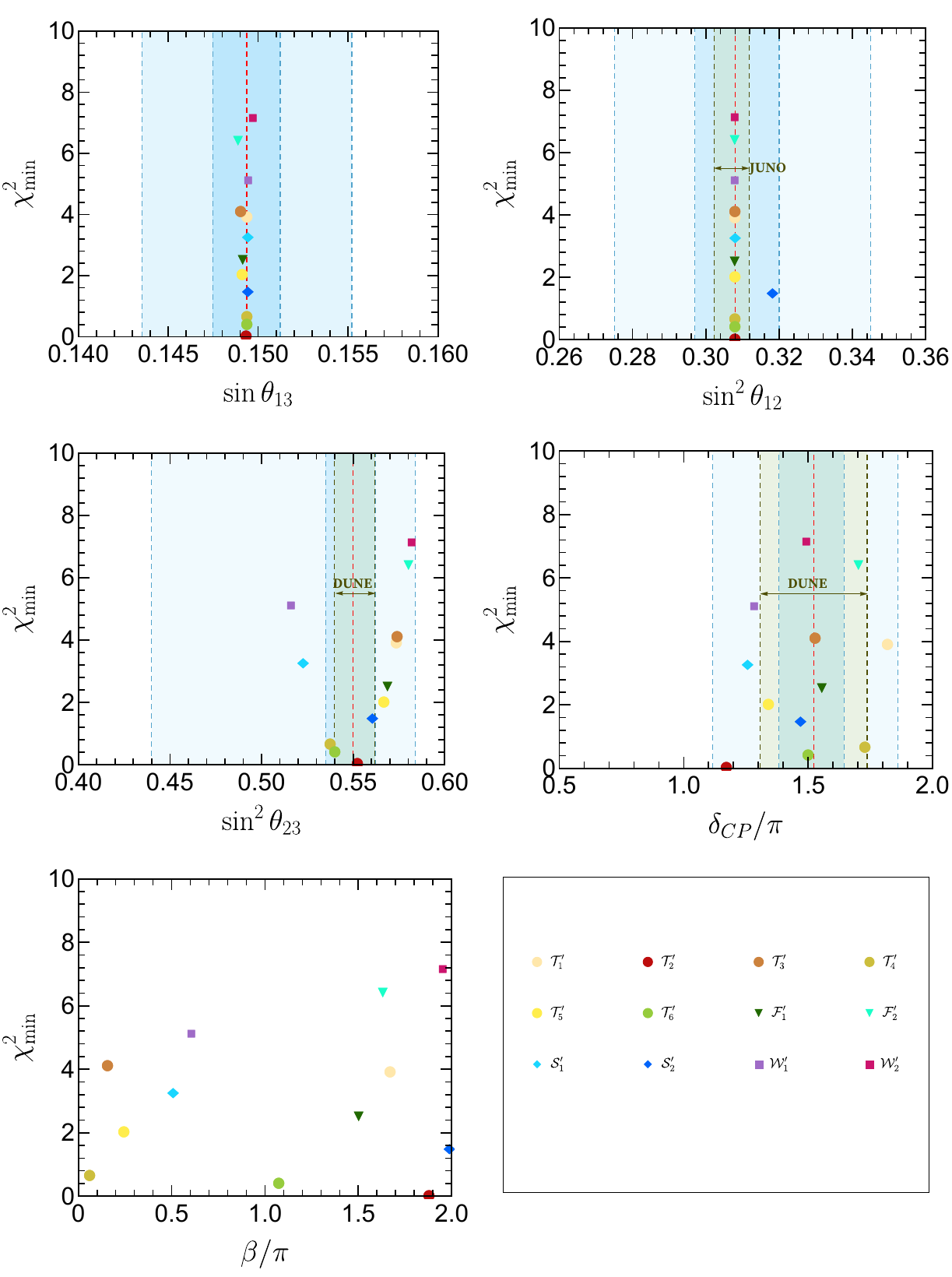}
	\end{tabular}
	\caption{\label{fig:bf_mixing IO} The the best fit $\chi^2$ for all 12 viable breaking patterns with IO neutrino mass spectrum, covering the three lepton mixing angles and CP violation phases.}
\end{figure}

Furthermore, we can pin down the minimal value of $\chi^2$ for each breaking pattern  relevant to three lepton mixing angles, and two CP violation phases. The numerical outcomes for the NO and IO cases are graphically summarized in figures~\ref{fig:bf_mixing} and ~\ref{fig:bf_mixing IO}, respectively.  Analysis of the figure reveals that almost half of the theoretically permissible breaking patterns yield optimal parameter values for the mixing angles and Dirac CP phase that reside within the experimentally constrained $1\sigma$ confidence intervals reported in Ref.~\cite{Esteban:2024eli}.  These breaking patterns can agree with the experimental data very well. Theoretically, advanced measurements of neutrino mass properties and lepton mixing parameters from next-generation neutrino oscillation experiments and cosmological surveys could enable rigorous testing of different breaking patterns. Particularly noteworthy is the JUNO experiment~\cite{JUNO:2022mxj} predicted to achieve sub-percent precision in determining the solar mixing angle $\theta_{12}$. Figures~\ref{fig:bf_mixing} and ~\ref{fig:bf_mixing IO} illustrate the anticipated $3\sigma$ confidence intervals for $\sin^2\theta_{12}$ following six years of operational data collection at JUNO, highlighting its potential to constrain theoretical models. The feasibility of these symmetry breaking patterns will undergo rigorous evaluation through upcoming high precision measurements of $\theta_{23}$ and $\delta_{CP}$ parameters conducted by next generation neutrino experiments DUNE~\cite{DUNE:2020ypp} and T2HK~\cite{Hyper-Kamiokande:2018ofw}. Figures~\ref{fig:bf_mixing} and ~\ref{fig:bf_mixing IO} also illustrate the angular resolution capabilities achieved after 15 years of data acquisition in the DUNE experiment. The collaborative interplay between JUNO and the long baseline neutrino experiments DUNE and T2HK demonstrates remarkable potential for investigating almost all of these breaking patterns.

\begin{center}
	\renewcommand{\tabcolsep}{1.0mm}
	\renewcommand{\arraystretch}{1.2}
	\small
	\begin{longtable}{|c|c|c|c|c|c|c|c|c|c|c|c|c|c|}
		 \caption{\label{tab:viable_best_fit}The best fit results for input parameters and lepton observables for the 54 viable phenomenologically interesting breaking patterns in table~\ref{tab:viable_BPs}.
		 } \\ \hline 
\multicolumn{14}{c}{ } \\[-0.17in] \hline
		\multicolumn{14}{|c|}{For the NO case}\\  	
		\hline
	 & $x$ &	$\eta/\pi$&	$\frac{m_{a}}{\text{meV}}$&$r$&$\chi^2_{\text{min}}$&$\sin^{2}\theta_{13}$&$\sin^{2}\theta_{12}$&$\sin^{2}\theta_{23}$&$\delta_{CP}/\pi$&$\beta/\pi$ &$\frac{m_{2}}{\text{meV}}$&$\frac{m_{3}}{\text{meV}}$ & $\frac{m_{ee}}{\text{meV}}$\\ \hline
	$\mathcal{T}_{1}$&2.59 & 1.85& 21.7 & 0.0595 & 2.97 & 0.02221 & 0.308 & 0.482 & 0.815 & 1.81 & 8.65 & 50.1 & 2.63\\ \hline
	$\mathcal{T}_{2}$&$-0.325$ & 0.619 & 27.3 & 0.441 & 3.03 & 0.02211 & 0.307 & 0.457 & 0.827 & 0.748 & 8.66 & 50.1 & 3.12 \\
	\hline
	$\mathcal{T}_{3}$&$	-0.252 $& 0.561& 24.1 & 0.397 & 3.53 & 0.02205 & 0.307 & 0.444 & 0.962 & 0.386 & 8.66 & 50.1 & 3.35\\ \hline	
	$\mathcal{T}_{4}$&0.343& 1.41 & 9.21 & 2.56 & 4.40 & 0.02216 & 0.318 & 0.465& 1.45 & 1.41 & 8.65 & 50.1 & 3.44 \\
	\hline
	$\mathcal{T}_{5}$&$-1.35$ & 0.882 & 11.2 & 1.22 & 8.10 & 0.02223 & 0.333 & 0.468 & 1.45 & 1.88 & 8.66 & 50.1 & 2.08 \\
	\hline	
	$\mathcal{T}_{6}$&
	0.168 & 0.202 & 29.9 & 0.583& 2.02 & 0.02203 & 0.307 & 0.449 & 1.07 & 1.62 & 8.66& 50.1 & 3.50 \\
	\hline
	$\mathcal{T}_{7}$&	1.35 & 2.00 & 27.7 & 0.0614& 6.45 & 0.02278 & 0.324 & 0.442 & 1.26 & 0.796 & 8.70 & 50.1 & 2.32 \\
	\hline
	$\mathcal{T}_{8}$&$
	-0.171$ & 1.07 & 5.55 & 5.13& 4.40 & 0.02216 & 0.318 & 0.465& 1.45 & 0.0909 & 8.65 & 50.1 & 1.58 \\
	\hline
	$\mathcal{T}_{9}$&
	0.166 & 0.191 & 27.7 & 0.679 & 0.658 & 0.02220 & 0.310 & 0.485 & 1.20 & 1.71 & 8.65 & 50.1 & 3.68 \\ \hline
	\rowcolor{highlight}
	$\mathcal{T}_{10}$&
	3.58 & 0.971 & 19.1 & 0.0755 & 3.73 & 0.02216 & 0.307 & 0.466 & 1.45 & 1.19 & 8.65 & 50.1 & 3.67 \\ \hline
	$\mathcal{T}_{11}$&$
	-1.30$ & 1.03 & 27.7 & 0.0652 & 4.91 & 0.02237 & 0.314 & 0.437 & 1.25 & 0.760 & 8.68 & 50.1 & 2.08 \\ \hline
	$\mathcal{T}_{12}$&
	2.28 & 1.23 & 5.34 & 0.477 & 4.42 & 0.02214 & 0.318 & 0.463 & 1.45 & 1.73 & 8.66 & 50.1 & 2.48 \\\hline
	$\mathcal{T}_{13}$&                                          
	3.99 & 2.00 & 10.6 & 0.0777 & 8.20 & 0.02217 & 0.318 & 0.500 & 1.50 & 1.00 & 8.65 & 50.1 & 3.80 \\
	\hline
	$\mathcal{T}_{14}$&$
	-0.361$ & 0 & 5.41 & 8.23& 1.68 & 0.02218 & 0.298& 0.461 & 1.01& 0.423& 8.65 & 50.1 & 2.95\\
	\hline
	$\mathcal{T}_{15}$&$-2.83$ & 0.0310 & 5.43 & 1.02 & 0.911 & 0.02217 & 0.304 & 0.475 & 0.991 & 1.62 & 8.65 & 50.1 & 3.09 \\
	\hline
	$\mathcal{T}_{16}$&
	10.0 & 1.89 & 7.85 & 0.0158 & 7.07 & 0.02259 & 0.318 & 0.489 & 1.48& 1.63 & 8.59 & 50.2 & 2.58 \\
	\hline
	$\mathcal{T}_{17}$&$
	-10.0$ & 1.79 & 7.45 & 0.0385& 8.01& 0.02254 & 0.318 & 0.438 & 1.41& 0.573& 8.60 & 50.2 & 2.56 \\
	\hline
	$\mathcal{T}_{18}$&$
	-3.81$ & 0.941 & 19.7 & 0.0646 & 8.97 & 0.02188 & 0.324 & 0.493 & 1.49 & 0.859 & 8.75 & 50.1 & 3.77 \\
	\hline
	$\mathcal{T}_{19}$&
	1.26& 1.19 & 20.6 & 0.123& 8.20 & 0.02216 & 0.308 & 0.463 & 1.59 & 0.116 & 8.65 & 50.1 & 2.11\\
	\hline
	$\mathcal{T}_{20}$&
	0.622& 0.294 & 21.7 & 0.442 & 4.53& 0.02225 & 0.306 & 0.453 & 1.44& 0.763 & 8.63 & 50.2 & 3.20 \\
	\hline
	$\mathcal{T}_{21}$&$
	-1.30$& 1.97& 27.7 & 0.0652 & 8.20 & 0.02235 & 0.314 & 0.437 & 0.749 & 1.24 & 8.68 & 50.1 & 2.07 \\
	\hline
	$\mathcal{T}_{22}$&
	0.283 & 1.38 & 15.1 & 2.81 & 8.64 & 0.02216 & 0.309 & 0.473 & 1.60 & 0.796 & 8.66& 50.1 & 3.73 \\
	\hline
	$\mathcal{T}_{23}$&	0.166 & 1.45 & 23.1 & 0.970 & 8.61& 0.02216& 0.311& 0.523 & 1.28 & 0.805 & 8.61 & 50.2& 2.38 \\
	\hline
	$\mathcal{T}_{24}$&$
	-0.0824$ & 0.324& 9.54 & 2.83 & 4.70 & 0.02208 & 0.318 & 0.457 & 1.44 & 1.55 & 8.67 & 50.1 & 3.15 \\
	\hline

	$\mathcal{T}_{25}$&	0.251 & 1.42 & 24.2 & 0.397 & 2.96 & 0.02212 & 0.309 & 0.444& 1.05 & 1.63 & 8.67 & 50.1& 3.43 \\
	\hline
	$\mathcal{T}_{26}$&$-0.285$ & 0.755 & 29.4 & 0.744 & 8.39 & 0.02215 & 0.307 & 0.470 & 1.60 & 0.659& 8.66 & 50.1 & 3.62\\
	\hline
	$\mathcal{T}_{27}$&
	2.60& 0.154 & 21.7 & 0.0594 & 0.435 & 0.02220 & 0.309& 0.482 & 1.19 & 0.190 & 8.65 & 50.1 & 2.64\\
	\hline

	$\mathcal{F}_{1}$&	$-0.285$ & 1.23& 2.91 & 4.22 & 3.53 & 0.02217 & 0.327 & 0.487 & 1.24 & 1.73 & 8.66 & 50.1 & 3.71 \\
	\hline	
	$\mathcal{F}_{2}$&$
	-0.286$ & 0.737 & 2.91& 4.22 & 6.69 & 0.02217 & 0.327 & 0.486& 0.763 & 0.236 & 8.66 & 50.1 & 3.67 \\
	\hline
	$\mathcal{F}_{3}$&
	4.79 & 0.461 & 2.91 & 0.368 & 8.09 & 0.02217 & 0.328& 0.488 & 1.48& 1.46 & 8.66 & 50.1 & 3.22 \\
	\hline
		\rowcolor{highlight}
	$\mathcal{F}_{4}$&
	5.03 & 1.97 & 2.91 & 0.334 & 3.86 & 0.02217& 0.327 & 0.486 & 1.00 & 1.97 & 8.66 & 50.1 & 3.88 \\
	\hline 
	$\mathcal{F}_{5}$&
	3.74 & 1.30 & 23.5 & 0.0627 & 4.15 & 0.02222 & 0.305& 0.457 & 1.44 & 0.664 & 8.64 & 50.1 & 2.95 \\
	\hline
	\rowcolor{highlight}
	$\mathcal{F}_{6}$&
	3.07 & 1.66 & 26.4& 0.0993 & 4.54 & 0.02229 & 0.304 & 0.478 & 1.47 & 1.61 & 8.63 & 50.1 & 2.62 \\
	\hline
	\rowcolor{highlight}
	$\mathcal{F}_{7}$&
	2.99 & 1.32 & 27.3 & 0.0980 & 9.02& 0.02179 & 0.320& 0.499 & 1.50& 0.423 & 8.71 & 50.1 & 2.68 \\
	\hline
	
	$\mathcal{S}_{1}$& $-0.583$ & 0.583 & 16.0 & 0.709& 2.40 & 0.02212 & 0.308 & 0.478 & 0.840 & 0.792 & 8.67 & 50.1 & 2.92 \\
	\hline
	\rowcolor{highlight}
	$\mathcal{S}_{2}$&	
	2.19 & 0.791& 13.9& 0.244 & 1.83 & 0.02207 & 0.312 & 0.459 & 0.925 & 1.85 & 8.66 & 50.1& 3.40 \\
	\hline
	$\mathcal{S}_{3}$&
	1.76 & 1.40 & 6.93 & 1.11 & 6.08 & 0.02218 & 0.331 & 0.468 & 0.835 & 2.00 & 8.66 & 50.1 & 3.50 \\
	\hline
	$\mathcal{S}_{4}$&
	1.76 & 1.60 & 6.93& 1.10 & 3.81 & 0.02218 & 0.331 & 0.469 & 1.17 & 0.00444 & 8.65 & 50.1 & 3.49 \\
	\hline
	$\mathcal{S}_{5}$&
	1.38& 1.83 & 7.95& 1.18 & 8.58 & 0.02218 & 0.334 & 0.470 & 1.47 & 1.44 & 8.66 & 50.1 & 3.41 \\
	\hline
	
	$\mathcal{W}_{1}$&	
	1.04 & 0.620 & 7.59& 1.40 & 4.72 & 0.02216 & 0.302 & 0.485 & 0.734 & 0.836 & 8.66 & 50.1 & 3.32\\
	\hline
	$\mathcal{W}_{2}$&
	0.391 & 0.666 & 4.85 & 5.79& 8.94 & 0.02287 & 0.292 & 0.439 & 1.22& 1.33 & 8.78 & 50.1 & 3.46\\
	\hline
	$\mathcal{W}_{3}$&$
	-0.918$ & 0.920 & 6.26 & 2.18 & 4.30 & 0.02216 & 0.312 & 0.466 & 1.47 & 1.71 & 8.66 & 50.1 & 2.38 \\ \hline
\multicolumn{14}{c}{ } \\[-0.17in] \hline
	\multicolumn{14}{|c|}{For the IO case}\\\hline
	 & $x$ &	$\eta/\pi$&	$\frac{m_{a}}{\text{meV}}$&$r$&$\chi^2_{\text{min}}$&$\sin^{2}\theta_{13}$&$\sin^{2}\theta_{12}$&$\sin^{2}\theta_{23}$&$\delta_{CP}/\pi$&$\beta/\pi$ &$\frac{m_{1}}{\text{meV}}$&$\frac{m_{2}}{\text{meV}}$ & $\frac{m_{ee}}{\text{meV}}$\\ \hline
	$\mathcal{T}_{1}^{\prime}$&0.543& 1.50 & 57.6 & 0.464 & 3.94 & 0.02232 & 0.308 & 0.574 & 1.82 & 1.67 & 49.1 & 49.8 & 42.9 \\ \hline
	$\mathcal{T}_{2}^{\prime}$&0.0938 & 0.00455& 20.1& 2.87 & 0.0483& 0.02230& 0.308 & 0.553 & 1.17 & 1.88 & 49.1 & 49.8 & 47.5\\ \hline
	$\mathcal{T}_{3}^{\prime}$&0.153 & 0.993& 18.0 & 2.71 & 4.13 & 0.02222 & 0.308 & 0.574& 1.53 & 0.158 & 49.1& 49.8 & 47.0 \\ \hline
	$\mathcal{T}_{4}^{\prime}$&0.208 & 1.83 & 18.4& 2.80 & 0.685 & 0.02232 & 0.308 & 0.538 & 1.73 & 0.0642& 49.1 & 49.8 & 48.0 \\ \hline
	$\mathcal{T}_{5}^{\prime}$&-1.42 & 0.75 & 25.0 & 0.757 & 2.05 & 0.02224 & 0.308 & 0.567 & 1.34 & 0.244 & 49.1& 49.8 & 45.2 \\ \hline
	$\mathcal{T}_{6}^{\prime}$&7.12 & 1.89 & 19.1 & 0.0564 & 0.435& 0.02232 & 0.308 & 0.540 & 1.50 & 1.08 & 49.1 & 49.8 & 19.0 \\ \hline
	$\mathcal{F}_{1}^{\prime}$&-0.165 & 0.130 & 18.9 & 2.82 & 2.54 & 0.02225 & 0.308 & 0.569 & 1.56 & 1.50 & 49.1 & 49.8 & 36.6  \\ \hline
	\rowcolor{highlight}	
	$\mathcal{F}_{2}^{\prime}$&7.09 & 1.93& 21.8 & 0.0573 & 6.44 & 0.02217& 0.308 & 0.580 & 1.70 & 1.64 & 49.1 & 49.8 & 41.7  \\ \hline
	$\mathcal{S}_{1}^{\prime}$&6.46 & 1.10 & 34.4 & 0.0461 & 3.28 & 0.02234 & 0.308 & 0.523 & 1.26 & 0.511& 49.1 & 49.8& 36.0 \\ \hline
	$\mathcal{S}_{2}^{\prime}$&0.382 & 0 & 18.7 & 1.87 & 1.51 & 0.02234& 0.318 & 0.561 & 1.47 & 1.99 & 49.1& 49.8 & 48.2 \\ \hline
	$\mathcal{W}_{1}^{\prime}$&0.122 & 0.0845 & 31.0 & 1.91 & 5.09 & 0.02234 & 0.308 & 0.516 & 1.28 & 0.609 & 49.1 & 49.8 & 31.5 \\ \hline
	$\mathcal{W}_{2}^{\prime}$&-3.55 & 0.995 & 21.8 & 0.0582 & 7.12 & 0.02240 & 0.308& 0.582 & 1.50 & 1.96 & 49.1 & 49.8 & 48.1\\ \hline
\multicolumn{14}{c}{ } \\[-0.17in] \hline
\end{longtable}{}
\end{center}

\begin{landscape}
\renewcommand{\tabcolsep}{0.4mm}
\renewcommand{\arraystretch}{1.1}
\center
\setlength\LTcapwidth{\textwidth}
\setlength\LTleft{-0.3in}
\setlength\LTright{0pt}
\begin{longtable}{|c|c|c|c|c||c|c|c|c|c|c|c|} 
\caption{\label{tab:viable_regions} The predicted ranges of the input parameters and lepton observables for all the 54 viable breaking patterns. Note  the mixing angle $\theta_{13}$ can take any values within their $3\sigma$ range, and the neutrino mass $m_{3}$ is predicted to lie in $[49.5\text{meV},50.8\text{meV}]$ for the NO case. The mixing angle $\theta_{12}$ can take any values
within their $3\sigma$ range, and the neutrino mass $m_{1}$ is predicted to lie in $[48.4\text{meV},49.8\text{meV}]$ for the IO case.  We would not show them here.
} \\\hline
\multicolumn{11}{c}{ } \\[-0.17in] \hline
\multicolumn{11}{|c|}{For the NO case}\\
\hline
&$x$&$\eta/\pi$&$m_{a}$/meV&$r$&	$\sin^{2}\theta_{12}$&$\sin^{2}\theta_{23}$&$\delta_{CP}/\pi$&$\beta/\pi$ &$ m_{2}$/meV&$m_{ee}$/meV\\				\hline
$\mathcal{T}_{1}$&
	[2.32,2.96] & [1.12,1.88] & [21.2,22.2] & [0.0441,0.0771] &  [0.275,0.345] & [0.467,0.500] & [0.777,1.22] & [0.169,1.83] & [8.32,8.97] & [2.48,2.83] \\
				\hline
$\mathcal{T}_{2}$&
	[$-$0.362,$-$0.287]& [0.359,0.682] & [24.6,27.8] & [0.406,0.550] &  [0.275,0.345] & [0.435,0.463] & [0.797,1.00] & [0.665,1.02]& [8.32,8.97]  & [1.26,3.91] \\

				\hline
$\mathcal{T}_{3}$&
	[$-$0.290,$-$0.213] & [0.325,0.731] & [22.6,25.1]& [0.366,0.513] & [0.276,0.345] & [0.435,0.451] & [0.904,1.09]&[0.255,0.585] & [8.32,8.97]  & [2.30,4.18] \\

				\hline
$\mathcal{T}_{4}$&
	[0.293,0.466] & [0,2)& [8.86,9.70] & [2.34,2.71] & [0.317,0.320] & [0.458,0.542] & [1.44,1.56] & [0,1.98] & [8.32,8.97]  & [1.38,3.96]\\
				\hline
$\mathcal{T}_{5}$&
	[$-$1.50,$-$1.32] & [0.696,1.43] & [10.8,11.7] & [1.00,1.31] & [0.330,0.345] & [0.435,0.583] & [1.40,1.63] & [0,2) & [8.32,8.97]  & [1.49,3.79]\\				
				\hline
$\mathcal{T}_{6}$&
	[0.149,0.184]& [0.183,0.573] & [23.0,30.5] & [0.558,1.03] &  [0.275,0.345] & [0.438,0.539] & [0.702,1.11] & [1.18,1.65] & [8.32,8.97]  & [1.73,4.10] \\

				\hline
$\mathcal{T}_{7}$&
	[1.16,1.46]& [0,2) & [27.3,28.1] & [0.0511,0.0800] & [0.285,0.345]& [0.435,0.450] & [1.21,1.28] & [0.726,0.858] & [8.32,8.97]  & [1.93,2.87]\\
				\hline
$\mathcal{T}_{8}$&
[$-$0.621,$-$0.143] & [0,2) & [4.25,5.94] & [4.20,5.45] & [0.317,0.320]& [0.435,0.585] & [1.40,1.63]& [0.0158,1.98]& [8.32,8.97] & [1.38,3.93]\\
				\hline
$\mathcal{T}_{9}$&
	[0.149,0.184] & [0.168,0.222] & [26.7,28.4] & [0.639,0.723] & [0.275,0.345]& [0.465,0.508] & [1.19,1.21]& [1.68,1.74] & [8.32,8.97]  & [3.24,4.11]\\
				\hline
	\rowcolor{highlight}			
$\mathcal{T}_{10}$&
	[3.15,4.16] & [0.889,1.18]& [18.3,20.9] & [0.0512,0.101] & [0.275,0.345]& [0.435,0.561] & [1.41,1.60] & [1.05,1.24] & [8.32,8.97] & [3.32,4.08]\\
				\hline
$\mathcal{T}_{11}$&
	[$-$1.41,$-$1.21] & [0.955,1.05] & [27.3,28.1] & [0.0590,0.0721]& [0.295,0.336] & [0.435,0.450] & [1.22,1.28] & [0.731,0.850] & [8.33,8.97]  & [1.94,2.80]\\
				\hline
$\mathcal{T}_{12}$&
[1.41,3.40]& [0.945,1.41]& [4.79,5.87]& [0.216,1.09] & [0.317,0.320]& [0.435,0.479] & [1.40,1.47]& [1.46,1.93] & [8.32,8.97]  & [2.00,3.54]\\ 
				\hline
$\mathcal{T}_{13}$&
	[3.58,5.13] & [0,2) & [9.80,11.0] & [0.0535,0.0931] & [0.317,0.320] & [0.444,0.579] & [1.42,1.62]& [0.989,1.01] & [8.32,8.97]  & [3.45,3.97] \\
				\hline
$\mathcal{T}_{14}$&
	[$-$0.379,$-$0.300] & [0,2) & [5.18,5.83] & [7.38,8.59] & [0.288,0.345]& [0.439,0.571] & [0.984,1.02] & [0.148,0.488] & [8.32,8.97] & [2.62,4.06]\\
				\hline
$\mathcal{T}_{15}$&
	[$-$3.34,$-$2.64] & [0,2) & [5.21,5.81]& [0.678,1.20] & [0.288,0.345] & [0.440,0.571] & [0.979,1.02] & [1.51,1.85] & [8.32,8.97]  & [2.60,4.03]\\
				\hline
$\mathcal{T}_{16}$&
	[5.50,15.0] & [1.75,1.95] & [6.49,8.54] & [0.00657,0.0692] & [0.317,0.320] & [0.436,0.585] & [1.41,1.63]& [1.45,1.70] & [8.32,8.97]  & [2.05,2.83]\\
				\hline
$\mathcal{T}_{17}$&
	[$-$15.0,$-$8.21]& [1.66,1.80] & [7.24,8.19]& [0.0149,0.0588] & [0.317,0.320] & [0.435,0.511] & [1.40,1.51] & [0.424,0.587] & [8.32,8.97] & [2.47,2.75]\\
			\hline
$\mathcal{T}_{18}$&
	[$-$4.16,$-$3.16] & [0.826,1.16] & [18.1,21.0] & [0.0520,0.102] & [0.275,0.345] & [0.443,0.585] & [1.41,1.63] & [0.736,0.941] &[8.32,8.97]  & [3.34,4.07]\\
			\hline
$\mathcal{T}_{19}$&
	[1.11,1.46] & [1.10,1.25] & [19.8,21.3] & [0.0898,0.160] & [0.275,0.345] & [0.445,0.480] & [1.56,1.61] & [0.0686,0.177] & [8.32,8.97]  & [1.85,2.40]\\
\hline
$\mathcal{T}_{20}$&
	[0.581,0.657] & [0.253,0.373] & [19.7,23.4]& [0.408,0.473] & [0.276,0.344] & [0.435,0.495] & [1.40,1.49] & [0.718,0.797]& [8.32,8.97]& [2.91,3.53]\\
			\hline
$\mathcal{T}_{21}$&
	[$-$1.46,$-$1.15] & [0,2) & [27.2,28.2] & [0.0515,0.0816] & [0.283,0.345] & [0.435,0.450] & [0.715,0.785] & [1.14,1.27] & [8.32,8.97]  & [1.93,2.87]\\
			\hline
$\mathcal{T}_{22}$&
	[0.255,0.311] & [1.33,1.44] & [14.0,16.3]&[2.48,3.18] & [0.275,0.345] & [0.460,0.494] & [1.58,1.62] & [0.704,0.881] & [8.32,8.97]  & [3.28,4.19]\\
			\hline
$\mathcal{T}_{23}$&
	[0.150,0.185] & [13.4,15.5]& [22.1,24.1] & [0.872,1.07]& [0.275,0.344]& [0.511,0.538] & [1.25,1.31] & [0.785,0.817] & [8.32,8.97] & [1.81,3.00]\\
			\hline
$\mathcal{T}_{24}$&
	[$-$0.250,0.0983] & [0.205,0.500] & [8.38,12.5] & [2.53,2.97] & [0.317,0.320] & [0.435,0.484] & [1.40,1.48] & [1.40,1.73] & [8.32,8.97]  & [2.79,3.69]\\
			\hline
$\mathcal{T}_{25}$&
	[0.213,0.290] & [1.27,1.67]& [22.6,25.2] & [0.365,0.511] & [0.276,0.345]& [0.435,0.451] & [0.914,1.10] & [1.42,1.75] & [8.32,8.97]  & [2.30,4.18]\\
			\hline
$\mathcal{T}_{26}$&
		[$-$0.312,$-$0.254] & [0.724,0.780] & [28.5,30.2]& [0.696,0.795] & [0.275,0.345] & [0.449,0.492] & [1.56,1.63] & [0.597,0.723] & [8.32,8.97]  & [3.21,4.09]\\
			\hline
$\mathcal{T}_{27}$&	
	[2.32,2.96]& [0.127,0.870]& [21.1,22.3] & [0.0441,0.0783] & [0.275,0.345] & [0.467,0.502] & [0.776,1.23] & [0.170,1.83] & [8.32,8.97]  & [2.47,2.84]\\
				\hline
$\mathcal{F}_{1}$&
	[$-$0.330,$-$0.269]& [0,2) & [2.81,2.99] & [4.04,4.43] & [0.327,0.330] & [0.485,0.512] & [1.23,1.27] & [0.0189,2) & [8.34,8.91]  & [1.47,4.04]\\
				\hline
$\mathcal{F}_{2}$&
[$-$0.325,0.325]& [0,2)& [2.81,2.96] & [4.08,4.43] & [0.327,0.330] &[0.486,0.514] & [0.731,1.77] & [0,1.98] & [8.36,8.82] & [1.51,4.01]\\
				\hline	
$\mathcal{F}_{3}$&[4.30,5.27] & [0,2) & [2.81,2.99] & [0.295,0.473]& [0.327,0.330]& [0.486,0.514] & [1.48,1.52] & [0.0120,1.99] & [8.34,8.92]  &[1.48,4.05]\\
				\hline
				\rowcolor{highlight}
$\mathcal{F}_{4}$&
[4.30,5.25] & [0,2) & [2.81,2.99]& [0.297,0.474]& [0.327,0.330] & [0.485,0.512] & [0.981,1.02] & [0,1.98]& [8.34,8.92]  & [1.48,4.05]\\
				\hline	
$\mathcal{F}_{5}$&
	[3.48,3.97] & [1.27,1.37] & [21.5,24.8] & [0.0565,0.0693] & [0.275,0.345] & [0.435,0.510] & [1.40,1.51] & [0.578,0.730] & [8.32,8.97]  & [2.64,3.30]\\
				\hline
				\rowcolor{highlight}
$\mathcal{F}_{6}$&
	[2.88,3.21] &[1.61,1.74] & [24.4,29.3] & [0.0905,0.108] & [0.275,0.345] & [0.435,0.550] & [1.42,1.58] & [1.49,1.67] & [8.32,8.97]  & [2.37,2.92]\\
				\hline
	\rowcolor{highlight}			
$\mathcal{F}_{7}$&
	[2.89,3.21] & [1.26,1.40] & [24.1,29.3] & [0.0901,0.107] & [0.275,0.345]& [0.448,0.578] & [1.42,1.60] & [0.312,0.515] & [8.32,8.97]  & [2.37,2.92]\\
				\hline
$\mathcal{S}_{1}$&
		[$-$0.661,$-$0.522] & [0.509,0.655]& [14.5,17.5] & [0.610,0.798] & [0.275,0.345] & [0.460,0.499] & [0.784,0.906]& [0.732,0.843] & [8.32,8.97]  & [2.63,3.25]\\
				\hline
\rowcolor{highlight}				
$\mathcal{S}_{2}$&
	[2.05,2.35]& [0.713,1.55] & [12.6,15.2]& [0.222,0.268] & [0.275,0.345] & [0.435,0.585] & [0.862,1.37] &[1.52,1.90] & [8.32,8.97]  & [3.02,3.76]\\
				\hline
$\mathcal{S}_{3}$&
	[1.65,1.99] & [1.11,1.46]& [6.58,7.78] & [0.764,1.28] & [0.330,0.335] & [0.435,0.585] & [0.689,0.856] & [0,2) & [8.32,8.97]  & [1.55,3.87]\\
				\hline
$\mathcal{S}_{4}$&
	[1.15,1.98] & [0,2) & [6.66,8.23] & [0.754,1.54] & [0.330,0.345] & [0.435,0.585] & [1.15,1.61] & [0,2) & [8.32,8.97]  & [1.54,4.09]\\
				\hline
$\mathcal{S}_{5}$&
	[1.15,1.97] & [0.146,1.98] & [6.60,8.23] & [0.801,1.56] & [0.330,0.345] & [0.435,0.584] & [1.39,1.86] & [0.0154,2.00] & [8.32,8.97]  & [1.83,4.10]\\
				\hline
$\mathcal{W}_{1}$&
	[0.858,1.34] & [0.337,0.869] & [7.23,8.42] & [0.825,1.78] & [0.301,0.310] & [0.479,0.530] & [0.689,0.744] & [0.505,1.15]& [8.32,8.97]  & [2.67,3.86]\\
				\hline
$\mathcal{W}_{2}$&
	[0.370,0.410] & [0.574,0.794] & [4.66,5.05] & [5.19,6.36] & [0.280,0.302] & [0.435,0.468] & [1.18,1.27] &[1.26,1.44] & [8.47,8.97]  & [3.22,3.72]\\
				\hline
$\mathcal{W}_{3}$&
	[$-$1.05,$-$0.876] & [0.763,0.970] & [6.01,6.45] & [1.61,2.38] & [0.303,0.345] & [0.435,0.556]& [1.44,1.53] & [1.47,1.78] & [8.32,8.97]  & [2.17,3.15]\\
				\hline

\multicolumn{11}{c}{ } \\[-0.17in] \hline
\multicolumn{11}{|c|}{For the IO case}\\
\hline
&$x$&$\eta/\pi$&$m_{a}$/meV&$r$&	$\sin^{2}\theta_{13}$&$\sin^{2}\theta_{23}$&$\delta_{CP}/\pi$&$\beta/\pi$ &$ m_{2}$/meV&$m_{ee}$/meV\\				\hline
$\mathcal{T}_{1}^{\prime}$&
	[0.532,0.554] & [1.495,1.497]& [56.61,58.53] & [0.454,0.475] & [0.0207,0.0239] &  [0.574,0.574] & [1.793,1.849] & [1.658,1.683] &  [49.2,50.5] & [42.3,43.5]\\
	\hline
	$\mathcal{T}_{2}^{\prime}$&[0.0906,0.0972]& [0.00376,0.00539] & [19.89,20.34] & [2.860,2.879]& [0.0209,0.0238]&  [0.551,0.554] & [1.168,1.179] & [1.875,1.884]& [49.2,50.4] & [46.9,48.1]\\
	\hline
	$\mathcal{T}_{3}^{\prime}$&[0.146,0.160] & [0.992,0.994] & [17.64,18.24]& [2.689,2.736] &[0.0207,0.0240]&  [0.572,0.577] & [1.524,1.538] & [0.153,0.162] &  [49.2,50.5] & [46.3,47.7]\\
	\hline
	$\mathcal{T}_{4}^{\prime}$&[0.199,0.216] & [1.827,1.842] & [18.09,18.68] & [2.779,2.819] &[0.0207,0.0237]  &  [0.537,0.538] & [1.725,1.733]& [0.0527,0.0746] &  [49.2,50.5] & [47.3,48.8]\\
\hline
$\mathcal{T}_{5}^{\prime}$&
[$-$1.434,$-$1.408] & [0.749,0.750] & [24.65,25.32] & [0.747,0.761] & [0.0216,0.0240] & [0.566,0.570] &[1.332,1.351] & [0.234,0.253] &  [49.2,50.5] & [44.6,45.8]\\	
\hline
$\mathcal{T}_{6}^{\prime}$&
[7.060,7.180] & [1.890,1.899] & [18.79,19.30] & [0.0552,0.0574]& [0.0220,0.0226]&  [0.5398,0.5404] & [1.165,1.635] & [0.834,1.170] &  [49.2,50.4] & [18.5,19.3]\\
\hline
$\mathcal{F}_{1}^{\prime}$&
[$-$0.172,$-$0.158] & [0.123,0.137] & [18.59,19.16] & [2.801,2.832] & [0.0206,0.0240] &  [0.567,0.572]& [1.529,1.587] & [1.481,1.523] &  [49.2,50.5] & [36.0,37.2]\\
\hline
\rowcolor{highlight}
$\mathcal{F}_{2}^{\prime}$&
[6.960,7.180]& [1.926,1.928] & [21.46,22.04] & [0.0558,0.0593] & [0.0216,0.0230]&  [0.579,0.582] & [1.686,1.721] & [1.620,1.647] &  [49.2,50.5] & [41.1,42.3]\\
\hline
$\mathcal{S}_{1}^{\prime}$&
[6.353,6.544] & [1.093,1.097] & [33.94,34.88] & [0.0450,0.0476] & [0.0218,0.0231] &  [0.522,0.524] & [1.231,1.285] & [0.495,0.532] &  [49.2,50.5] & [35.5,36.5]\\
\hline
$\mathcal{S}_{2}^{\prime}$&
[0.372,0.392] & [0,0.000571] & [18.40,18.97] & [1.831,1.909] & [0.0207,0.0240] & [0.558,0.564] & [1.450,1.487]& [1.991,1.993] &  [49.2,50.5] & [47.5,49.0]\\
\hline
$\mathcal{W}_{1}^{\prime}$&
[0.117,0.128]& [0.0795,0.08934] & [30.46,31.56] & [1.896,1.919] & [0.0207,0.0240]&  [0.515,0.517] & [1.242,1.327]& [0.584,0.640] &  [49.2,50.5] & [31.0,32.0]\\
\hline
$\mathcal{W}_{2}^{\prime}$&
[$-$3.600,$-$3.502] & [0.9947,0.9953] & [21.50,22.11] & [0.0565,0.0597] & [0.0219,0.0230] &  [0.581,0.583] & [1.492,1.497] & [1.952,1.957] & [49.2,50.5] & [47.4,48.8]\\
\hline				
\multicolumn{11}{c}{ } \\[-0.17in] \hline
\end{longtable}{}
\end{landscape}

\section{\label{sec:viable_model_x} Typical interesting breaking patterns }

By performing a $\chi^{2}$ analysis on the formulated lepton flavor symmetry breaking patterns for the representation assignments $L\sim \bm{3_{0}}$, $\phi_{\text{atm}}\sim \bm{\bar{3}_{0}}$ and $\phi_{\text{sol}}\sim \bm{\bar{3}_{1}}$  within the flavor symmetry $\Delta(96)\rtimes H_{CP}$ under the tri-direct CP approach, we have derived that 54 symmetry breaking patterns yield predictions that agree  with the experimental data very well. It is beyond the scope of our study to explore all the viable cases in detail and to present a complete graphical treatment of the  predictions for each breaking pattern. In what follows we consider six (five for NO and  one for IO) representative cases of them  and present their predictions for the lepton mixing angles, CP violation phases and neutrino masses, in which the quality of the results can be thoroughly appreciated.

\subsection{The breaking pattern $\mathcal{F}_{6}$}

For this symmetry breaking pattern, the corresponding residual flavor symmetry combination is given by $(G_{l},G_{atm},G_{sol})=(Z_{4}^{c},Z_{2}^{a^2bd^2},Z_{2}^{abc})$, with the residual CP transformations of the atmospheric  and solar neutrino sectors $(X^{\text{atm}}_{\bm{\bar{3}_{0}}}, X^{\text{sol}}_{\bm{\bar{3}_{1}}})=(\mathbb{1}_{3},\rho_{\bm{\bar{3}_{1}}}(d^2))$. 	In this case, the diagonalization matrix $U_{l}$ of the charged lepton mass matrix is the identity matrix and the VEV alignment of flavor fields $\phi_{\text{atm}}$ and $\phi_{\text{sol}}$ can be obtained from table~\ref{tab:invariant_VEVs}, i.e. 
\begin{eqnarray}
		\langle\phi_{\text{atm}}\rangle =(0,1,1)^{T}v_{\phi_{a}},\qquad
		\langle\phi_{\text{sol}}\rangle =(-1,i,ix)^{T}v_{\phi_{s}}\,,
\end{eqnarray}
where the VEV magnitudes $v_{\phi_a}$ and $v_{\phi_s}$ are real. Then the low-energy effective light neutrino mass matrix reads as
\begin{equation}\label{eq:mnu_withx_1}
m_{\nu }= m_{a}\left[
\begin{pmatrix}
0 & 0 & 0 \\
0 & 1 & 1 \\
0 & 1 & 1 
\end{pmatrix}
+ re^{i \eta } 
\begin{pmatrix}
1 & -i & -i x \\
-i & -1 & -x \\
-i x & -x & -x^2 \\
\end{pmatrix}\right]\,,
\end{equation}
where the parameters $m_{a}$, $r$ and $\eta$  are defined in Eq.~\eqref{eq:mareta_def}. Note that an overall unphysical phase (the phase of $m_{a}$) of $m_{\nu }$ can be omitted and it will be neglected for the other cases in the following. The light neutrino mass matrix $m_{\nu}$ in Eq.~\eqref{eq:mnu_withx_1} can be simplified into a block diagonal form $m^{\prime}_{\nu }$ by performing a unitary transformation $U_{\nu1}$ with $m^{\prime}_{\nu }=U_{\nu1}^{T}m_{\nu }U_{\nu1}$, where the unitary matrix $U_{\nu1}$ and the mass matrix $m^{\prime}_{\nu }$ are 
\begin{eqnarray} 
\nonumber &&U_{\nu1}=\begin{pmatrix}
		\frac{i (x-1)}{\sqrt{(x-1)^2+2}} & -\frac{\sqrt{2}}{\sqrt{(x-1)^2+2}} & 0  \\
		-\frac{1}{\sqrt{(x-1)^2+2}} & \frac{i (x-1)}{\sqrt{2 (x-2) x+6}} & \frac{1}{\sqrt{2}}   \\
		\frac{1}{\sqrt{(x-1)^2+2}}  & -\frac{i (x-1)}{\sqrt{2 (x-2) x+6}}  & \frac{1}{\sqrt{2}} 
			\end{pmatrix}, \\
&& m_{\nu}^{\prime}=\frac{m_{a}}{2}\begin{pmatrix}
0 & 0 & 0 \\
0 &  [(x-2) x+3]re^{i \eta }    &  i (x+1) \sqrt{(x-2) x+3} re^{i \eta }  \\
0 &  i(x+1) \sqrt{(x-2) x+3} re^{i \eta }   &    4 -(x+1)^2 re^{i \eta }
\end{pmatrix}\,.
\end{eqnarray}
Furthermore, $m_{\nu}^{\prime}$ can be further diagonalized by the unitary matrix $U_{\nu2}$ in Eq.~\eqref{eq:Unu2} following the procedures presented in appendix~\ref{sec:Diag_mnup}. Hence the resulting PMNS matrix is taken to be 
\begin{equation}
U_{PMNS}=U_{\nu1}U_{\nu2}\,.
\end{equation}
Then  one can straightforwardly extract the following results for the lepton mixing angles and CP invariants
\begin{eqnarray}
\nonumber
\sin^{2}\theta_{13}&=&\frac{2\sin^{2}\theta}{x^{2}-2x+3},\qquad  
\sin^{2}\theta_{12}=\frac{2\cos^{2}\theta}{x^2-2 x+2+\cos 2 \theta },\  \\
\nonumber
\sin^{2}\theta_{23}&=&\frac{1}{2}-\frac{ (x-1) \sqrt{(x-2) x+3} \sin 2 \theta \sin \psi}{2 (x^2-2 x+2+\cos2 \theta )},\\
\label{eq:mixing_F6x} J_{CP}&=&\frac{(1-x) \sin2 \theta  \cos \psi }{2(x^{2}-2x+3)^{3/2}} , \qquad  I_{1 }=\frac{\sin^{2}2 \theta\sin (\rho-\sigma)}{(x^{2}-2x+3)^2}\,.
\end{eqnarray}	
The  mixing matrix indicates the following sum rule among the Dirac CP phase $\delta_{CP}$ and mixing angles:
\begin{equation}\label{eq:CP_sum_rules}
\cos\delta_{CP}=\frac{\csc \theta _{13} \cot 2 \theta _{23} [(x-2) x-(x^2-2 x+2) \cos 2 \theta _{13}]}{2|x-1| \sqrt{ (x^{2}-2x+3) \cos ^2\theta _{13}-(x-1)^2}}\,.
\end{equation}
All mixing parameters and neutrino masses are functions of $m_{a}$, $x$, $r$ and $\eta$. Experimental measurements can be matched by choosing suitable values for these parameters. Our numerical analysis of the breaking pattern $\mathcal{F}_{6}$ provides best fit values, summarized in table~\ref{tab:viable_best_fit}. After scanning the four free parameters of the breaking pattern and enforcing the experimental $3\sigma$ ranges for the mixing angles and Dirac CP phase~\cite{Esteban:2024eli}, we present the predicted ranges for neutrino mixing parameters and the absolute neutrino masses in table~\ref{tab:viable_regions}.

\begin{figure}[t!]
\centering
\begin{tabular}{c}
\includegraphics[width=0.52\linewidth]{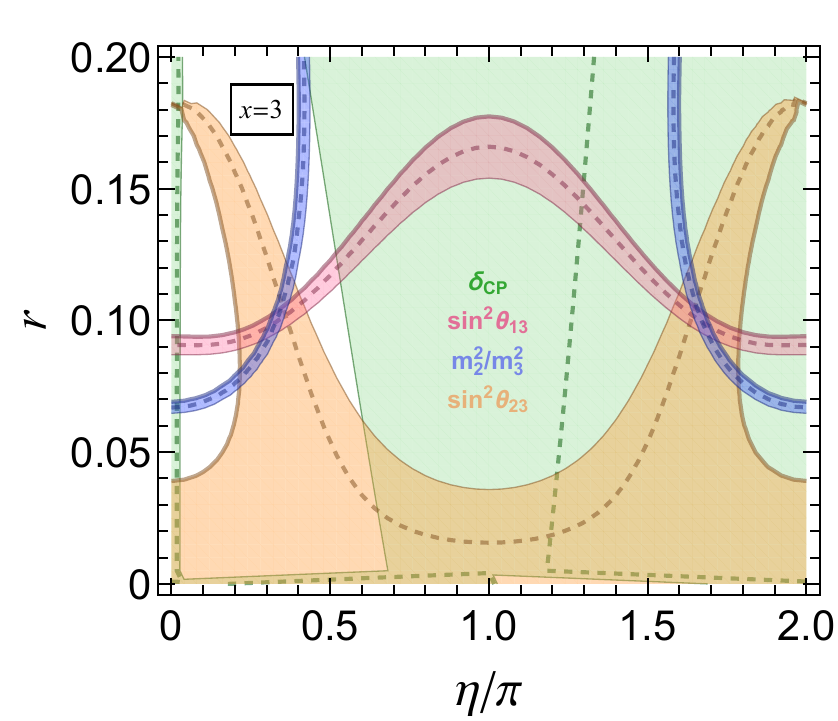}\,\,\,
\includegraphics[width=0.46\linewidth]{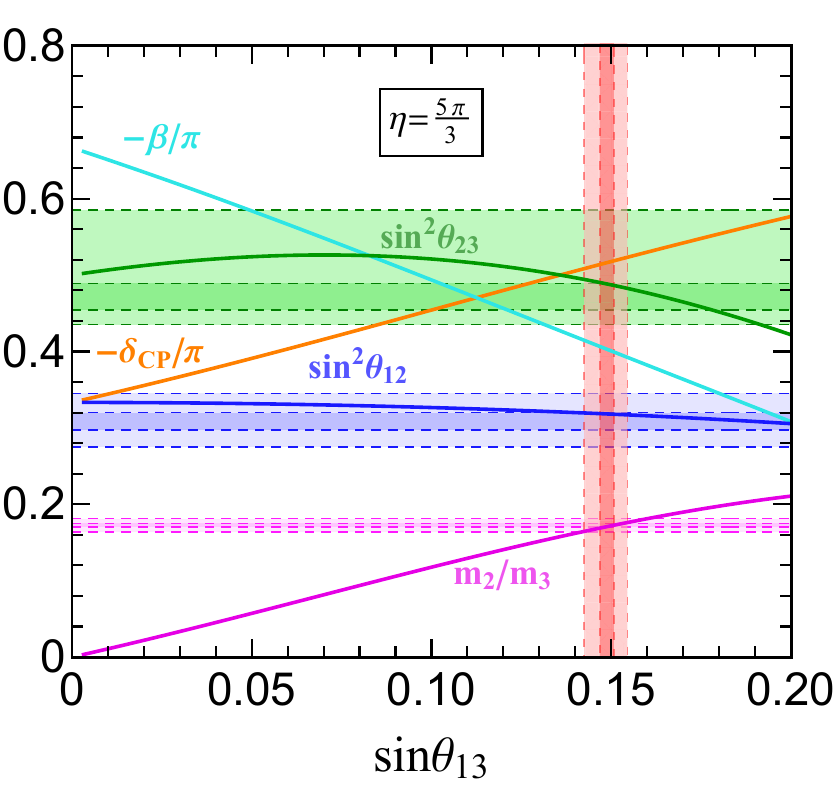}
\end{tabular}
\caption{\label{fig:F6_x3} Contour plots of $\sin^{2}\theta_{23}$, $\sin^{2}\theta_{13}$, $\delta_{CP}$ and $m_{2}^2/m_{3}^2$ in the $\eta$/$\pi-r$ plane for breaking pattern $\mathcal{F}_{6}$ with $x=3$. On the right panel, we show the predictions for mixing parameters and mass ratio as functions of $\sin\theta_{13}$ for $\mathcal{F}_{6}$ with $x=3$ and $\eta=5\pi/3$.
 }
\end{figure}

To provide concrete illustrations, specific benchmark values for the parameters $x$ and $\eta$ may be assigned via the method of vacuum alignment. In this section, we focus on a phenomenologically compelling example of  $x=3$.  This configuration yields the following form for $U_{\nu1}$: 
\begin{equation} 
U_{\nu1}=	\begin{pmatrix}
i \sqrt{\frac{2}{3}} & -\frac{1}{\sqrt{3}} & 0 \\
-\frac{1}{\sqrt{6}} & \frac{i}{\sqrt{3}} & \frac{1}{\sqrt{2}} \\
\frac{1}{\sqrt{6}} & -\frac{i}{\sqrt{3}} & \frac{1}{\sqrt{2}} 
\end{pmatrix}\,.
\end{equation}
The first column of PMNS matrix is in common with that of the tri-bimaximal mixing matrix, and the so-called TM1 mixing matrix is obtained. The expressions for the three lepton mixing angles and CP invariants derived from Eq.~\eqref{eq:mixing_F6x} simplify to:
\begin{eqnarray}
\nonumber && \sin^{2}\theta_{13}=\frac{\sin^{2}\theta}{3},\qquad
			\sin^{2}\theta_{12}=\frac{2\cos^{2}\theta}{5+\cos2\theta},
			\qquad\sin^{2}\theta_{23}=\frac{1}{2}-\frac{\sqrt{6} \sin 2\theta \sin\psi}{5+\cos 2 \theta }\\
&& J_{CP}=-\frac{\sin2\theta\sin\psi}{6\sqrt{6}},\qquad
I_{1 }=\frac{1}{36} \sin^{2}2 \theta\sin (\rho-\sigma)\,.
\end{eqnarray}
Furthermore, the sum rule presented in Eq.~\eqref{eq:CP_sum_rules} reduces to the form shown in Eq.~\eqref{eq:delta_CP_sumrule1}, a relationship that remains valid for all TM1 mixing matrices. All mixing parameters and the mass ratio $m_{2}/m_{3}$ are now determined by two input parameters $r$ and $\eta$. Figure~\ref{fig:F6_x3} displays contour plots in the $\eta/\pi-r$ plane, showing the $3\sigma$ confidence regions for the mixing parameters $\sin^{2}\theta_{13}$, $\sin^{2}\theta_{23}$, CP violation phase $\delta_{CP}/\pi$ and the squared mass ratio $m_{2}^2/m_{3}^2$. 

From the left panel of figure~\ref{fig:F6_x3}, it can be seen that the values of observables are significantly influenced by the phase parameter $\eta$, with specific $\eta$ values enabling the model predictions for both neutrino mixing angles and mass parameters to achieve excellent agreement with experimental constraints. The best fit values of the neutrino oscillation parameters and mass spectrum for some  benchmark values of $\eta$ are systematically presented in table~\ref{tab:viable_BP}. Here, we fix the parameter $\eta=\frac{5\pi}{3}$ for simplicity. The corresponding correlations between all aforementioned physical observables and the reactor mixing angle $\sin\theta_{13}$  are displayed in figure~\ref{fig:F6_x3}. Through comprehensive numerical scanning of the $m_{a}-r$ parameter space, we identify the experimentally viable ranges:
\begin{equation}
m_{a}\in [26.345\,\text{meV}, 27.210\,\text{meV}], \qquad  r\in[0.0954,0.104]\,,
\end{equation}
which remain consistent with current neutrino oscillation data at the $3\sigma$ confidence level~\cite{Esteban:2024eli}. Additionally, our calculations reveal highly constrained predictions for neutrino mass parameters and mixing angles, specifically:
\begin{eqnarray}
\nonumber && 0.02030\leq \sin^2\theta_{13}\leq0.02388, \qquad 0.317\leq \sin^2\theta_{12}\leq0.320, \qquad 0.483\leq \sin^2\theta_{23}\leq0.494, \\
\nonumber &&1.477\leq\delta_{CP}/\pi\leq1.492, \qquad  1.585\leq\beta/\pi\leq1.608, \qquad  2.587\,\text{meV}\leq m_{ee}\leq2.801\,\text{meV}, \\
\label{eq:model_prediction_F6} &&m_{1}=0, \qquad 8.319\,\text{meV}\leq m_2\leq8.942\,\text{meV}, \qquad 49.508\,\text{meV}\leq m_3\leq50.774\,\text{meV}\,.
\end{eqnarray}
Therefore the breaking pattern $\mathcal{F}_{6}$ with $x=3$ and $\eta=5\pi/3$ is very predictive and it should be easily excluded by precise measurement of $\theta_{12}$, $\theta_{23}$ and $\delta_{CP}$ in forthcoming neutrino facilities.

\begin{table}[t!]
\begin{center}
\renewcommand{\tabcolsep}{0.5mm}
\renewcommand{\arraystretch}{1.2}
\small
\begin{tabular}{|c|c|c|c|c|c|c|c|c|c|c|c|c|c|c|}\hline\hline
	\multicolumn{15}{|c|}{For the NO case}\\\hline
& $\bm{\hat{v}}_{\text{fix}}$ & $x$& $\eta/\pi$ & $\frac{m_{a}}{\text{meV}}$ & $r$ & $\chi^{2}_{\text{min}}$ & $\sin^{2}\theta_{13}$ & $\sin^{2}\theta_{12}$ & $\sin^{2}\theta_{23}$ & $\delta_{CP}/\pi$ & $\beta/\pi$ & $\frac{m_{2}}{\text{meV}}$ & $\frac{m_{3}}{\text{meV}}$ & $\frac{m_{\text{ee}}}{\text{meV}}$ \\ \hline
\multirow{2}*{$\mathcal{F}_{6}$} &	\multirow{2}*{$\left(\sqrt{\frac{2}{3}},\frac{1}{\sqrt{6}},\frac{1}{\sqrt{6}}\right)^{T}$}  & \multirow{2}*{$3$}& $\frac{5}{3}$ & 26.772 & 0.100 & 6.421 &0.02244  & 0.318 & 0.487 & 1.483 & 1.599 & 8.609 & 50.158 & 2.688 \\ \cline{4-15}
 & & & $\frac{17}{10}$ & $27.939$ & $0.0949$ & $21.102$ & $0.02104$ & $0.319$ & $0.517$ & $1.524$ & $1.551$ & $8.884$ & $50.006$ & $2.650$\\ \hline

\multirow{4}*{$\mathcal{F}_{7}$} &	\multirow{2}*{$\left(\sqrt{\frac{2}{3}},\frac{1}{\sqrt{6}},\frac{1}{\sqrt{6}}\right)^{T}$}  & \multirow{2}*{$3$}& $\frac{4}{3}$ & 26.786 & 0.100 & 11.598& 0.02230& 0.318& 0.512& 1.517 & 0.402 &8.590 & 50.170 & 2.681 \\ \cline{4-15}
 & & & $\frac{13}{10}$ & 27.947 & 0.0946 & 14.092 & 0.02093 & 0.319 & 0.483 & 1.476 & 0.449  & 8.867 & 50.016 & 2.645\\  \cline{2-15}
&	\multirow{2}*{$\frac{1}{5}\left(\sqrt{17},2,2\right)^{T}$}  & \multirow{2}*{$\frac{2+\sqrt{17}}{2}$}& $\frac{4}{3}$ & 26.725 & 0.0975 & 11.301 & 0.02181 & 0.305 & 0.513 & 1.518 & 0.401 & 8.690 & 50.111 & 2.607 \\ \cline{4-15}
 & & & $\frac{13}{10}$ & 27.899 & 0.0920& 22.560 & 0.02037 & 0.306 & 0.484& 1.477 & 0.449 &  8.955 & 49.967 & 2.566\\  \hline

\multirow{5}*{$\mathcal{F}_{4}$} &	$\multirow{5}*{$\frac{1}{\sqrt{38}}\left(
5,2,3
\right)^{T}$}$ &$\multirow{5}*{$5$}$& $0$ & $2.907$ & $0.338$ & $4.003$ & $0.02239$ & $0.327$ & $0.486$ & $1$ & $0$ & $8.653$ & $50.131$ & $3.889$ \\ \cline{4-15}
&	$~$ &$~$& $\frac{1}{3}$ & $2.909$ & $0.338$ & $7.287$ & $0.02123$ & $0.328$ & $0.492$ & $0.985$ & $0.334$ & $8.664$ & $50.125$ & $3.505$ \\ \cline{4-15}
&	$~$ &$~$& $\frac{5}{3}$ & $2.909$ & $0.338$ & $7.078$ & $0.02123$ & $0.328$ & $0.492$ & $1.015$ & $1.666$ & $8.664$ & $50.125$ & $3.505$ \\ \cline{4-15}
&	$~$ &$~$& $\frac{1}{4}$ & $2.909$ & $0.338$ & $4.957$ & $0.02171$ & $0.328$ & $0.489$ & $0.988$ & $0.251$ & $8.661$ & $50.127$ & $3.669$ \\ \cline{4-15}
&	$~$ &$~$& $\frac{7}{4}$ & $2.909$ & $0.338$ & $4.788$ & $0.02171$ & $0.328$ & $0.489$ & $1.012$ & $1.749$ & $8.661$ & $50.127$ & $3.669$ \\ \hline
$\multirow{2}*{$\mathcal{T}_{10}$}$&	$\multirow{2}*{$\left(\sqrt{\frac{37}{57}},\sqrt{\frac{10}{57}},\sqrt{\frac{10}{57}}\right)^{T}$}$ &$\multirow{2}*{$4$}$& $1$ & $19.289$ & $0.0602$ & $14.058$ & $0.02270$ & $0.336$ & $0.482$ & $1.474$ & $1.165$ & $8.465$ & $50.244$ & $3.868$\\\cline{4-15}
& & & $\frac{11}{10}$ & $20.031$ & $0.0577$ & 21.972 & $0.02183$ & $0.336$ & $0.525$ & $1.537$ & $1.111$ & $8.783$ & $50.061$ & $3.851$\\ \hline
$\multirow{2}*{$\mathcal{S}_{2}$}$&	$\multirow{2}*{$\frac{1}{3}\left(\sqrt{6},1,\sqrt{2}\right)^{T}$}$ &$\multirow{2}*{$\frac{1+\sqrt{11}}{2}$}$& $\frac{3}{4}$ & $14.432$ & $0.237$ & $9.469$ &0.02133 & $0.319$ & $0.472$ & $0.893$ & $1.875$ & $8.890$ & $50.003$ & $3.422$\\ \cline{4-15}
& & & $\frac{4}{5}$ & $13.849$ & $0.249$ &3.219 &0.02237  & $0.318$ & $0.459$ & $0.932$ & $1.837$ & $8.562$ & $50.186$ & $3.448$ \\\hline \hline 
	\multicolumn{15}{|c|}{For the IO case}\\\hline 
	& $\bm{\hat{v}}_{\text{fix}}$ & $x$& $\eta/\pi$ & $\frac{m_{a}}{\text{meV}}$ & $r$ & $\chi^{2}_{\text{min}}$ & $\sin^{2}\theta_{13}$ & $\sin^{2}\theta_{12}$ & $\sin^{2}\theta_{23}$ & $\delta_{CP}/\pi$ & $\beta/\pi$ & $\frac{m_{1}}{\text{meV}}$ & $\frac{m_{2}}{\text{meV}}$ & $\frac{m_{\text{ee}}}{\text{meV}}$ \\
	\hline
	$\multirow{2}*{$\mathcal{F}_{2}^{\prime}$}$&	$\multirow{2}*{$\frac{1}{2\sqrt{11}}\left(1,5,3\sqrt{2}\right)^{T}$}$ &$\multirow{2}*{$7$}$& $\frac{52}{27}$ & $21.847$ & $0.0587$ & $14.672$ &0.02273 & $0.300$ & $0.581$ & $1.700$ & $1.632$ & $49.284$ & $49.991$ & $41.839$\\ \cline{4-15}
	& & & $\frac{79}{41}$ & $21.670$ & $0.0587$ &17.345 &0.02273  & $0.318$ & $0.581$ & $1.708$ & $1.639$ & $48.829$ & $49.646$ & $41.484$ \\
 \hline\hline
\end{tabular}
\caption{\label{tab:viable_BP}The best fit values of the five interesting breaking patterns $\mathcal{F}_{6}$, $\mathcal{F}_{7}$, $\mathcal{F}_{4}$, $\mathcal{T}_{10}$, $\mathcal{S}_{2}$ and $\mathcal{F}_{2}^{\prime}$ for some benchmark values of parameters  $x$ and $\eta$, where $\bm{\hat{v}}_{\text{fix}}$ represents the absolute values of the first (third) of PMNS matrix for the NO (IO) case. 
}
\end{center}
\end{table}

\subsection{The breaking pattern $\mathcal{F}_{7}$}

For this breaking pattern, the residual flavor symmetry combination is $(G_{l},G_{atm},G_{sol})=(Z_{4}^{c},Z_{2}^{a^2bd^2},Z_{2}^{ab})$ with the residual CP transformation $\mathbb{1}_{3}$ in both the atmospheric  and the solar neutrino sectors. Then the corresponding VEV alignments of flavor fields $\phi_{\text{atm}}$ and $\phi_{\text{sol}}$ are
\begin{eqnarray}
		\langle\phi_{\text{atm}}\rangle =(0,1,1)^{T}v_{\phi_{a}},\qquad
		\langle\phi_{\text{sol}}\rangle =(1,1,x)^{T}v_{\phi_{s}}\,.
\end{eqnarray}
The light neutrino mass matrix $m_{\nu}$ reads as 
\begin{equation}
m_{\nu }= m_{a}\left[
		\begin{pmatrix}
			0 & 0 & 0 \\
			0 & 1 & 1 \\
			0 & 1 & 1 \\
		\end{pmatrix}
		+re^{i \eta }  
		\begin{pmatrix}
			1 & 1 & x \\
			1 & 1 & x \\
			x & x & x^2 \\
		\end{pmatrix}\right]\,,
\end{equation}
which can be simplified to a block diagonal form $m^{\prime}_{\nu}$ by performing a $U_{\nu1}$ transformation with 
\begin{equation}
U_{\nu1}=
\begin{pmatrix}
\frac{x-1}{\sqrt{(x-1)^2+2}} & \frac{\sqrt{2}}{\sqrt{(x-1)^2+2}} & 0 \\
\frac{1}{\sqrt{(x-1)^2+2}} & \frac{1-x}{\sqrt{2 (x-2) x+6}} & \frac{1}{\sqrt{2}} \\
-\frac{1}{\sqrt{(x-1)^2+2}} & \frac{x-1}{\sqrt{2 (x-2) x+6}} & \frac{1}{\sqrt{2}} \\
\end{pmatrix}\,.
\end{equation}
Then we can obtain the three nonzero elements $y$, $z$ and $w$ of $m^{\prime}_{\nu}$:
\begin{equation}
y=\frac{m_{a}}{2} r [(x-1)^2 +2]e^{i \eta }, ~~ z=\frac{m_{a}}{2}  r (x+1) \sqrt{(x-1)^2 +2}e^{i \eta }, ~~ w=\frac{m_{a}}{2} \left[4+  r (x+1)^2e^{i \eta }\right]\,.
\end{equation}
The neutrino mass matrix $m^{\prime}_{\nu}$ can be diagonalized by performing the unitary transformation $U_{\nu2}$. As a consequence, the lepton mixing matrix takes the following form
\begin{equation}
U_{PMNS}=P_{132}U_{\nu1}U_{\nu2}\,.
\end{equation}
Then the analytical expressions for the three lepton mixing angles and CP invariants can be readily derived. When treating $m_{a}$,  $x$, $r$ and $\eta$ as free parameters, we present the best fit values of  the neutrino masses and mixing parameters  in table~\ref{tab:viable_best_fit}. Following a thorough exploration of the parameter space, we determine the predicted ranges for input parameters and observables, as systematically compiled in table~\ref{tab:viable_regions}.

To provide concrete illustrations, we analyze selected benchmark values for parameters $x$ and $\eta$, and summarize the resulting mixing parameters and neutrino masses in table~\ref{tab:viable_BP}. Notably, the benchmark choice $x=3$ yields the solar flavon alignment $\langle\phi_{\text{sol}}\rangle =(1,1,3)^{T}v_{\phi_{s}}$, aligning with the littlest seesaw model with CSD(3)  as initially presented in  Ref.~\cite{King:2015dvf}. This configuration corresponds to the TM1 pattern in the lepton mixing matrix.

\begin{figure}[t!]
\centering
\begin{tabular}{c}
\includegraphics[width=0.48\linewidth]{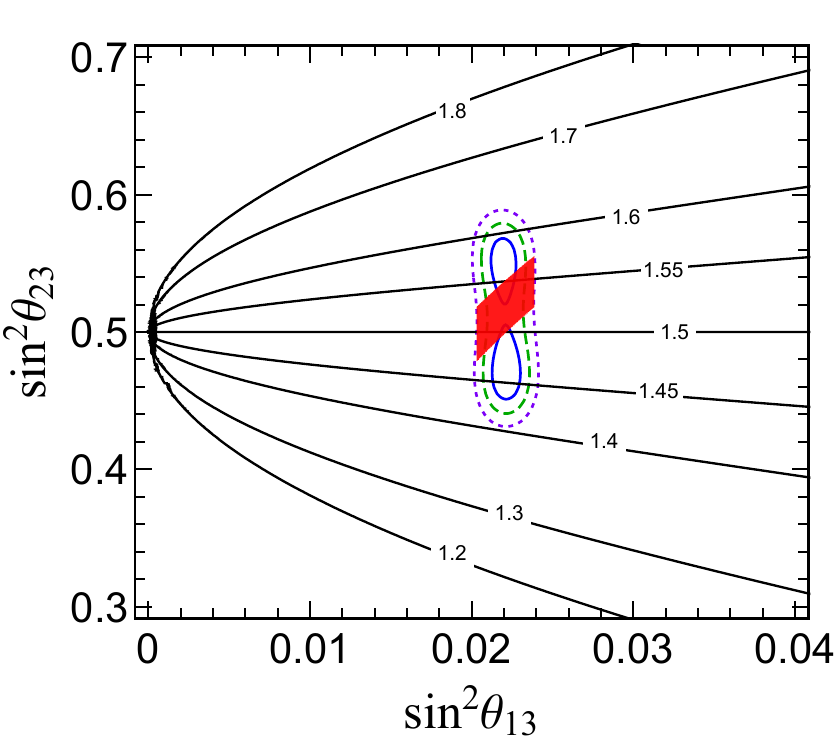}
\includegraphics[width=0.5\linewidth]{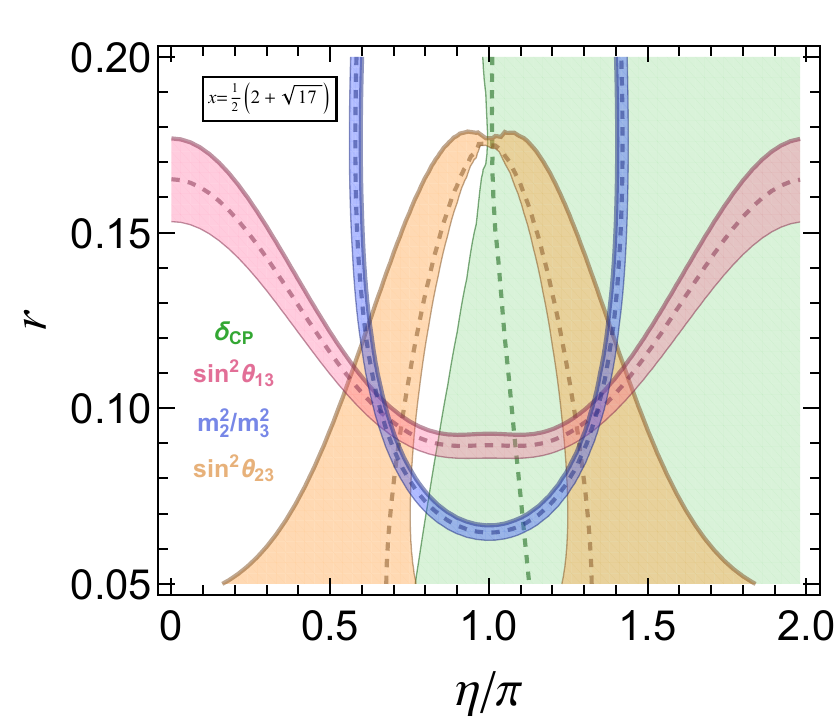}
\end{tabular}
\caption{\label{fig:contour_CP_F7} Contour plot of $\delta_{CP}/\pi$ within $\sin^2\theta_{13}-\sin^2\theta_{23}$ plane, and the dimensionless observable quantities $\sin^{2}\theta_{23}$, $\sin^{2}\theta_{13}$, $\delta_{CP}$ and $m_{2}^2/m_{3}^2$ in the $\eta$/$\pi-r$ plane for the breaking pattern $\mathcal{F}_{7}$ with $x=(2+\sqrt{17})/2\approx3$.
 }
\end{figure}

Moreover, we observe that for the benchmark value $x=(2+\sqrt{17})/2\approx3$, the absolute values of the first column in the PMNS matrix are determined to be

\begin{equation}\label{eq:fix_column_F7}
\frac{1}{5}\left(\sqrt{17},2,2\right)^{T},
\end{equation}
a result we find particularly intriguing. For this case, the three lepton mixing angles and the CP invariants are given by
\begin{eqnarray}
		\nonumber
		\sin^{2}\theta_{13}&=&\frac{8\sin ^2\theta }{25},\quad 
		\sin^{2}\theta_{12}=\frac{8 \cos ^2\theta }{21+4\cos 2 \theta },\quad 
		\sin^{2}\theta_{23}=\frac{1}{2}+\frac{5 \sqrt{17} \sin 2 \theta  \cos \psi }{42+8 \cos 2 \theta },\\
	J_{CP}&=&-\frac{2\sqrt{17}}{125}\sin 2\theta  \sin \psi \,, \qquad	I_{1 }=\frac{16}{625}\sin ^2 2 \theta  \sin (\rho -\sigma )\,.
\end{eqnarray}	
The corresponding mixing matrix indicates the following sum rule among the Dirac CP phase $\delta_{CP}$ and mixing angles
\begin{equation}\label{eq:F7_sum_rules}
\cos^2\theta_{12}\cos^2\theta_{13}=\frac{17}{25}\,, \qquad \cos\delta_{CP}=\frac{(13-21 \cos 2 \theta _{13}) \csc \theta _{13} \cot 2 \theta _{23}}{2 \sqrt{425 \cos ^2\theta _{13}-289}}\,.
\end{equation}
The left panel of figure~\ref{fig:contour_CP_F7} illustrates the contour map of $\delta_{CP}/\pi$ in the $\sin^2\theta_{13}-\sin^2\theta_{23}$ plane. With the parameter $x$ held constant, all mixing parameters and the mass ratio $m_{2}/m_{3}$ become fully determined by the two input parameters $r$ and $\eta$. The red shaded region in the left panel represents the viable parameter space, where $r$ and $\eta$ are numerically scanned across their experimentally allowed ranges while satisfying $3\sigma$ constraints on the five  dimensionless observable quantities in table~\ref{tab:bf_13sigma_data}. This constrained parameter space reveals that both the atmospheric mixing angle $\theta_{23}$ and the Dirac CP violation phase $\delta_{CP}$ are predicted to lie in narrow intervals near their theoretically maximal values. The right panel of figure~\ref{fig:contour_CP_F7} displays contour bands corresponding to those dimensionless observable quantities except $\theta_{12}$ within the $\eta/\pi-r$ parameter plane. It shows that the input parameters $r$ and $\eta$ are restricted to a narrow region of the parameter space. The solar mixing angle $\theta_{12}$ is governed by the reactor angle $\theta_{13}$ through the sum rule derived in Eq.~\eqref{eq:F7_sum_rules}, establishing a predictive relationship between these two oscillation parameters.

\begin{figure}[t!]
\centering
\begin{tabular}{c}
\includegraphics[width=0.5\linewidth]{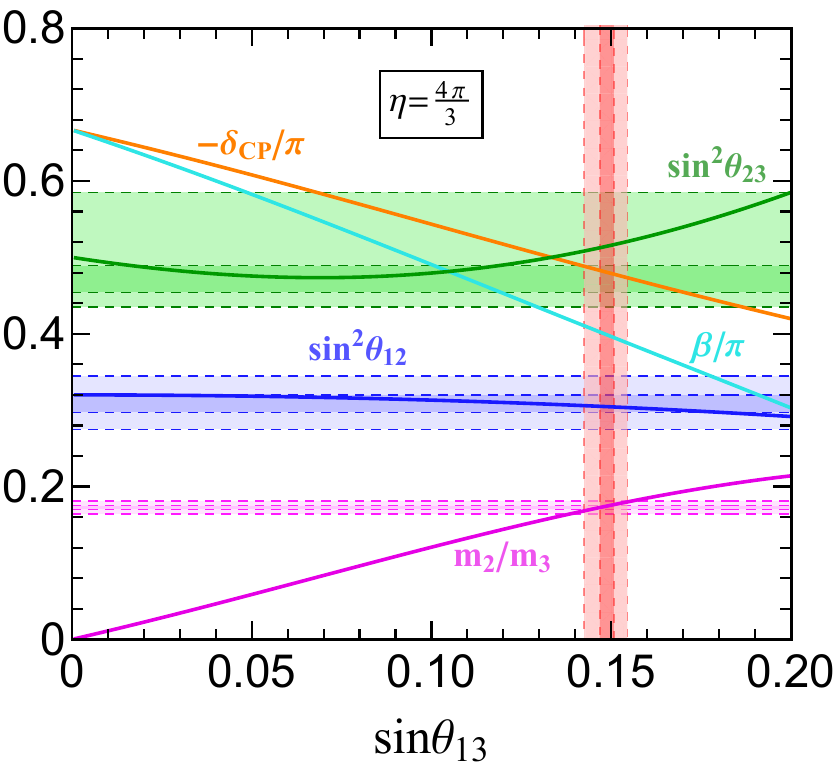}
\end{tabular}
\caption{\label{fig:Mix_par_F7} The predicted values of the breaking pattern $\mathcal{F}_{7}$ with $x=(2+\sqrt{17})/2$ and $\eta=4\pi/3$ for the lepton mixing parameters and mass ratio $m_2/m_3$ as a function of $\sin\theta_{13}$.
 }
\end{figure}

In the present analysis, we direct our attention to the parametrization $x=(2+\sqrt{17})/2$. This specific choice generates an interesting mixing matrix structure characterized by its first column component explicitly given in Eq.~\eqref{eq:fix_column_F7}. The phase parameter $\eta$ is constrained to $\eta=4\pi/3$, a value that proves theoretically accessible for implementation within concrete model frameworks. This optimized parameter selection yields a phenomenologically successful and predictive description of lepton mixing parameters and neutrino masses. All lepton mixing parameters and the neutrino mass ratio $m_{2}/m_{3}$ are now determined by a single dimensionless parameter $r$. Upon elimination of the free parameter $r$, sum rules emerge among these quantities. While their explicit analytical forms are omitted here, we present the numerical correlations graphically. Figure~\ref{fig:Mix_par_F7} illustrates the functional dependence of $\sin^2\theta_{12}$, $\sin^2\theta_{23}$, the Dirac CP phase $\delta_{CP}$, the Majorana CP phase $\beta$ and mass ratio $m_2/m_3$ on $\sin\theta_{13}$. Furthermore, we numerically scan over the parameter space of $m_{a}$ and $r$. We find that the input parameters, the neutrino masses and mixing parameters are predicted to lie in the following rather narrow regions
\begin{eqnarray}
\nonumber && 26.280\,\text{meV} \leq m_{a}\leq 27.162\,\text{meV}, \qquad  0.0941\leq r \leq0.102\,, \\
\nonumber &&0.02031\leq \sin^2\theta_{13}\leq0.02387, \qquad 0.303\leq \sin^2\theta_{12}\leq0.306, \qquad  0.508\leq \sin^2\theta_{23}\leq0.521, \\
\nonumber &&1.512\leq\delta_{CP}/\pi\leq1.527, \qquad  0.388\leq\beta/\pi\leq0.410, \qquad  2.491\,\text{meV}\leq m_{ee}\leq2.705\,\text{meV}, \\
\label{eq:model_prediction_F7} &&m_{1}=0, \qquad 8.331\,\text{meV}\leq m_2\leq8.972\,\text{meV}, \qquad 49.508\,\text{meV}\leq m_3\leq50.774\,\text{meV}\,.
\end{eqnarray}
These predictions could be tested by the JUNO experiment~\cite{JUNO:2022mxj} in combination with DUNE~\cite{DUNE:2020ypp} or T2HK~\cite{Hyper-Kamiokande:2018ofw}.

\subsection{The breaking pattern $\mathcal{F}_{4}$}

For this breaking pattern, the preserved flavor symmetry groups are $(G_{l},G_{atm},G_{sol})=(Z_{4}^{c},Z_{3}^{ac},Z_{2}^{bc^3d^3})$, with the atmospheric and solar neutrino sectors exhibiting residual CP transformations characterized by $\rho_{\bm{\bar{3}_{0}}}(d^{2})$ and $\rho_{\bm{\bar{3}_{1}}}(d^{2})$, respectively. The resulting vacuum alignments for the flavon fields yield:
\begin{eqnarray}
\langle\phi_{\text{atm}}\rangle =(1,-i,-i)^{T}v_{\phi_{a}},\qquad \langle\phi_{\text{sol}}\rangle =(x,i,ix)^{T}v_{\phi_{s}}\,.
\end{eqnarray}
The corresponding light neutrino mass matrix $m_{\nu}$ can be simplified to a block diagonal form $m^{\prime}_{\nu}$ by performing a unitary transformation $U_{\nu1}$, where the unitary matrix is 
\begin{equation}
U_{\nu1}=
\begin{pmatrix}
\frac{x-1}{\sqrt{2(3 x^2+1)}} & \frac{1}{\sqrt{3}} & \frac{3 x+1}{\sqrt{6(3 x^2+1)}}  \\
-\frac{2 i x}{\sqrt{2(3 x^2+1)}} & \frac{i}{\sqrt{3}} & -\frac{2 i}{\sqrt{6(3 x^2+1)}}  \\
\frac{i (x+1)}{\sqrt{2(3 x^2+1)}} & \frac{i}{\sqrt{3}} & \frac{i-3 i x}{\sqrt{6(3 x^2+1)}}  \\
\end{pmatrix}\,,
\end{equation}
As demonstrated in Eq.~\eqref{eq:yzw}, the nonzero entries $y$, $z$ and $w$ of $m^{\prime}_{\nu}$ take the following form
\begin{equation}
y=\frac{m_{a}}{3}  (9+re^{i \eta }), \qquad  z=-\frac{m_{a}}{3} \sqrt{6 x^2+2}\,r e^{i \eta }, \qquad  w=\frac{2m_{a}}{3} (3 x^2+1)re^{i \eta }\,.
\end{equation}
Following the diagonalization procedure outlined in appendix~\ref{sec:Diag_mnup}, we find that the neutrino mass matrix $m^{\prime}_{\nu}$ can be diagonalized through the unitary transformation $U_{\nu2}$. Consequently, the PMNS matrix is obtained as
\begin{equation}
U_{PMNS}=P_{213}U_{\nu1}U_{\nu2}\,.
\end{equation}
The lepton mixing angles and CP invariants $J_{CP}$ and $I_{1}$ can be directly extracted through this mixing matrix, though we refrain from providing their explicit mathematical expressions here. By treating $m_{a}$, $x$, $r$ and $\eta$  as free parameters, we determine the best fit neutrino masses and mixing angles. Please see table~\ref{tab:viable_best_fit}. A full parameter scan yields allowed ranges for input parameters, mixing angles, CP phases, and neutrino masses, and we summarize them in table~\ref{tab:viable_regions}.

To provide illustrative examples, specific benchmark values for the parameters $x$ and $\eta$ are selected, with corresponding numerical results for mixing parameters and neutrino masses cataloged in table~\ref{tab:viable_BP}. With these fixed, the neutrino mass matrix $m_{\nu}$ depends only on $m_{a}$ and $r$, which can be determined by $\Delta m^2_{21}$ and $\Delta m^2_{31}$. This allows predictions for all three mixing angles and two CP violation phases. For the benchmark value $x=5$, the solar flavon alignment is $\langle\phi_{\text{sol}}\rangle =(5,i,5i)^{T}v_{\phi_{s}}$ and the absolute value of the first column of PMNS matrix is fixed to be:
\begin{eqnarray}
\frac{1}{\sqrt{38}}\left(
5,2,3
\right)^{T}\,,
\end{eqnarray}
which leads to the following  sum rules among mixing angles and Dirac CP phase
\begin{equation}\label{eq:F4_sum_rules}
\cos^2\theta_{12}\cos^2\theta_{13}=\frac{25}{38}\,, \qquad \cos\delta_{CP}=\frac{(37-63 \cos 2 \theta _{13})  \cot 2 \theta _{23}-10 \cos^2 \theta _{13} \csc2 \theta _{23}}{20\sin\theta_{13} \sqrt{19 \cos 2 \theta _{13}-6}}\,.
\end{equation}
By incorporating the experimentally determined $3\sigma$  ranges of $\theta_{13}$ and $\theta_{23}$ given in table~\ref{tab:bf_13sigma_data}, we obtain the prediction for the solar mixing angle and Dirac CP phase:
\begin{equation}
 0.326\leq\sin^2\theta_{12}\leq 0.328, \qquad 0.687\leq\delta_{CP}/\pi\leq1.313\,.
\end{equation}
These precise constraints indicate that the theoretical framework can be tested by future experiments like JUNO, DUNE, or T2HK, which are sensitive enough to detect CP violation and measure neutrino mixing angles.

\begin{figure}[t!]
\centering
\begin{tabular}{c}
\includegraphics[width=0.5\linewidth]{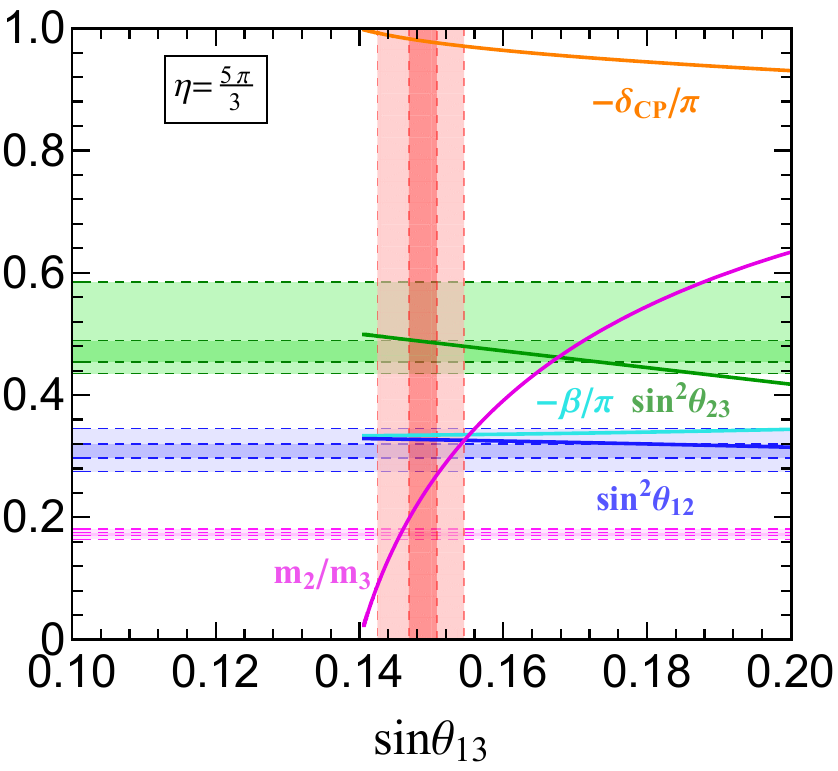}
\end{tabular}
\caption{\label{fig:Mix_par_F4} The predicted values of the breaking pattern $\mathcal{F}_{4}$ with $x=5$ and $\eta=5\pi/3$ for the lepton mixing parameters and mass ratio $m_2/m_3$ as a function of $\sin\theta_{13}$.
 }
\end{figure}

We now focus on the simple values $x=5$ and $\eta=5\pi/3$, which successfully predict lepton mixing and neutrino masses. Moreover, the corresponding  vacuum alignments of the flavon fields $\phi_{\text{atm}}$ and $\phi_{\text{sol}}$, and the phase $\eta$ should be easily achievable in a realistic model. Now, all lepton mixing parameters and $m_{2}/m_{3}$ depend solely on the dimensionless parameter $r$, which can be determined by $\theta_{13}$ measurements. In figure~\ref{fig:Mix_par_F4}, we show correlations between $\theta_{13}$ and other observables. If $m_{2}/m_{3}$ is constrained to its $3\sigma$ range, the  input parameters and all mixing parameters fall within narrow predicted regions.
\begin{eqnarray}
\nonumber && 2.794\,\text{meV} \leq m_{a}\leq 3.012\,\text{meV}, \qquad  0.322\leq r \leq 0.356\,, \\
\nonumber &&0.02112\leq \sin^2\theta_{13}\leq0.02134, \qquad 0.3278\leq \sin^2\theta_{12}\leq0.3279, \qquad  0.491\leq \sin^2\theta_{23}\leq0.492, \\
\nonumber &&1.014\leq\delta_{CP}/\pi\leq1.016, \qquad  1.6659\leq\beta/\pi\leq1.6660, \qquad  3.385\,\text{meV}\leq m_{ee}\leq3.613\,\text{meV}, \\
\label{eq:model_prediction_F4} &&m_{1}=0, \qquad 8.321\,\text{meV}\leq m_2\leq8.970\,\text{meV}, \qquad 49.508\,\text{meV}\leq m_3\leq50.774\,\text{meV}\,.
\end{eqnarray}
Note that the  Dirac CP violation phase $\delta_{CP}$ is predicted to lie in narrow intervals near $\pi$ for the breaking pattern $\mathcal{F}_{4}$ with  $x=5$ and $\eta=5\pi/3$. It yields distinct predictions for $\delta_{CP}$ compared to the original Littlest Seesaw model~\cite{King:2015dvf} and  the breaking patterns $\mathcal{F}_{6}$ and $\mathcal{F}_{7}$.

\subsection{The breaking pattern $\mathcal{T}_{10}$}

The residual symmetries in atmospheric neutrino and solar neutrino sectors require that the VEVs of the flavons  $\phi_{\textrm{atm}}$ and $\phi_{\textrm{sol}}$ are taken to be
\begin{eqnarray}
	\langle\phi_{\text{atm}}\rangle =(1,0,i)^{T}v_{\phi_{a}},\qquad
	\langle\phi_{\text{sol}}\rangle =(1,-1,x)^{T}v_{\phi_{s}}\,.
\end{eqnarray}
Applying the seesaw formula, we obtain the effective light neutrino mass matrix
\begin{equation}
	m_{\nu }= m_{a}\left[
	\begin{pmatrix}
		1 & 0 & i \\
		0 & 0 & 0 \\
		i & 0 & -1 \\
	\end{pmatrix}
	+re^{i \eta }  
	\begin{pmatrix}
		1 & -1 & x \\
		-1 & 1 & -x \\
		x & -x & x^2 
	\end{pmatrix}\right]\,,
\end{equation}
which can be transformed into a block diagonal matrix structure $m^{\prime}_{\nu}$ by unitary matrix $U_{\nu1}$ with
\begin{equation}
	U_{\nu1}=
	\begin{pmatrix}
		\frac{i}{\sqrt{x^2+3}} & \frac{1-i x}{\sqrt{2(x^2+3)}} & \frac{1}{\sqrt{2}} \\
		\frac{-x+i}{\sqrt{x^2+3}} & -\frac{2}{\sqrt{2(x^2+3)}} & 0 \\
		-\frac{1}{\sqrt{x^2+3}} & \frac{x+i}{\sqrt{2(x^2+3)}} & -\frac{i}{\sqrt{2}} \\
	\end{pmatrix}\,.
\end{equation}
Then we can obtain the three nonzero elements $y$, $z$ and $w$ of $m^{\prime}_{\nu}$:
\begin{equation}
	y=\frac{m_{a}}{2} r (x^2+3)e^{i \eta }, ~~ z=\frac{m_{a}}{2}  r (1-ix) \sqrt{x^2+3}e^{i \eta }, ~~ w=\frac{m_{a}}{2} \left[4-  r (x+i)^2e^{i \eta }\right]\,.
\end{equation}
Similarly, the neutrino mass matrix $m^{\prime}_{\nu}$ can be further diagonalized by performing the unitary transformation $U_{\nu2}$. Therefore, we can write the lepton mixing matrix as
\begin{equation}
		U_{PMNS}=P_{321}U_{l}^{(1)\dagger}U_{\nu1}U_{\nu2}\,.
\end{equation}
The lepton mixing angles and the Jarlskog invariant can be naturally derived from the PMNS matrix.  By comprehensively scanning over the parameter space of $m_{a}$, $x$, $r$ and $\eta$, we  present the predicted ranges for input parameters, three lepton mixing angles, Dirac CP phase, Majorana CP phase and neutrino masses  in table~\ref{tab:viable_regions}. Furthermore, we obtain the best fit values of the input and output parameters, as shown in table~\ref{tab:viable_best_fit}.

\begin{figure}[t!]
	\centering
	\begin{tabular}{c}
		\includegraphics[width=0.5\linewidth]{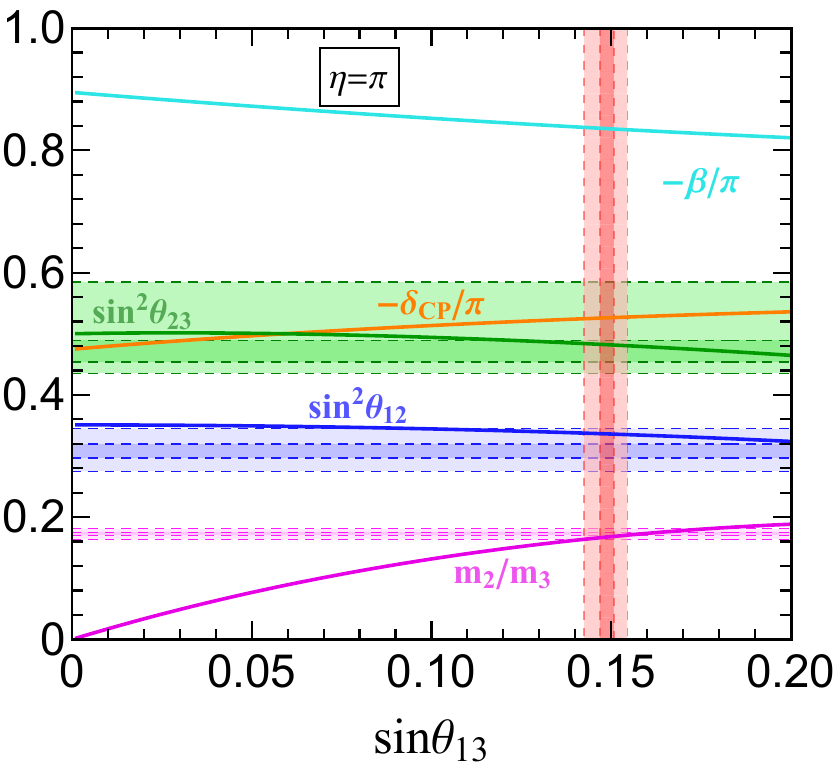}
	\end{tabular}
	\caption{\label{fig:Mix_par_T10} The predicted values of the breaking pattern $\mathcal{T}_{10}$ with $x=4$ and $\eta=\pi$ for the lepton mixing parameters and mass ratio $m_2/m_3$ as a function of $\sin\theta_{13}$.}
\end{figure}

To provide concrete outcomes for mixing parameters and neutrino masses, we select different values for parameters $x$ and $\eta$. The  corresponding best fit values of observables are summarized in table~\ref{tab:viable_BP}. Subsequently, we turn to the parametrization $x=4$ and  $\eta=\pi$, which is an interesting case that successfully fits the experiment. It yields the solar flavon alignment $\langle\phi_{\text{sol}}\rangle =(1,-1,4)^{T}v_{\phi_{s}}$, and the absolute value of the first column of PMNS matrix is fixed to be:  
\begin{equation}\label{eq:fix_column_T10}
	\frac{1}{\sqrt{57}}\left(\sqrt{37},\sqrt{10},\sqrt{10}\right)^{T},
\end{equation}
which leads to the following  sum rule among the Dirac CP phase $\delta_{CP}$ and mixing angles
\begin{equation}\label{eq:T10_sum_rules}
	\cos^2\theta_{12}\cos^2\theta_{13}=\frac{37}{57}\,, \qquad \cos\delta_{CP}=\frac{(27-47 \cos 2 \theta _{13})\cot 2 \theta _{23} \csc \theta_{13} }{\sqrt{74} \sqrt{57 \cos 2 \theta _{13}-17}}\,.
\end{equation}
In this case, the output parameters such as lepton mixing angles and the absolute  masses are all solely determined by two input parameters $m_{a}$ and $r$. By comprehensively scanning over the parameter space of $m_{a}$ and $r$, we constrain the corresponding input parameters, neutrino masses and mixing parameters  to the ranges below:
\begin{eqnarray}
	\nonumber && 18.845\,\text{meV} \leq m_{a}\leq 19.830\,\text{meV}, \qquad  0.0565\leq r \leq 0.0620\,, \\
	\nonumber &&0.02038\leq \sin^2\theta_{13}\leq0.02387, \qquad 0.335\leq \sin^2\theta_{12}\leq0.337, \qquad  0.480\leq \sin^2\theta_{23}\leq0.484, \\
	\nonumber &&1.473\leq\delta_{CP}/\pi\leq1.475, \qquad  1.162\leq\beta/\pi\leq1.166, \qquad  3.737\,\text{meV}\leq m_{ee}\leq3.992\,\text{meV}, \\
	\label{eq:model_prediction_T10} &&m_{1}=0, \qquad 8.319\,\text{meV}\leq m_2\leq8.660\,\text{meV}, \qquad 49.508\,\text{meV}\leq m_3\leq50.774\,\text{meV}\,.
\end{eqnarray}
After  eliminating the input parameter $r$, we can relate mixing parameters and mass ratios to the reactor mixing angle $\sin\theta_{13}$,  and  we present them in the figure~\ref{fig:Mix_par_T10}.

\subsection{The breaking pattern $\mathcal{S}_{2}$ }

The atmospheric and solar residual symmetries uniquely fix the vacuum alignments of $\phi_{\text{atm}}$ and $\phi_{\text{sol}}$ as follows
\begin{eqnarray}
	\langle\phi_{\text{atm}}\rangle =(1,-i,-i)^{T}v_{\phi_{a}},\qquad
	\langle\phi_{\text{sol}}\rangle =(1,1,x)^{T}v_{\phi_{s}}\,.
\end{eqnarray}
Based on Eq. \eqref{eq:mnu}, the neutrino mass matrix $m_{\nu}$ can be written in the following form 
\begin{equation}
	m_{\nu }= m_{a}\left[
	\begin{pmatrix}
		1 & -i & -i \\
		-i & -1 & -1 \\
		-i & -1 & -1 \\
	\end{pmatrix}
	+re^{i \eta }  
	\begin{pmatrix}
		1 & 1 & x \\
		1 & 1 & x \\
		x & x & x^2 \\
	\end{pmatrix}\right]\,,
\end{equation}
The diagonalization of $m_{\nu}$ can be achieved using the method provided by Eq. \eqref{eq:mnup}, yielding a block diagonal form $m^{\prime}_{\nu}$. The unitary matrix $U_{\nu1}$ is taken to be 
\begin{equation}
U_{\nu1}=
	\begin{pmatrix}
 \frac{i (1-x)}{\sqrt{2 \left(x^2-x+2\right)}} & \frac{i x+2+i}{\sqrt{6 \left(x^2-x+2\right)}} & \frac{1}{\sqrt{3}} \\
 \frac{-x-i}{\sqrt{2 \left(x^2-x+2\right)}} & \frac{-x+2-i}{\sqrt{6 \left(x^2-x+2\right)}} & \frac{i}{\sqrt{3}} \\
 \frac{1+i}{\sqrt{2 \left(x^2-x+2\right)}} & \frac{2 x-1-i}{\sqrt{6 \left(x^2-x+2\right)}} & \frac{i}{\sqrt{3}} \\
	\end{pmatrix}\,.
\end{equation}
Then we can obtain the three nonzero elements $y$, $z$ and $w$ of $m^{\prime}_{\nu}$:
\begin{equation}
	y=\frac{2m_{a}}{3} r (x^2-x+2)e^{i \eta }, ~~ z=\frac{m_{a}}{3}  r (1+i+ix) \sqrt{2 \left(x^2-x+2\right)}e^{i \eta }, ~~ w=\frac{m_{a}}{3} \left[ 9-r (x+1-i)^2e^{i \eta }\right]\,.
\end{equation}
The neutrino mass matrix $m^{\prime}_{\nu}$ can be further diagonalized by the unitary transformation $U_{\nu2}$. Thus, the PMNS matrix can be obtained in the following form via the systematic procedure described in Eq. \eqref{eq:genral_UPMNS}
\begin{equation}
U_{PMNS}=U_{l}^{(3)\dagger}U_{\nu1}U_{\nu2}\,.
\end{equation}
The explicit expressions for the mixing parameters can then be easily derived. We present the best fit values of neutrino mixing parameters and neutrino masses with input parameters $m_{a}$, $x$, $r$ and $\eta$ all being free parameters, as listed in table~\ref{tab:viable_best_fit}. The specific ranges of various parameters can be obtained and the results are presented in table \ref{tab:viable_regions}.

\begin{figure}[t!]
	\centering
	\begin{tabular}{c}
		\includegraphics[width=0.51\linewidth]{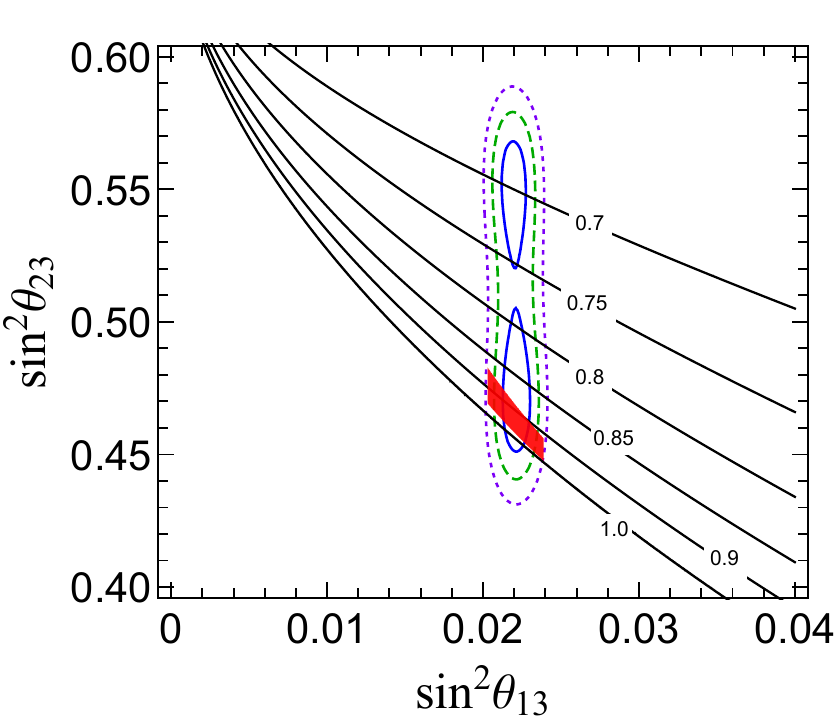}
		\includegraphics[width=0.5\linewidth]{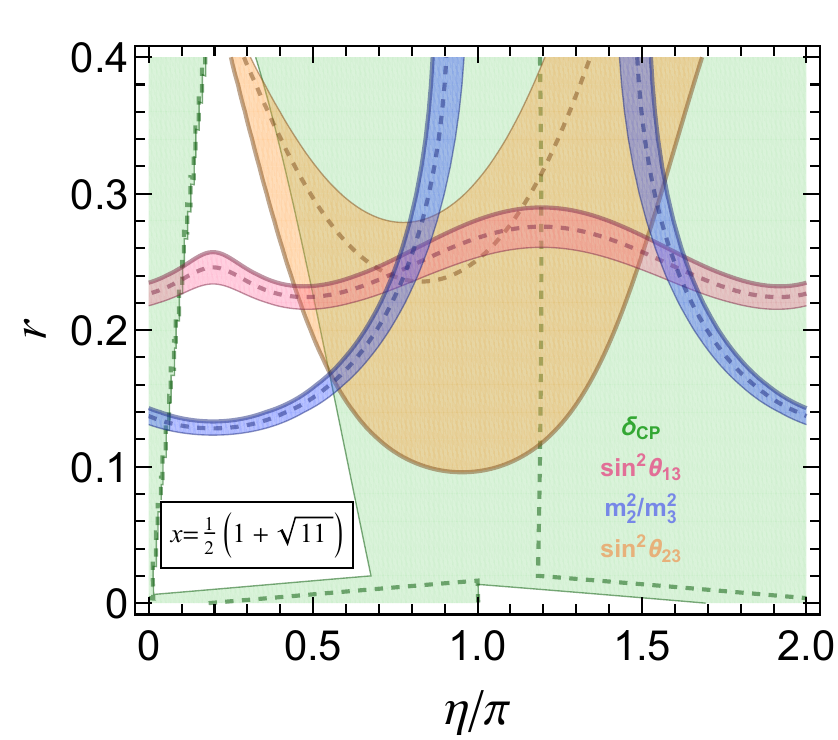}
	\end{tabular}
	\caption{\label{fig:contour_CP_S2} Contour plot of $\delta_{CP}/\pi$ within $\sin^2\theta_{13}-\sin^2\theta_{23}$ plane, and the dimensionless observable quantities $\sin^{2}\theta_{23}$, $\sin^{2}\theta_{13}$, $\delta_{CP}$ and $m_{2}^2/m_{3}^2$ in the $\eta$/$\pi-r$ plane for the breaking pattern $\mathcal{S}_{2}$ with $x=(1+\sqrt{11})/2$.
	}
\end{figure}

As an illustrative example, we choose the solar alignment parameter $x=(1+\sqrt{11})/2$ which gives  relatively simple values of the first column of the PMNS matrix
\begin{equation}\label{eq:fix_column_S2}
	\frac{1}{3}\left(\sqrt{6},1,\sqrt{2}\right)^{T},
\end{equation}
which is very interesting and has not so far appeared in the literature. Then we can extract the expressions of the mixing angles and CP invariants as follows
\begin{eqnarray}
	\nonumber
	&&\sin^{2}\theta_{13}=\frac{\sin ^2\theta }{3},~~
	\sin^{2}\theta_{12}=\frac{2\cos^2\theta }{5+\cos 2 \theta },~~  
	\sin^{2}\theta_{23}=\frac{2}{3}\left[2-\frac{6-\sin 2 \theta  (\cos \psi-\sqrt{11} \sin \psi )}{5+\cos 2 \theta }\right],\\
	&&J_{CP}=-\frac{1}{54}\sin2\theta(\sin \psi+\sqrt{11} \cos \psi ) \,, \qquad	I_{1 }=\frac{1}{36}\sin ^2 2 \theta  \sin (\rho -\sigma )\,.
\end{eqnarray}	
Furthermore, we can derive the following sum rule between the mixing angles $\theta_{12}$ and $\theta_{13}$ 
\begin{equation}\label{eq:S2_sum_rules}
	\cos^2\theta_{12}\cos^2\theta_{13}=\frac{2}{3}\,, 
\end{equation}
which is the same as the sum rule of TM1 mixing matrix. Additionally, we can express the Dirac CP phase $\delta_{CP}$ in terms of the mixing angles
\begin{equation}\label{eq:S2_Dirac_phase}
	 \cos\delta_{CP}=\frac{3 (3-5 \cos 2 \theta _{13}) \cos 2 \theta _{23}-2\cos^2 \theta _{13}}{12\sin\theta _{13}\sin2\theta _{23} \sqrt{3 \cos 2 \theta _{13}-1}}\,.
\end{equation}
The left panel of figure~\ref{fig:contour_CP_S2} illustrates the contour map of $\delta_{CP}/\pi$ in the $\sin^2\theta_{13}-\sin^2\theta_{23}$ plane. The red region represents the resulting area obtained by scanning parameters $m_{a}$, $\eta$ and $r$ within a reasonable range, followed by constraining all observable quantities according to their $3\sigma$ experimental ranges. From the figure, it is evident that the Dirac CP phase $\delta_{CP}$ is concentrated around $0.9\pi$.  Moreover, we plot the contour regions for the $3\sigma$ experimental ranges  of mixing angles $\theta_{13}$, $\theta_{23}$, the Dirac CP phase $\delta_{CP}$ and mass ratio $m_{2}/m_{3}$ in the plane $\eta$ and $r$ in the right panel of figure~\ref{fig:contour_CP_S2}.

\begin{figure}[t!]
	\centering
	\begin{tabular}{c}
		\includegraphics[width=0.5\linewidth]{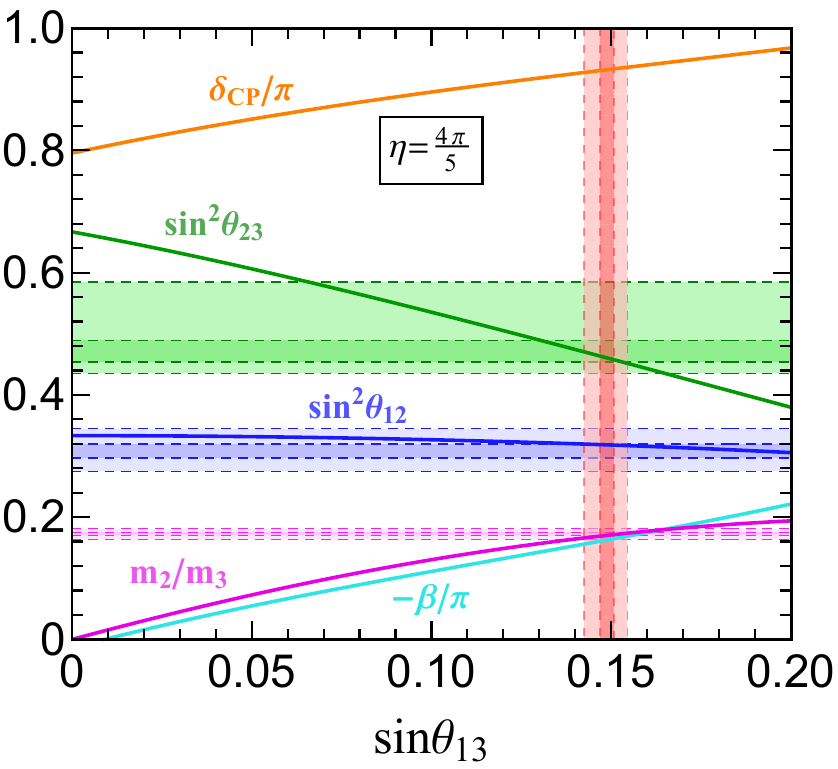}
	\end{tabular}
	\caption{\label{fig:Mix_par_S2} The predicted values of the breaking pattern $\mathcal{S}_{2}$ with $x=(1+\sqrt{11})/2$ and $\eta=4\pi/5$ for the lepton mixing parameters and mass ratio $m_2/m_3$ as a function of $\sin\theta_{13}$.
	}
\end{figure}

The numerical results for lepton mixing angles and neutrino masses arising from two additional special $\eta$ are tabulated in table \ref{tab:viable_BP}. In the following analysis,  the phase parameter $\eta$ is constrained to $\eta=4\pi/5$, a value that  can be easily realized in a concrete model. All lepton mixing parameters and the neutrino mass ratio $m_{2}/m_{3}$ are now determined by a single dimensionless parameter $r$. This leads to interesting predictions for mixing parameters and neutrino masses. Figure~\ref{fig:Mix_par_S2} illustrates the functional correlation of lepton mixing parameters, the Dirac CP phase $\delta_{CP}$, the Majorana CP phase $\beta$ and mass ratio $m_2/m_3$ on $\sin\theta_{13}$. Upon further scanning of parameters $m_{a}$ and $r$, it is found that the input parameters and output parameters are constrained to 
\begin{eqnarray}
	\nonumber && 13.578\,\text{meV} \leq m_{a}\leq 14.202\,\text{meV}, \qquad  0.236\leq r \leq 0.255\,, \\
	\nonumber &&0.02030\leq \sin^2\theta_{13}\leq0.02333, \qquad 0.317\leq \sin^2\theta_{12}\leq0.320, \qquad  0.454\leq \sin^2\theta_{23}\leq0.470, \\
	\nonumber &&0.927\leq\delta_{CP}/\pi\leq0.935, \qquad  1.833\leq\beta/\pi\leq1.844, \qquad  3.305\,\text{meV}\leq m_{ee}\leq3.550\,\text{meV}, \\
	\label{eq:model_prediction_S2} &&m_{1}=0, \qquad 8.319\,\text{meV}\leq m_2\leq8.763\,\text{meV}, \qquad 49.508\,\text{meV}\leq m_3\leq50.774\,\text{meV}\,.
\end{eqnarray}
We see that the model is very predictive and the neutrino mixing parameters are determined to vary in narrow regions.

\subsection{The breaking pattern $\mathcal{F}_{2}^{\prime}$ }

From table \ref{tab:invariant_VEVs}, we find that the VEVs of $\phi_{\text{atm}}$ and $\phi_{\text{sol}}$ are proportional to 
\begin{eqnarray}
	\langle\phi_{\text{atm}}\rangle =(1,-1,-1)^{T}v_{\phi_{a}},\qquad
	\langle\phi_{\text{sol}}\rangle =(1,i,ix)^{T}v_{\phi_{s}}\,.
\end{eqnarray}
Consequently, the light neutrino mass matrix $m_{\nu}$ takes the form
\begin{equation}
	m_{\nu }= m_{a}\left[
	\begin{pmatrix}
		1 & -1 & -1 \\
		-1 & 1 & 1 \\
		-1 & 1 & 1 \\
	\end{pmatrix}
	+re^{i \eta }  
	\begin{pmatrix}
		1 & i & i x \\
		i & -1 & -x \\
		i x & -x & -x^2 \\
	\end{pmatrix}\right]\,,
\end{equation}
Using the form specified for the IO case in Eq. \eqref{eq:mnup}, $m_{\nu}$ can be transformed into a block-diagonal matrix $m^{\prime}_{\nu}$ via the unitary transformation $U_{\nu1}$, where
\begin{equation}
	U_{\nu1}=
	\begin{pmatrix}
		\frac{1}{\sqrt{3}} & -\frac{i (x+1+2 i)}{ \sqrt{6(x-1) x+12}} & -\frac{i (x-1)}{\sqrt{2 (x-1) x+4}} \\
		-\frac{1}{\sqrt{3}} & \frac{i x+1-2 i}{ \sqrt{6(x-1) x+12}} & \frac{-1-i x}{\sqrt{2 (x-1) x+4}} \\
		-\frac{1}{\sqrt{3}} & \frac{1+i-2 i x}{\sqrt{6(x-1) x+12}} & \frac{1+i}{\sqrt{2 (x-1) x+4}} \\
	\end{pmatrix}\,.
\end{equation}
The entries of $m^{\prime}_{\nu}$ are given by
\begin{equation}
	y=\frac{m_{a}}{3}\left[ 9-r (x+1+i)^2e^{i \eta }\right], ~~ z=\frac{m_{a}}{3}  r (1-i-ix) \sqrt{2 \left(x^2-x+2\right)}e^{i \eta }, ~~ w=\frac{2m_{a}}{3} r (x^2-x+2)e^{i \eta} \,.
\end{equation}
Furthermore, $m^{\prime}_{\nu}$ can be diagonalized by the unitary matrix $U_{\nu2}$ given in Eq. \eqref{eq:Unu2}. Thus, the PMNS matrix can be written as
\begin{equation}
	U_{PMNS}=P_{321}U_{\nu1}U_{\nu2}\,.
\end{equation}
The explicit expressions for the mixing parameters can then be easily derived. When input parameters are treated as free parameters, we present the best fit values of the other output parameters in table~\ref{tab:viable_best_fit}. By comprehensively scanning over space of input parameters, the phenomenologically viable ranges for $x$, $\eta$, $m_{a}$, $r$, three lepton mixing angles, Dirac CP phase, Majorana CP phase and neutrino masses are presented in table \ref{tab:viable_regions}.

For the benchmark value $x$ = 7, the absolute value of the third column of PMNS matrix is fixed to be:
\begin{equation}\label{eq:fix_column_S2}
	\frac{1}{2\sqrt{11}}\left(1,5,3\sqrt{2}\right)^{T},
\end{equation}
 Then the expressions of the mixing angles and the Jarlskog invariant $J_{CP}$ can be read as
\begin{eqnarray}
	\nonumber
	&&\sin^{2}\theta_{13}=\frac{1 }{44},\qquad 
	\sin^{2}\theta_{12}=\frac{1}{258} \left[129-2 \sqrt{22} \sin 2 \theta (\cos \psi-13 \sin \psi )-41  \cos 2\theta \right], \\
	&&	\sin^{2}\theta_{23}=\frac{25}{43},\qquad J_{CP}=\frac{1}{968} \left[22  \cos 2\theta-\sqrt{22} \sin \theta  \cos \theta  (4 \sin \psi +11 \cos \psi )\right] \,.
\end{eqnarray}	
Furthermore, we can derive the following sum rule between the mixing angles $\theta_{13}$ and $\theta_{23}$ 
\begin{equation}\label{eq:F2P_sum_rules}
	\cos^2\theta_{13}\cos^2\theta_{23}=\frac{9}{22}\,. 
\end{equation}

In this case, the best fit values of the observables corresponding to different values of $\eta$ are summarized in table \ref{tab:viable_BP}.
In the following analysis,  the phase parameter $\eta$ is constrained to $\eta=52\pi/27$. All lepton mixing parameters and the neutrino mass ratio are now determined by a single dimensionless parameter $r$. By combining the experimental data presented in table \ref{tab:bf_13sigma_data} and further scanning the parameters $m_{a}$ and $r$, we find that the observables are constrained to
\begin{eqnarray}
	\nonumber && 21.700\,\text{meV} \leq m_{a}\leq 21.898\,\text{meV}, \qquad  0.0586\leq r \leq 0.0587\,, \\
	\nonumber &&\sin^2\theta_{13}=0.02273, \qquad 0.276\leq \sin^2\theta_{12}\leq0.332, \qquad   \sin^2\theta_{23}=0.581, \\
	\nonumber &&1.688\leq\delta_{CP}/\pi\leq1.716, \qquad  1.622\leq\beta/\pi\leq1.644, \qquad  13.801\,\text{meV}\leq m_{ee}\leq16.544\,\text{meV}, \\
	\label{eq:model_prediction_S2} &&m_{3}=0, \qquad 48.927\,\text{meV}\leq m_1\leq49.432\,\text{meV}, \qquad 49.643\,\text{meV}\leq m_2\leq50.128\,\text{meV}\,.
\end{eqnarray}
For the IO scenario, we can see that $\sin^2\theta_{13}$ and $\sin^2\theta_{23}$ are constant values.

\section{\label{sec:conclusion} Conclusion }

The type I seesaw mechanism elegantly explains the exceptionally small of neutrino masses on a theoretical level, yet its experimental verification remains challenging. The Majorana masses of RHNs could be extremely heavy, which may lie beyond the reach of the most advanced particle colliders. Furthermore, the seesaw mechanism introduces more free parameters than measurable quantities and is hard to test experimentally. In the present work, guided by the principles of minimality and symmetry, we explore a highly constrained and predictive variant of the seesaw model: the minimal seesaw model with two right-handed neutrinos based on the tri-direct CP approach. In this kind of models, the neutrino masses, mixing angles and CP violation phases are determined from just three or four real input parameters.

We emphasise that the model independent tri-direct CP approach is a quite predictive scheme for constructing neutrino mass models based on discrete flavor symmetry and CP symmetry. In the tri-direct CP approach, we have analyzed all possible breaking patterns which arise from the high energy symmetry $\Delta(96)\rtimes H_{CP}$ based on the two right-handed neutrino seesaw mechanism. The symmetry $\Delta(96)\rtimes H_{CP}$ is spontaneously broken down to an Abelian subgroup $G_{l}$ (non $Z_2$ subgroups) in the charged lepton sector, to $G_{\text{atm}}\rtimes H^{\text{atm}}_{CP}$ in the atmospheric neutrino sector and to $G_{\text{sol}}\rtimes H^{\text{sol}}_{CP}$  in the solar neutrino sector. In this work, we assign the left-handed lepton doublets $L$ to transform as the  $\Delta(96)$ triplet $\bm{3_{0}}$. The flavon field $\phi_{\text{atm}}$,  coupling to $N_{\text{atm}}$ and  $L$, transforms as the conjugate triplet $\bm{\bar{3}_{0}}$. The flavon $\phi_{\text{sol}}$, coupling to $N_{\text{sol}}$ and  $L$, transforms as either  $\bm{\bar{3}_{0}}$ or $\bm{\bar{3}_{1}}$. For the former assignment, the neutrino masses, lepton mixing angles and CP violation phases of each breaking pattern are determined by three real input parameters $|m_{a}|$, $r$ and $\eta$. After performing an exhaustive numerical analysis for all possible independent breaking patterns,  we find out only one phenomenologically interesting mixing pattern which agrees with the experimental data very well. The corresponding representative combination of residual flavor symmetries is $\{G_{l},G_{\text{atm}},G_{\text{sol}}\}=(Z_{3}^{ac},Z_{2}^{a^2b},Z_{3}^{ad^2})$, and both the atmospheric and solar neutrino sectors exhibit trivial residual CP transformations represented by the identity matrix $\mathbb{1}_{3}$. The residual symmetries lead to the vacuum alignments $\langle\phi_{\text{atm}}\rangle =(0,1,-1)^{T}v_{\phi_a}$ and $\langle\phi_{\text{sol}}\rangle =(1,-1,1)^{T}v_{\phi_s}$. As a consequence, the lepton mixing matrix is predicted to be the TM1 pattern, the neutrino masses are NO and the lightest neutrino is massless with $m_{1}=0$.

When the flavon $\phi_{\text{sol}}$ is considered to transform as $\bm{\bar{3}_{1}}$, a new real parameter $x$ must be included. We have analyzed all possible symmetry breaking patterns in a model independent way, including the corresponding predictions for lepton mixing parameters and neutrino masses with both NO and IO mass hierarchy. We find 54 (42 for NO and 12 for IO) phenomenologically interesting mixing patterns with $\chi^{2}_{\text{min}}$ less than 10: 33 feature the charged lepton residual symmetry $G_{l}$ as $Z^{ac}_{3}$ (labeled $\mathcal{T}_{i}$ and $\mathcal{T}_{i}^{\prime}$), 9 exhibit $Z^{c}_{4}$ (labeled  $\mathcal{F}_{i}$ and $\mathcal{F}_{i}^{\prime}$), 7 demonstrate $Z^{abd^2}_{4}$ (labeled $\mathcal{S}_{i}$ and $\mathcal{S}_{i}^{\prime}$), and 5 display $Z^{abd}_{8}$ (labeled $\mathcal{W}_{i}$ and $\mathcal{W}_{i}^{\prime}$). The corresponding combination of residual flavor symmetries of each viable breaking pattern is summarized in table~\ref{tab:viable_BPs}. The best fit values of input parameters, mixing parameters and neutrino masses are presented in table \ref{tab:viable_best_fit}. Furthermore, after a comprehensive scan of the parameter space, the predicted ranges of input and output parameters of the 54 breaking patterns are derived, as shown in table~\ref{tab:viable_regions}.

Guided by the analysis of viable breaking patterns, we present detailed numerical results of the five NO cases $\mathcal{F}_{6}$, $\mathcal{F}_{7}$, $\mathcal{F}_{4}$, $\mathcal{T}_{10}$,  $\mathcal{S}_{2}$ and one IO case $\mathcal{F}_{2}^{\prime}$ for some benchmark values of $x$ and $\eta$. In the breaking pattern $\mathcal{F}_{6}$,  the benchmark value $x=3$ leads to $\langle\phi_{\text{sol}}\rangle =(-1,i,3i)^{T}v_{\phi_{s}}$ and  the so-called TM1 mixing matrix. For the breaking pattern $\mathcal{F}_{7}$, the benchmark values of $x$ are taken to be $x=3$ and $x=(2+\sqrt{17})/2\approx3$. The former case  is exactly the littlest seesaw model with CSD(3) which was originally proposed in Ref.~\cite{King:2015dvf}. For the latter case, the first column of PMNS matrix is predicted to be $\frac{1}{5}(\sqrt{17},2,2)^{T}$. The breaking pattern $\mathcal{F}_{4}$ shows the predictions for benchmark values $x = 5$ and $\eta=5\pi/3$. In this case, the absolute value of the first column of PMNS matrix is fixed to be  $\frac{1}{\sqrt{38}}(5,2,3)^{T}$. For the breaking pattern $\mathcal{T}_{10}$, to enhance the predictive power, we focus on $x = 4$ and $\eta = \pi$, and derive that the corresponding first column of the PMNS matrix is $\frac{1}{\sqrt{57}}(\sqrt{37},\sqrt{10},\sqrt{10})^{T}$. We consider the case of $x = (1+\sqrt{11})/2$ in the breaking pattern $\mathcal{S}_{2}$. In the same way, the first column of the PMNS matrix is obtained as $\frac{1}{3}(\sqrt{6},1,\sqrt{2})^{T}$. While the TM1 mixing scheme is well established, the other four matrices presented here have  not so far appeared in the literature. For the breaking pattern $\mathcal{F}_{2}^{\prime}$, when $x$ = 7, the third column of the PMNS matrix corresponds to $\frac{1}{2\sqrt{11}}(1,5,3\sqrt{2})^{T}$. Our analysis reveals that these six distinct breaking patterns are highly effective in predicting lepton mixing parameters and neutrino mass values when $x$ and $\eta$ assume specific benchmark values. Crucially, all the lepton mixing parameters and the neutrino masses are restricted in rather narrow regions, and can easily be tested by the forthcoming neutrino experiments. In particular the predictions of solar mixing angles, atmospheric mixing angles and the normal neutrino mass hierarchy will be tested quite soon.

\section*{Acknowledgements}

This work is supported by Natural Science Basic Research Program of Shaanxi (Program No. 2024JC-YBQN-0004), and the National Natural Science Foundation of China under Grant No. 12247103. 
 
\newpage

\section*{Appendix}

\begin{appendix}

\section{\label{sec:Delta96_group_theory}Group theory of $\Delta(96)$}

The group $\Delta(96)$, a member of the celebrated $\Delta(6n^2)$ family for $n=4$~\cite{Escobar:2008vc,Ding:2014ora}, exhibits a structure $(Z_{4}\times Z_{4})\rtimes S_{3}$. It can be generated using four generators $a$, $b$, $c$ and $d$ subject to the following multiplication rules~\cite{Ding:2014ssa,Ding:2012xx}:
\begin{eqnarray}
\nonumber && a^{3} = b^{2} = (ab)^{2} = c^{4} = d^{4} = 1,\quad cd = dc \\
\nonumber && aca^{-1}=c^{-1}d^{-1},\quad ada^{-1}=c,  \quad  bcb^{-1}=d^{-1},\quad bdb^{-1}=c^{-1} , 
\end{eqnarray}
Note that the generator $c=ada^{-1}$ is not independent. $\Delta(96)$ has 96 elements and each group element $g$  can be written as a product of powers of $a$, $b$, $c$ and $d$ 
\begin{equation}
g=a^{\alpha}b^{\beta}c^{\gamma}d^{\delta}, \qquad \alpha=0,1,2, \quad \beta=0,1,\quad \gamma,\delta=0,1,2,3\,.
\end{equation}	
The 96 elements of $\Delta(96)$ are divided into ten conjugacy classes:
\begin{eqnarray}
\nonumber&&\hskip-0.3in1C_1: \{1\}\,,\\
\nonumber&&\hskip-0.3in3C_2: \{c^2, d^2, c^2d^2\},\\
\nonumber&&\hskip-0.3in12C_2: \{ab, abc, abc^2, abc^3, a^2b,a^2bd, a^2bd^2, a^2bd^3, b, bcd, bc^2d^2, bc^3d^3\},\\
\nonumber&&\hskip-0.3in32C_3: \{a, ac, ac^2, ac^3, ad, ad^2, ad^3, acd, acd^2, acd^3, ac^2d, ac^2d^2, ac^2d^3, ac^3d, ac^3d^2, ac^3d^3,  a^2, a^2c,  \\
\nonumber	&&\hskip-0.3in\qquad \quad \ a^2c^2, a^2c^3, a^2d, a^2d^2, a^2d^3, a^2cd, a^2cd^2, a^2cd^3, a^2c^2d, a^2c^2d^2, a^2c^2d^3, a^2c^3d, a^2c^3d^2, a^2c^3d^3\},\\
\nonumber&&\hskip-0.3in3C^{(1)}_4: \{cd^2, cd^3, c^2d^3\},\\	
\nonumber&&\hskip-0.3in3C^{(2)}_4: \{c^2d,c^3d, c^3d^2\},\\ 
\nonumber&&\hskip-0.3in6C_4: \{c, d, cd, c^3, d^3,c^3d^3\},\\
\nonumber&&\hskip-0.3in12C_4: \{abd^2, abcd^2, abc^2d^2, abc^3d^2,a^2bc^2, a^2bc^2d, a^2bc^2d^2, a^2bc^2d^3, bc^2, bc^3d, bd^2, bcd^3\},\\
\nonumber&&\hskip-0.3in12C^{(1)}_8: \{abd, abcd, abc^2d, abc^3d, a^2bc^3,a^2bc^3d, a^2bc^3d^2, a^2bc^3d^3, bc, bc^2d,bc^3d^2, bd^3\},\\
\label{eq:Delta96_CC}&&\hskip-0.3in12C^{(2)}_{8}: \{abd^3, abcd^3, abc^2d^3, abc^3d^3, a^2bc,a^2bcd, a^2bcd^2, a^2bcd^3, bc^3, bd, bcd^2, bc^2d^3\}\,,
\end{eqnarray}
where $kC_{n}$ denotes a conjugacy class which contains $k$ elements with order $n$. The group $\Delta$(96) has fifteen $Z_{2}$ subgroups, sixteen $Z_{3}$ subgroups, seven $K_{4}$ subgroups, twelve $Z_{4}$ subgroups and six $Z_{8}$ subgroups. In terms of the generators $a$, $b$, $c$ and $d$,  the concrete forms of these Abelian subgroups are as follows:
\begin{itemize}[leftmargin=1.5em]
\item{$Z_{2}$ subgroups}
\begin{eqnarray}
\begin{aligned}
&Z_{2}^{c^2}=\left\{1,c^{2}\right\}, &&Z_{2}^{d^2}=\left\{1,d^{2}\right\},&&Z_{2}^{c^{2}d^{2}}=\left\{1,c^{2}d^{2}\right\},\\
&Z_{2}^{ab}=\left\{1,ab\right\}, &&Z_{2}^{abc}=\left\{1,abc\right\}, &&Z_{2}^{abc^{2}}=\left\{1,abc^{2}\right\}, &&Z_{2}^{abc^{3}}=\left\{1,abc^{3}\right\},\\
&Z_{2}^{a^{2}b}=\left\{1,a^{2}b\right\}, &&Z_{2}^{a^{2}bd}=\left\{1,a^{2}bd\right\}, &&Z_{2}^{a^{2}bd^{2}}=\left\{1,a^{2}bd^{2}\right\}, &&Z_{2}^{a^{2}bd^{3}}=\left\{1,a^{2}bd^{3}\right\}, \\
&Z_{2}^{b}=\left\{1,b\right\}, &&Z_{2}^{bcd}=\left\{1,bcd\right\}, &&Z_{2}^{bc^{2}d^{2}}=\left\{1,bc^{2}d^{2}\right\}, &&Z_{2}^{bc^{3}d^{3}}=\left\{1,bc^{3}d^{3}\right\}\,.
\end{aligned}
\end{eqnarray}

The three $Z_{2}$ in the first line are conjugate to each other, and the other twelve $Z_{2}$ subgroups  are related with each other by group conjugation.
\item{$Z_{3}$ subgroups}
\begin{equation}
\begin{aligned}
&Z_{3}^{a} =\{1,a,a^{2}\}, && Z_{3}^{ac}=\{1,ac,a^{2}cd\}, &&&& Z_{3}^{ac^{2}}=\{1,ac^{2},a^{2}c^{2}d^{2}\}, \\
&Z_{3}^{ac^{3}}= \{1,ac^{3},a^{2}c^{3}d^{3}\}, && Z_{3}^{ad}=\{1,ad,a^{2}c^{3}\}, &&&& {Z}_{3}^{ad^{2}}=\{1,ad^{2},a^{2}c^{2}\}\,, \\
&Z_{3}^{ad^{3}} =\{1,ad^{3},a^{2}c\}, && Z_{3}^{acd}=\{1,acd,a^{2}d\}, &&&& Z_{3}^{acd^{2}}=\{1,acd^{2},a^{2}c^{3}d\}, \\
&Z_{3}^{acd^{3}} =\{1,acd^{3},a^{2}c^{2}d\}, && Z_{3}^{ac^{2}d}=\{1,ac^{2}d,a^{2}cd^{2}\}, &&&& Z_{3}^{ac^{2}d^{2}}=\{1,ac^{2}d^{2},a^{2}d^{2}\}, \\
&Z_{3}^{ac^{2}d^{3}}=\{1,ac^{2}d^{3},a^{2}c^{3}d^{2}\},&& Z_{3}^{ac^{3}d}=\{1,ac^{3}d,a^{2}c^{2}d^{3}\}, &&&& Z_{3}^{ac^{3}d^{2}}=\{1,ac^{3}d^{2},a^{2}cd^{3}\}, \\
&Z_{3}^{ac^{3}d^{3}} =\{1,ac^{3}d^{3},a^{2}d^{3}\}\,.
\end{aligned}
\end{equation}
All the sixteen $Z_{3}$ subgroups are conjugate to each other.
\item{$Z_{4}$ subgroups}
\begin{equation}
\begin{aligned}
&Z_{4}^{cd^{2}}=\{1,cd^{2},c^{2},c^{3}d^{2}\}, &&Z_{4}^{cd^{3}}=\{1,cd^{3},c^{2}d^{2},c^{3}d\},  &&Z_{4}^{c^{2}d^{3}}=\{1,c^{2}d^{3},d^{2},c^{2}d\},\\
&Z_{4}^{c}=\{1,c,c^{2},c^{3}\}, &&Z_{4}^{d}=\{1,d,d^{2},d^{3}\},  &&Z_{4}^{cd}=\{1,cd,c^{2}d^{2},c^{3}d^{3}\},\\
&Z_{4}^{abd^{2}}=\{1,abd^{2},c^{2},abc^{2}d^{2}\}, &&Z_{4}^{abcd^{2}}=\{1,abcd^{2},c^{2},abc^{3}d^{2}\}, &&Z_{4}^{a^{2}bc^{2}}=\{1,a^{2}bc^{2},d^{2},a^{2}bc^{2}d^{2}\},\\&Z_{4}^{a^{2}bc^{2}d}=\{1,a^{2}bc^{2}d,d^{2},a^{2}bc^{2}d^{3}\}, &&Z_{4}^{bc^{2}}=\{1,bc^{2},c^{2}d^{2},bd^{2}\}, &&Z_{4}^{bc^{3}d}=\{1,bc^{3}d,c^{2}d^{2},bcd^{3}\}\,.\end{aligned}
\end{equation}
The twelve $Z_{4}$ subgroups fall into three categories applying similarity transformations belonging to  $\Delta(96)$:  the three $Z_{4}$ in the first line, the three $Z_{4}$ in the second line, and the remaining six subgroups.
\item{$K_{4}$ subgroups}
\begin{equation}
\begin{aligned}
&K_{4}^{(ab,c^{2})}=\{1,ab,c^{2},abc^{2}\}, && K_{4}^{(abc,c^{2})}=\{1,abc,c^{2},abc^{3}\},\\ 
&K_{4}^{(a^{2}b,d^{2})}=\{1,a^{2}b,d^{2},a^{2}bd^{2}\}, &&K_{4}^{(a^{2}bd,d^{2})}=\{1,a^{2}bd,d^{2},a^{2}bd^{3}\},\\
 & K_{4}^{(b,c^{2}d^{2})}=\{1,b,c^{2}d^{2},bc^{2}d^{2}\}, && K_{4}^{(bcd,c^{2}d^{2})}=\{1,bcd,c^{2}d^{2},bc^{3}d^{3}\},\\
&K_{4}^{(c^{2},d^{2})}=\{1,c^{2},d^{2},c^{2}d^{2}\} \,.\end{aligned}
\end{equation}
$K ^{(c^2,d^2)} _{4}$ is a normal subgroup of $\Delta$(96), and the remaining $K_{4}$ subgroups are conjugate to each other.
\item{$Z_{8}$ subgroups}
\begin{equation}
\begin{aligned}
&Z_{8}^{abd} =\{1,abd,cd^{2},abcd^{3},c^{2},abc^{2}d,c^{3}d^{2},abc^{3}d^{3}\}, \\
&Z_{8}^{abcd} =\{1,abcd,cd^{2},abc^{2}d^{3},c^{2},abc^{3}d,c^{3}d^{2},abd^{3}\}, \\
&Z_{8}^{a^{2}bc^{3}} =\{1,a^{2}bc^{3},c^{2}d^{3},a^{2}bcd^{3},d^{2},a^{2}bc^{3}d^{2},c^{2}d,a^{2}bcd\}, \\
&Z_{8}^{a^{2}bc^{3}d} =\{1,a^{2}bc^{3}d,c^{2}d^{3},a^{2}bc,d^{2},a^{2}bc^{3}d^{3},c^{2}d,a^{2}bcd^{2}\}, \\
&Z_{8}^{bc} =\{1,bc,cd^{3},bc^{2}d^{3},c^{2}d^{2},bc^{3}d^{2},c^{3}d,bd\}, \\
&Z_{8}^{bc^{2}d} =\left\{1,bc^{2}d,cd^{3},bc^{3},c^{2}d^{2},bd^{3},c^{3}d,bcd^{2}\right\}\,.
\end{aligned}
\end{equation}
All the six $Z_{8}$ subgroups are conjugate to each other.
\end{itemize}

The number of irreducible representations of a group is equal to the number of its conjugacy class. Therefore $\Delta(96)$  has ten irreducible representations: two one-dimensional representations $\bm{1_{m}}$, one two-dimensional representation $\bm{2}$, six three-dimensional representations $\bm{3_{m}}$ and $\bm{\bar{3}_{m}}$ and $\bm{\hat{3}_{m}}$ and one six-dimensional representation $\bm{6}$, where $m=0,1$.  The representation matrices for the generators $a$, $b$, $c$ and $d$  in different irreducible representations are listed in table~\ref{tab:Delta96_Reps}. It should be noted that the representations $\bm{\bar{3}_{m}}$ are complex conjugate representations of $\bm{3_{m}}$. Together with the representation $\bm{6}$, these triplet representations $\bm{3_{m}}$ and $\bm{\bar{3}_{m}}$ form faithful representations of $\Delta(96)$, while $\bm{\hat{3}_{m}}$ are not. Then the Kronecker products between different irreducible representations can be easily obtained:
\begin{eqnarray}
\nonumber &&\hskip-0.3in \bm{1_{m}}\otimes \bm{1_{n}}= \bm{1_{[m+n]}},\quad \bm{1_{m}}\otimes \bm{2}= \bm{2},\quad \bm{1_{m}}\otimes \bm{3_{n}}= \bm{3_{[m+n]}},\quad \bm{1_{m}}\otimes \bm{\bar{3}_{n}}= \bm{\bar{3}_{[m+n]}},  \\
\nonumber &&\hskip-0.3in \bm{1_{m}}\otimes \bm{\hat{3}_{n}}= \bm{\hat{3}_{[m+n]}}, \quad \bm{1_{m}}\otimes \bm{6}= \bm{6}, \quad \bm{2}\otimes\bm{2}=\bm{1_{0}}\oplus\bm{1_{1}}\oplus\bm{2}, \quad \bm{2}\otimes\bm{3_{m}}=\bm{3_{0}}\oplus\bm{3_{1}}, \\
\nonumber &&\hskip-0.3in \bm{2}\otimes\bm{\bar{3}_{m}}=\bm{\bar{3}_{0}}\oplus\bm{\bar{3}_{1}}, \quad \bm{2}\otimes\bm{\hat{3}_{m}}=\bm{\hat{3}_{0}}\oplus\bm{\hat{3}_{1}}, \quad \bm{2}\otimes\bm{6}=\bm{6_{i}}\oplus\bm{6_{ii}}, \quad \bm{3_{m}}\otimes \bm{3_{n}}= \bm{\bar{3}_{0}}\oplus\bm{\bar{3}_{1}}\oplus\bm{\hat{3}_{[m+n+1]}}, \\
\nonumber &&\hskip-0.3in \bm{3_{m}}\otimes \bm{\bar{3}_{n}}= \bm{1_{[m+n]}}\oplus\bm{2}\oplus\bm{6}, \quad \bm{3_{m}}\otimes \bm{\hat{3}_{n}}= \bm{\bar{3}_{[m+n+1]}}\oplus\bm{6}, \quad  \bm{3_{m}}\otimes \bm{6}= \bm{3_{0}}\oplus\bm{3_{1}}\oplus\bm{\hat{3}_{0}}\oplus\bm{\hat{3}_{1}}\oplus\bm{6}, \\
\nonumber &&\hskip-0.3in \bm{\bar{3}_{m}}\otimes \bm{\bar{3}_{n}}= \bm{3_{0}}\oplus\bm{3_{1}}\oplus\bm{\hat{3}_{[m+n+1]}}, ~~ \bm{\bar{3}_{m}}\otimes \bm{\hat{3}_{n}}= \bm{3_{[m+n+1]}}\oplus\bm{6}, ~~  \bm{\bar{3}_{m}}\otimes \bm{6}= \bm{\bar{3}_{0}}\oplus\bm{\bar{3}_{1}}\oplus\bm{\hat{3}_{0}}\oplus\bm{\hat{3}_{1}}\oplus\bm{6}, \\
\nonumber &&\hskip-0.3in \bm{\hat{3}_{m}}\otimes \bm{\hat{3}_{n}}= \bm{1_{[m+n]}}\oplus\bm{2}\oplus\bm{\hat{3}_{0}}\oplus\bm{\hat{3}_{1}}, \quad  \bm{\hat{3}_{m}}\otimes \bm{6}= \bm{3_{0}}\oplus\bm{3_{1}}\oplus\bm{\bar{3}_{0}}\oplus\bm{\bar{3}_{1}}\oplus\bm{6}, \\
 &&\hskip-0.3in \bm{6}\otimes \bm{6}= \bm{1_{0}}\oplus\bm{1_{1}}\oplus\bm{2_{S}}\oplus\bm{2_{A}}\oplus\bm{3_{0}}\oplus\bm{3_{1}}\oplus\bm{\bar{3}_{0}}\oplus\bm{\bar{3}_{1}}\oplus\bm{\hat{3}_{0}}\oplus\bm{\hat{3}_{1}}\oplus\bm{6_{S}}\oplus\bm{6_{A}}\,,
\end{eqnarray}
where $m,n=0,1$, and we have defined $[m+n]\equiv m+n$ (mod 2). The symbol $\bm{6_{i}}$ and $\bm{6_{ii}}$ stand for two six-dimensional representations $\bm{6}$ that appear in the Kronecker products. The subscript ``$S$'' (``$A$'')  refers to symmetric (antisymmetric) combinations.  In the following, we list the CG coefficients in our basis. The notation $\alpha_{i}$ ($\beta_{i}$) are denoted the elements of the first (second) representation.  Furthermore, we shall adopt the following notations facilitating the expressions of CG coefficients
\begin{equation}
	P_{2}=\begin{pmatrix}
		0&1\\
		-1&0
	\end{pmatrix},\qquad  
	P_{6}=\begin{pmatrix}
		\mathbb{1}_{3}&\mathbb{0}_{3}\\
		\mathbb{0}_{3}&-\mathbb{1}_{3}
	\end{pmatrix}
\end{equation}
Then the results are summarized in table~\ref{tab:Delta96_CG}.

\begin{table}[t!]
\begin{center}
\renewcommand{\tabcolsep}{1.2mm}
\begin{tabular}{|c|c|c|c|c|c|}\hline\hline
~~  &  $a$  &   $b$  &  $c$  &   $d$ \\ \hline
&   &  &  &    \\ [-0.16in]
$\bm{1_{m}}$ & $1$   &  $(-1)^{m}$ &$1$   &  $1$    \\[0.03in] \hline
&   &   &  &    \\ [-0.15in]
$\bm{2}$ & $-\frac{1}{2}\begin{pmatrix}
 1 & \sqrt{3} \\
 -\sqrt{3} & 1 \\
\end{pmatrix}$
& $\begin{pmatrix}
				1 & 0 \\
				0 & -1 \\
			\end{pmatrix}$
& $\mathbb{1}_{2}$
& $\mathbb{1}_{2}$ \\[0.1in] \hline
&   &  &  &   \\ [-0.15in]
			
$\bm{3_{m}}$ &  $\begin{pmatrix}
				0 & 1 & 0 \\
				0 & 0 & 1 \\
				1 & 0 & 0 \\
			\end{pmatrix}$
& $(-1)^{m+1}\begin{pmatrix}
				0 & 0 & 1 \\
				0 & 1 & 0 \\
				1 & 0 & 0 \\
			\end{pmatrix}$
& $\begin{pmatrix}
				i & 0 & 0 \\
				0 & -i & 0 \\
				0 & 0 & 1 \\
			\end{pmatrix}$
& $	\begin{pmatrix}
				1 & 0 & 0 \\
				0 & i & 0 \\
				0 & 0 & -i \\
			\end{pmatrix}$ \\ [0.1in]\hline
	&   &  &  &    \\ [-0.15in]
			
$\bm{\bar{3}_{m}}$ & 
			$	\begin{pmatrix}
				0 & 1 & 0 \\
				0 & 0 & 1 \\
				1 & 0 & 0 \\
			\end{pmatrix}$
			& 
			$(-1)^{m+1}\begin{pmatrix}
				0 & 0 & 1 \\
				0 & 1 & 0 \\
				1 & 0 & 0 \\
			\end{pmatrix}$
			& 
			$	\begin{pmatrix}
				-i & 0 & 0 \\
				0 & i & 0 \\
				0 & 0 & 1 \\
			\end{pmatrix}$
			& 
			$\begin{pmatrix}
				1 & 0 & 0 \\
				0 & -i & 0 \\
				0 & 0 & i \\
			\end{pmatrix}$
			\\ [0.1in] \hline
	&   &  &  &    \\ [-0.16in]		
$\bm{\hat{3}_{m}}$  & $\begin{pmatrix}
				0 & 1 & 0 \\
				0 & 0 & 1 \\
				1 & 0 & 0 \\
			\end{pmatrix}$
&$(-1)^{m+1}\begin{pmatrix}
				0 & 0 & 1 \\
				0 & 1 & 0 \\
				1 & 0 & 0 \\
			\end{pmatrix}$
& $\begin{pmatrix}
				-1 & 0 & 0 \\
				0 & -1 & 0 \\
				0 & 0 & 1 \\
			\end{pmatrix}$
& $\begin{pmatrix}
				1 & 0 & 0 \\
				0 & -1 & 0 \\
				0 & 0 & -1 \\
			\end{pmatrix}$
			\\ [0.1in]\hline
	&   &  &  &    \\ [-0.16in]					
$\bm{6}$ & $	\begin{pmatrix}
				\rho_{\bm{3_{0}}}(a) & \mathbb{0}_{3}\\
				\mathbb{0}_{3} & 	\rho^{T}_{\bm{3_{0}}}(a)
			\end{pmatrix}$
& $	\begin{pmatrix}
				\mathbb{0}_{3} & \mathbb{1}_{3}\\
				\mathbb{1}_{3} & \mathbb{0}_{3}
			\end{pmatrix}$
& $\diag(i,-1,i,-i,-i,-1)$
& $\diag(i,i,-1,-i,-1,-i)$\\[0.12in] \hline\hline
\end{tabular}
\caption{\label{tab:Delta96_Reps}The representation matrices for the $\Delta(96)$  generators $a$, $b$, $c$ and $d$ in our chosen basis.  }
\end{center}
\end{table}

\begin{small}
\renewcommand{\arraystretch}{1.13}
\renewcommand{\tabcolsep}{0.9mm}
	\setlength\LTcapwidth{\textwidth}
	\setlength\LTleft{-0.0in}
	\setlength\LTright{0pt}
	\begin{longtable}{|c|c|c|c|c|c|c|c|c|c|c|c|}
		\caption{\label{tab:Delta96_CG} Tensor products and the corresponding CG coefficients for the $\Delta(96)$.
}\\
	\midrule
\specialrule{0em}{1.0pt}{1.0pt}

\endfirsthead

\multicolumn{12}{c}
{{\bfseries \tablename\ \thetable{} -- continued from previous page}} \\
\hline

\endhead

\caption[]{continues on next page}\\
\endfoot

\endlastfoot

\hline
		
		\multicolumn{3}{|c}{$\bm{1_{m}}\otimes\bm{1_{n}}=\bm{1_{[m+n]}} $}& \multicolumn{3}{|c}{$\bm{1_{m}}\otimes\bm{2}=\bm{2}$} & \multicolumn{3}{|c}{$\bm{1_{m}}\otimes\bm{3_{n}}=\bm{3_{[m+n]}} $}&\multicolumn{3}{|c|}{$\bm{1_{m}}\otimes\bm{\bar{3}_{n}}=\bm{\bar{3}_{[m+n]}}$}  \\ \hline
		
		\multicolumn{3}{|c}{   $ \begin{array}{l}
				\bm{1_{[m+n]}}:~  \alpha_{1}\beta _{1}   \end{array} $ } &
		\multicolumn{3}{|c}{$ \begin{array}{l}
				\bm{2}:
			P_{2}^{m}		\begin{pmatrix} \alpha_{1}\beta _{1} \\ \alpha_{1}\beta _{2}
			\end{pmatrix}
			\end{array} $} &
		\multicolumn{3}{|c}{ $ \begin{array}{l}
				\bm{3_{[m+n]}}:~\begin{pmatrix}\alpha_{1} \beta_{1}\\
					\alpha_{1} \beta_{2}\\
				\alpha_{1} \beta_{3}	
				\end{pmatrix}
			\end{array} $  } &
		\multicolumn{3}{|c|}{ $ \begin{array}{l}
				\bm{\bar{3}_{[m+n]}}:~\begin{pmatrix}\alpha_{1} \beta_{1}\\
					\alpha_{1} \beta_{2}\\
					\alpha_{1} \beta_{3}	
				\end{pmatrix}
			\end{array} $  } \\ \hline

		\multicolumn{3}{|c}{$\bm{1_{m}}\otimes\bm{\hat{3}_{n}}=\bm{\hat{3}_{[m+n]}}$}& \multicolumn{3}{|c}{$\bm{1_{m}}\otimes\bm{6}=\bm{6}$} & \multicolumn{3}{|c}{$\bm{2}\otimes\bm{2}=\bm{1_{0}}\oplus\bm{1_{1}}\oplus\bm{2}$}&\multicolumn{3}{|c|}{$\bm{2}\otimes\bm{3_{m}}=\bm{3_{0}}\oplus\bm{3_{1}} $}  \\ \hline
		
		\multicolumn{3}{|c}{ $ \begin{array}{l}
				\bm{\hat{3}_{[m+n]}}:~\begin{pmatrix}\alpha_{1} \beta_{1}\\
					\alpha_{1} \beta_{2}\\
					\alpha_{1} \beta_{3}	
				\end{pmatrix}
			\end{array} $ } &
		\multicolumn{3}{|c}{  $ \begin{array}{l} \bm{6}:~P_{6}^{m}\begin{pmatrix} 
				\alpha_{1} \beta_{1}\\
				\alpha_{1} \beta_{2}\\
				\alpha_{1} \beta_{3}\\
				\alpha_{1} \beta_{4}\\
				\alpha_{1} \beta_{5}\\
				\alpha_{1} \beta_{6}
				\end{pmatrix}\\
			\end{array} $ } &
		\multicolumn{3}{|c}{  $ \begin{array}{l}
				\bm{1_{0}}: \alpha_{1} \beta_{1}+\alpha_{2} \beta_{2}	\\
\bm{1_{1}}:			\alpha_{1} \beta_{2}-\alpha_{2} \beta_{1}					\\
			\bm{2}:
			\begin{pmatrix} 
		\alpha_{1} \beta_{1}-\alpha_{2} \beta_{2}\\
		-\alpha_{1} \beta_{2}-\alpha_{2} \beta_{1}\\	
			\end{pmatrix}	
			\end{array} $ } &
		\multicolumn{3}{|c|}{  $ \begin{array}{l}
				\bm{3_{m}}:\quad~~ \begin{pmatrix} 
					(\alpha_{1}-\sqrt{3}\alpha_{2})\beta_{1}\\
					-2\alpha_{1}\beta_{2}\\
					(\alpha_{1}+\sqrt{3}\alpha_{2})\beta_{3}
				\end{pmatrix}\\
			\bm{3_{[m+1]}}:\begin{pmatrix} 
					(\alpha_{2}+\sqrt{3}\alpha_{1})\beta_{1}\\
					-2\alpha_{2}\beta_{2}\\
					(\alpha_{2}-\sqrt{3}\alpha_{1})\beta_{3}
				\end{pmatrix}	
			\end{array} $ } \\ \hline

	\multicolumn{4}{|c}{$\bm{2}\otimes \bm{\bar{3}_{m}}=\bm{\bar{3}_{0}}\oplus\bm{\bar{3}_{1}}$}& \multicolumn{4}{|c}{$\bm{2}\otimes\bm{\hat{3}_{m}}=\bm{\hat{3}_{0}}\oplus\bm{\hat{3}_{1}}$} & \multicolumn{4}{|c|}{$\bm{3_{m}}\otimes\bm{3_{n}}=\bm{\bar{3}_{0}}\oplus\bm{\bar{3}_{1}}\oplus\bm{\hat{3}_{[m+n+1]}}$} \\ \hline
		
		\multicolumn{4}{|c}{  $ \begin{array}{l}
	\bm{\bar{3}_{m}}:\quad~~ \begin{pmatrix} 
		(\alpha_{1}-\sqrt{3}\alpha_{2})\beta_{1}\\
		-2\alpha_{1}\beta_{2}\\
		(\alpha_{1}+\sqrt{3}\alpha_{2})\beta_{3}
	\end{pmatrix}\\
	\bm{\bar{3}_{[m+1]}}:\begin{pmatrix} 
		(\alpha_{2}+\sqrt{3}\alpha_{1})\beta_{1}\\
		-2\alpha_{2}\beta_{2}\\
		(\alpha_{2}-\sqrt{3}\alpha_{1})\beta_{3}
	\end{pmatrix}	
			\end{array} $ } &
		\multicolumn{4}{|c}{ $ \begin{array}{l}
		\bm{\hat{3}_{m}}:\quad~~ \begin{pmatrix} 
			(\alpha_{1}-\sqrt{3}\alpha_{2})\beta_{1}\\
			-2\alpha_{1}\beta_{2}\\
			(\alpha_{1}+\sqrt{3}\alpha_{2})\beta_{3}
		\end{pmatrix}\\
		\bm{\hat{3}_{[m+1]}}:\begin{pmatrix} 
			(\alpha_{2}+\sqrt{3}\alpha_{1})\beta_{1}\\
			-2\alpha_{2}\beta_{2}\\
			(\alpha_{2}-\sqrt{3}\alpha_{1})\beta_{3}
		\end{pmatrix}	
			\end{array} $  } &
		\multicolumn{4}{|c|}{ $ \begin{array}{l} \bm{\bar{3}_{[m+n]}}:\quad~ \begin{pmatrix} 
		\alpha_{2}\beta_{3}-\alpha_{3}\beta_{2}\\
		\alpha_{3}\beta_{1}-\alpha_{1}\beta_{3}\\
		\alpha_{1}\beta_{2}-\alpha_{2}\beta_{1}
		\end{pmatrix}\\
		\bm{\bar{3}_{[m+n+1]}}:\begin{pmatrix} 
		\alpha_{2}\beta_{3}+\alpha_{3}\beta_{2}\\
		\alpha_{3}\beta_{1}+\alpha_{1}\beta_{3}\\
		\alpha_{1}\beta_{2}+\alpha_{2}\beta_{1}
				\end{pmatrix}\\
				\bm{\hat{3}_{[m+n+1]}}:\begin{pmatrix} 
					\alpha_{1}\beta_{1}\\
					\alpha_{2}\beta_{2}\\
					\alpha_{3}\beta_{3}
				\end{pmatrix}	
			\end{array} $  } \\ \hline		
		
		\multicolumn{4}{|c}{$\bm{\bar{3}_{m}}\otimes\bm{\bar{3}_{n}}=\bm{3_{0}}\oplus\bm{3_{1}}\oplus\bm{\hat{3}_{[m+n-1]}}$}& \multicolumn{4}{|c}{$\bm{3_{m}}\otimes\bm{\bar{3}_{n}}=\bm{1_{[m+n]}}\oplus\bm{2}\oplus\bm{6}$}& \multicolumn{4}{|c|}{$\bm{3_{m}}\otimes\bm{\hat{3}_{n}}=\bm{\bar{3}_{[m+n+1]}}\oplus\bm{6}$}\\ \hline
		
		\multicolumn{4}{|c}{ $ \begin{array}{l}
		\bm{3_{[n+m]}}:\quad~ \begin{pmatrix} 
			\alpha_{2}\beta_{3}-\alpha_{3}\beta_{2}\\
			\alpha_{3}\beta_{1}-\alpha_{1}\beta_{3}\\
			\alpha_{1}\beta_{2}-\alpha_{2}\beta_{1}
		\end{pmatrix}\\
		\bm{3_{[n+m+1]}}:\begin{pmatrix} 
			\alpha_{2}\beta_{3}+\alpha_{3}\beta_{2}\\
			\alpha_{3}\beta_{1}+\alpha_{1}\beta_{3}\\
			\alpha_{1}\beta_{2}+\alpha_{2}\beta_{1}
		\end{pmatrix}\\
		\bm{\hat{3}_{[m+n+1]}}:\begin{pmatrix} 
			\alpha_{1}\beta_{1}\\
			\alpha_{2}\beta_{2}\\
			\alpha_{3}\beta_{3}
		\end{pmatrix}	
			\end{array} $} &
		\multicolumn{4}{|c}{  $ \begin{array}{l} \bm{1_{[m+n]}}:~~
		\alpha_{1}\beta_{1}+\alpha_{2}\beta_{2}+\alpha_{3}\beta_{3}				\\
				\bm{2}:P_{2}^{m+n}\begin{pmatrix} 
				\alpha_{1}\beta_{1}-2\alpha_{2}\beta_{2}+
				\alpha_{3}\beta_{3}	\\
				\sqrt{3}	(\alpha_{3}\beta_{3}-\alpha_{1}\beta_{1})\\
				\end{pmatrix}
				\\
				\bm{\hat{\bm{6}}}:P_{6}^{m+n}\begin{pmatrix} 
					\alpha_{1}\beta_{3}\\
					\alpha_{2}\beta_{1}\\
					\alpha_{3}\beta_{2}\\
					\alpha_{3}\beta_{1}\\
					\alpha_{2}\beta_{3}\\
					\alpha_{1}\beta_{2}\\
				\end{pmatrix} 	
			\end{array} $ } &
		\multicolumn{4}{|c|}{  $ \begin{array}{l}
				\bm{\bar{3}_{[m+n+1]}}:\begin{pmatrix} 
					\alpha_{1}\beta_{1}\\
					\alpha_{2}\beta_{2}\\
					\alpha_{3}\beta_{3}
				\end{pmatrix} \\
				\bm{6}:P_{6}^{m+n}	\begin{pmatrix} 
					\alpha_{2}\beta_{1}\\
					\alpha_{3}\beta_{2}\\
					\alpha_{1}\beta_{3}\\
					\alpha_{2}\beta_{3}\\
					\alpha_{1}\beta_{2}\\
					\alpha_{3}\beta_{1}
				\end{pmatrix}
			\end{array} $ }  \\ \hline

		\multicolumn{4}{|c}{$\bm{\bar{3}_{m}}\otimes \bm{\hat{3}_{n}}=\bm{3_{[m+n+1]}}\oplus\bm{6}$}& \multicolumn{4}{|c}{$\bm{\hat{3}_{m}}\otimes\bm{\hat{3}_{n}}=\bm{1_{[m+n]}}\oplus\bm{2}\oplus \bm{\hat{3}_{0}}\oplus\bm{\hat{3}_{1}}$}& \multicolumn{4}{|c|}{$\bm{2}\otimes\bm{6}=\bm{6_{i}}\oplus\bm{6_{ii}}$} \\ \hline
		
		\multicolumn{4}{|c}{  $ \begin{array}{l}
				\bm{3_{[m+n+1]}}:\begin{pmatrix} 
					\alpha_{1}\beta_{1}\\
					\alpha_{2}\beta_{2}\\
					\alpha_{3}\beta_{3}
				\end{pmatrix} \\
				\bm{6}:P_{6}^{m+n}	\begin{pmatrix} 
					\alpha_{2}\beta_{3}\\
					\alpha_{3}\beta_{1}\\
					\alpha_{1}\beta_{2}\\
					\alpha_{2}\beta_{1}\\
					\alpha_{1}\beta_{3}\\
					\alpha_{3}\beta_{2}
				\end{pmatrix}
			\end{array} $ } &
		\multicolumn{4}{|c}{  $ \begin{array}{l} \bm{1_{[m+n]}}:\alpha_{1}\beta_{1}+\alpha_{2}\beta_{2}+\alpha_{3}\beta_{3} \\
				\bm{2}:P_{2}^{m+n}	\begin{pmatrix} 
					\alpha_{1}\beta_{1}-2\alpha_{2}\beta_{2}+\alpha_{3}\beta_{3}\\
\sqrt{3}	(\alpha_{3}\beta_{3}-\alpha_{1}\beta_{1})
				\end{pmatrix}\\
				\bm{\hat{3}_{[m+n]}}: \begin{pmatrix} 
					\alpha_{2}\beta_{3}-\alpha_{3}\beta_{2}\\
					\alpha_{3}\beta_{1}-\alpha_{1}\beta_{3}\\
					\alpha_{1}\beta_{2}-\alpha_{2}\beta_{1}
				\end{pmatrix}\\
				\bm{\hat{3}_{[m+n+1]}}:\begin{pmatrix} 
					\alpha_{2}\beta_{3}+\alpha_{3}\beta_{2}\\
					\alpha_{3}\beta_{1}+\alpha_{1}\beta_{3}\\
					\alpha_{1}\beta_{2}+\alpha_{2}\beta_{1}
				\end{pmatrix}	
			\end{array} $ } &
		\multicolumn{4}{|c|}{  $ \begin{array}{l}
				\bm{6_{i}}:~~
		\begin{pmatrix}-2\alpha_{1}\beta_{1}\\ (\alpha_{1}+\sqrt{3}\alpha_{2})\beta_{2}\\
			(\alpha_{1}-\sqrt{3}\alpha_{2})\beta_{3}\\	
			-2\alpha_{1}\beta_{4}\\ (\alpha_{1}-\sqrt{3}\alpha_{2})\beta_{5}\\
			(\alpha_{1}+\sqrt{3}\alpha_{2})\beta_{6}	
		\end{pmatrix} \\
		\bm{6_{ii}}:
		\begin{pmatrix}
			2\alpha_{2}\beta_{1}\\	
			(\sqrt{3}\alpha_{1}-\alpha_{2})\beta_{2}	\\
			-(\sqrt{3}\alpha_{1}+\alpha_{2})\beta_{3}\\
			-2\alpha_{2}\beta_{4}\\
			(\sqrt{3}	\alpha_{1}+\alpha_{2})\beta_{5}	\\
			(\alpha_{2}	-\sqrt{3}\alpha_{1})\beta_{6}
		\end{pmatrix}
			\end{array} $ } \\ \hline

		\multicolumn{4}{|c|}{$\bm{3_{m}}\otimes\bm{6}=\bm{3_{0}}\oplus\bm{3_{1}}\oplus\bm{\hat{3}_{0}}\oplus\bm{\hat{3}_{1}}\oplus\bm{6}$}& \multicolumn{4}{|c}{$\bm{\bar{3}_{m}}\otimes\bm{6}=\bm{\bar{3}_{0}}\oplus\bm{\bar{3}_{1}}\oplus\bm{\hat{3}_{0}}\oplus\bm{\hat{3}_{1}}\oplus\bm{6}$}& \multicolumn{4}{|c|}{$\bm{\hat{3}_{m}}\otimes\bm{6}=\bm{3_{0}}\oplus\bm{3_{1}}\oplus\bm{\bar{3}_{0}}\oplus\bm{\bar{3}_{1}}\oplus\bm{6}$}  \\ \hline
		
		\multicolumn{4}{|c}{ $ \begin{array}{l}
				\bm{3_{m}}:\quad~~ \begin{pmatrix} 
					\alpha_{3}\beta_{1}+\alpha_{2}\beta_{6}\\
					\alpha_{1}\beta_{2}+\alpha_{3}\beta_{5}\\
					\alpha_{2}\beta_{3}+\alpha_{1}\beta_{4}
				\end{pmatrix}\\
				\bm{3_{[m+1]}}:\begin{pmatrix} 
					\alpha_{2}\beta_{6}-\alpha_{3}\beta_{1}\\
					\alpha_{3}\beta_{5}-\alpha_{1}\beta_{2}\\
					\alpha_{1}\beta_{4}-\alpha_{2}\beta_{3}
				\end{pmatrix}\\
				\bm{\hat{3}_{m}}:~\quad~ \begin{pmatrix} 
					\alpha_{3}\beta_{2}+\alpha_{2}\beta_{4}\\
					\alpha_{1}\beta_{3}+\alpha_{3}\beta_{6}\\
					\alpha_{2}\beta_{1}+\alpha_{1}\beta_{5}
				\end{pmatrix}\\
				\bm{\hat{3}_{[m+1]}}:~ \begin{pmatrix} 
					\alpha_{2}\beta_{4}-\alpha_{3}\beta_{2}\\
					\alpha_{3}\beta_{6}-\alpha_{1}\beta_{3}\\
					\alpha_{1}\beta_{5}-\alpha_{2}\beta_{1}
				\end{pmatrix}\\
				\bm{6}:P_{6}^{m}\begin{pmatrix} 
					\alpha_{3}\beta_{3}\\
					\alpha_{1}\beta_{1}\\
					\alpha_{2}\beta_{2}\\
					-\alpha_{1}\beta_{6}\\
					-\alpha_{3}\beta_{4}\\
					-\alpha_{2}\beta_{5}
				\end{pmatrix}
			\end{array} $   } &
		\multicolumn{4}{|c}{  $ \begin{array}{l} 	\bm{\bar{3}_{m}}:~\quad~ \begin{pmatrix} 
					\alpha_{2}\beta_{2}+\alpha_{3}\beta_{4}\\
					\alpha_{3}\beta_{3}+\alpha_{1}\beta_{6}\\
					\alpha_{1}\beta_{1}+\alpha_{2}\beta_{5}
				\end{pmatrix}\\
				\bm{\bar{3}_{[m+1]}}:\begin{pmatrix} 
					\alpha_{2}\beta_{2}-\alpha_{3}\beta_{4}\\
					\alpha_{3}\beta_{3}-\alpha_{1}\beta_{6}\\
					\alpha_{1}\beta_{1}-\alpha_{2}\beta_{5}
				\end{pmatrix}\\
				\bm{\hat{3}_{m}}:~\quad~ \begin{pmatrix} 
					\alpha_{2}\beta_{1}+\alpha_{3}\beta_{6}\\
					\alpha_{3}\beta_{2}+\alpha_{1}\beta_{5}\\
					\alpha_{1}\beta_{3}+\alpha_{2}\beta_{4}
				\end{pmatrix}\\
				\bm{\hat{3}_{[m+1]}}:~ \begin{pmatrix} 
					\alpha_{2}\beta_{1}-\alpha_{3}\beta_{6}\\
					\alpha_{3}\beta_{2}-\alpha_{1}\beta_{5}\\
					\alpha_{1}\beta_{3}-\alpha_{2}\beta_{4}
				\end{pmatrix}\\
				\bm{6}:P_{6}^{m}\begin{pmatrix} 
					\alpha_{1}\beta_{2}\\
					\alpha_{2}\beta_{3}\\
					\alpha_{3}\beta_{1}\\
					-\alpha_{3}\beta_{5}\\
					-\alpha_{2}\beta_{6}\\
					-\alpha_{1}\beta_{4}
				\end{pmatrix}
			\end{array} $ } &
		\multicolumn{4}{|c|}{  $ \begin{array}{l}
				\bm{3_{m}}:~\quad~ \begin{pmatrix} 
					\alpha_{3}\beta_{3}+\alpha_{2}\beta_{5}\\
					\alpha_{1}\beta_{1}+\alpha_{3}\beta_{4}\\
					\alpha_{2}\beta_{2}+\alpha_{1}\beta_{6}
				\end{pmatrix}\\
				\bm{3_{[m+1]}}:\begin{pmatrix} 
					\alpha_{2}\beta_{5}-\alpha_{3}\beta_{3}\\
					\alpha_{3}\beta_{4}-\alpha_{1}\beta_{1}\\
					\alpha_{1}\beta_{6}-\alpha_{2}\beta_{2}
				\end{pmatrix}\\
				\bm{\bar{3}_{m}}:~\quad~ \begin{pmatrix} 
					\alpha_{2}\beta_{3}+\alpha_{3}\beta_{5}\\
					\alpha_{3}\beta_{1}+\alpha_{1}\beta_{4}\\
					\alpha_{1}\beta_{2}+\alpha_{2}\beta_{6}
				\end{pmatrix}\\
				\bm{\bar{3}_{[m+1]}}:~ \begin{pmatrix} 
					\alpha_{2}\beta_{3}-\alpha_{3}\beta_{5}\\
					\alpha_{3}\beta_{1}-\alpha_{1}\beta_{4}\\
					\alpha_{1}\beta_{2}-\alpha_{2}\beta_{6}
				\end{pmatrix}\\
				\bm{6}:P_{6}^{m}\begin{pmatrix} 
					\alpha_{2}\beta_{4}\\
					\alpha_{3}\beta_{6}\\
					\alpha_{1}\beta_{5}\\
					-\alpha_{2}\beta_{1}\\
					-\alpha_{1}\beta_{3}\\
					-\alpha_{3}\beta_{2}
				\end{pmatrix}
			\end{array} $  }  \\ \hline

		\multicolumn{12}{|c|}{$\bm{6}\otimes \bm{6}=\bm{1_{0}}\oplus \bm{1_{1}}\oplus\bm{2_{S}}\oplus\bm{2_{A}}\oplus\bm{3_{0}}\oplus \bm{3_{1}}\oplus \bm{\bar{3}_{0}}\oplus\bm{\bar{3}_{1}}\oplus\bm{\hat{3}_{0}}\oplus\bm{\hat{3}_{1}}\oplus\bm6_{S}\oplus\bm{6_{A}}$}  \\ \hline
		\multicolumn{12}{|c|}{ $ \begin{array}{l}
				\bm{1_{0}}:~\alpha_{4} \beta_{1}+\alpha_{6} \beta_{2}+\alpha_{5} \beta_{3}+\alpha_{1} \beta_{4}+\alpha_{3} \beta_{5}+\alpha_{2} \beta_{6}\\
				\bm{1_{1}}:~\alpha_{1} \beta_{4}+\alpha_{3} \beta_{5}+\alpha_{2} \beta_{6}-\alpha_{4} \beta_{1}-\alpha_{6} \beta_{2}-\alpha_{5} \beta_{3} \\
				\bm{2_{S}}:~\begin{pmatrix}
					2\alpha_{4} \beta_{1}-\alpha_{6} \beta_{2}-\alpha_{5} \beta_{3}+2\alpha_{1} \beta_{4}-\alpha_{3} \beta_{5}-\alpha_{2} \beta_{6}\\
					\sqrt{3}(\alpha_{3} \beta_{5}+\alpha_{5} \beta_{3}-\alpha_{2} \beta_{6}-\alpha_{6} \beta_{2})
				\end{pmatrix} \\
				\bm{2_{A}}:~\begin{pmatrix}
					\sqrt{3}\left(	\alpha_{2} \beta_{6}-\alpha_{6} \beta_{2}-\alpha_{3} \beta_{5}+\alpha_{5} \beta_{3}\right)\\
					2\alpha_{1} \beta_{4}-\alpha_{2} \beta_{6}-\alpha_{3} \beta_{5}-2\alpha_{4} \beta_{1}+\alpha_{5} \beta_{3}+\alpha_{6} \beta_{2}
			\end{pmatrix}\end{array} $ } \\		[0.55in]

	\multicolumn{3}{|c}{ } &	\multicolumn{3}{c}{ $ \begin{array}{l}
				\bm{3_{0}}:~\begin{pmatrix} 
					\alpha_{2}\beta_{4}-\alpha_{4}\beta_{2}\\
					\alpha_{3}\beta_{6}-\alpha_{6}\beta_{3}\\
					\alpha_{1}\beta_{5}-\alpha_{5}\beta_{1}
				\end{pmatrix} \\
\bm{3_{1}}:~\begin{pmatrix} 
					\alpha_{4}\beta_{2}+\alpha_{2}\beta_{4}\\
					\alpha_{6}\beta_{3}+\alpha_{3}\beta_{6}\\
					\alpha_{5}\beta_{1}+\alpha_{1}\beta_{5}
				\end{pmatrix}\\
				\bm{\bar{3}_{0}}:~\begin{pmatrix} 
					\alpha_{1}\beta_{6}-\alpha_{6}\beta_{1}\\
					\alpha_{2}\beta_{5}-\alpha_{5}\beta_{2}\\
					\alpha_{3}\beta_{4}-\alpha_{4}\beta_{3}
				\end{pmatrix}\\
				\bm{\bar{3}_{1}}:~\begin{pmatrix} 
					\alpha_{6}\beta_{1}+\alpha_{1}\beta_{6}\\
					\alpha_{5}\beta_{2}+\alpha_{2}\beta_{5}\\
					\alpha_{4}\beta_{3}+\alpha_{3}\beta_{4}
				\end{pmatrix}\\
				\bm{\hat{3}_{0}}:~\begin{pmatrix} 
					\alpha_{3}\beta_{3}-\alpha_{5}\beta_{5}\\
					\alpha_{1}\beta_{1}-\alpha_{4}\beta_{4}\\
					\alpha_{2}\beta_{2}-\alpha_{6}\beta_{6}
				\end{pmatrix}
			\end{array} $	} &
		\multicolumn{6}{c|}{ $ \begin{array}{l}
				
				\bm{\hat{3}_{1}}:~\begin{pmatrix} 
					\alpha_{3}\beta_{3}+\alpha_{5}\beta_{5}\\
					\alpha_{1}\beta_{1}+\alpha_{4}\beta_{4}\\
					\alpha_{2}\beta_{2}+\alpha_{6}\beta_{6}
				\end{pmatrix}\\	
				\bm{6_{S}}:~\begin{pmatrix} 
					\alpha_{5}\beta_{6}+\alpha_{6}\beta_{5}\\
					\alpha_{4}\beta_{5}+\alpha_{5}\beta_{4}\\
					\alpha_{6}\beta_{4}+\alpha_{4}\beta_{6}\\
					\alpha_{2}\beta_{3}+\alpha_{3}\beta_{2}\\
					\alpha_{1}\beta_{2}+\alpha_{2}\beta_{1}\\
					\alpha_{3}\beta_{1}+\alpha_{1}\beta_{3}
				\end{pmatrix}\\
				\bm{6_{A}}:~\begin{pmatrix} 
				\alpha_{5}\beta_{6}-\alpha_{6}\beta_{5}\\
					\alpha_{4}\beta_{5}-\alpha_{5}\beta_{4}\\
					\alpha_{6}\beta_{4}-\alpha_{4}\beta_{6}\\
					\alpha_{2}\beta_{3}-\alpha_{3}\beta_{2}\\
					\alpha_{1}\beta_{2}-\alpha_{2}\beta_{1}\\
					\alpha_{3}\beta_{1}-\alpha_{1}\beta_{3}
				\end{pmatrix}\\
			\end{array} $ } \\	\hline

\specialrule{0em}{1.0pt}{1.0pt}
		\midrule
	\end{longtable}
\end{small}

\begin{center}
\renewcommand{\arraystretch}{1.2}
\renewcommand{\tabcolsep}{0.9mm}
	\setlength\LTcapwidth{\textwidth}
	\begin{longtable}{|c|c|c||c|c|c|c|c|c|c|c|c|}
		\caption{\label{tab:invariant_VEVs}Residual  symmetry restrictions on flavon vacuum alignments for $\phi_{\text{atm}}$ and $\phi_{\text{sol}}$ transforming under either the $\bm{\bar{3}_{0}}$ or $\bm{\bar{3}_{1}}$. Both flavons develop null VEVs $(0,0,0)^T$ when $K_{4}$, $Z_{8}$ and the three $Z_{4}$ subgroups $Z_{4}^{cd^2}$, $Z_{4}^{cd^3}$ and $Z_{4}^{c^2d^3}$ are preserved. $\phi_{\text{sol}}\sim\bm{\bar{3}_{1}}$ similarly shows vanishing VEVs under the other nine $Z_{4}$ subgroups. Vacuum alignment for $\phi_{\text{sol}}\sim\bm{\bar{3}_{1}}$ under all $Z_{3}$ subgroups and the three $Z_{2}$ subgroups whose generators do not include generator $b$  match those in the $\bm{\bar{3}_{0}}$ case. While only one representative residual CP transformation is explicitly presented, all compatible CP-invariant VEV configurations can be generated through global phase factors $e^{\pm i\frac{\pi}{4}}$ or $i$ multiplication. These phase contributions become physically irrelevant through parameter redefinition $\eta\to\eta\pm\frac{\pi}{2}$ or $\eta+\pi$. Residual CP symmetry enforces real VEV magnitudes $v_{\phi_a}$ and $v_{\phi_s}$, with $x$ remaining a free real parameter. 
}\\
\midrule
\specialrule{0em}{1.0pt}{1.0pt}

\endfirsthead

\multicolumn{6}{c}
{{\bfseries \tablename\ \thetable{} -- continued from previous page}} \\
\hline

\endhead

\caption[]{continues on next page}\\
\endfoot

\endlastfoot

\hline

	\multicolumn{6}{|c|}{ Invariant vacuum alignment in the representation $\bm{\bar{3}_{0}}$} \\  \hline 

	$G_{\text{atm}}$ ($G_{\text{sol}}$) & $X^{\text{atm}(\text{sol})}_{\bm{\bar{3}_{0}}}$ & $\langle\phi_{\text{atm}}\rangle/v_{\phi_a}$ $\left(\langle\phi_{\text{sol}}\rangle/v_{\phi_s}\right)$ & $G_{\text{atm}}$ ($G_{\text{sol}}$) & $X^{\text{atm}(\text{sol})}_{\bm{\bar{3}_{0}}}$ & $\langle\phi_{\text{atm}}\rangle/v_{\phi_a}$ $\left(\langle\phi_{\text{sol}}\rangle/v_{\phi_s}\right)$ \\ \hline
	$Z_{2}^{c^2}$, $Z_{4}^{c}$, & \multirow{2}*{$\mathbb{1}_{3}$} & \multirow{2}*{$(0,0,1)^{T}$} &$Z_{2}^{d^2}$, $Z_{4}^{d}$, & \multirow{2}*{$\mathbb{1}_{3}$} & \multirow{2}*{$(1,0,0)^{T}$}\\ 
	$Z_{4}^{abd^2}$, $Z_{4}^{abcd^2}$& && $Z_{4}^{a^2bc^2}$, $Z_{4}^{a^2bc^2d}$&&\\ \hline

   $Z_{2}^{c^2d^2}$, $Z_{4}^{cd}$,  &\multirow{2}*{$\mathbb{1}_{3}$} & \multirow{2}*{$(0,1,0)^{T}$} & $\multirow{2}*{$Z_{2}^{ab}$}$& \multirow{2}*{$\mathbb{1}_{3}$} & \multirow{2}*{$(1,-1,0)^{T}$}\\ 
 $Z_{4}^{bc^2}$, $Z_{4}^{bc^3d}$ & &&& &\\ \hline

	$\multirow{1}*{$Z_{2}^{abc}$}$&$\rho_{\bm{\bar{3}_{0}}}(d^2)$&$(1,i,0)^{T}$&$\multirow{1}*{$Z_{2}^{abc^2}$}$&$\mathbb{1}_{3}$&$(1,1,0)^{T}$\\				\hline
	$\multirow{1}*{$Z_{2}^{abc^3}$}$&$\rho_{\bm{\bar{3}_{0}}}(d^2)$ &$(1,-i,0)^{T}$&$\multirow{1}*{$Z_{2}^{a^2b}$}$&$\mathbb{1}_{3}$&$(0,1,-1)^{T}$\\ 					\hline
    $\multirow{1}*{$Z_{2}^{a^2bd}$}$&$\rho_{\bm{\bar{3}_{0}}}(c^2d^2)$ &$(0,1,i)^{T}$&$\multirow{1}*{$Z_{2}^{a^2bd^2}$}$&$\mathbb{1}_{3}$ &$(0,1,1)^{T}$\\ \hline
	$\multirow{1}*{$Z_{2}^{a^2bd^3}$}$&$\rho_{\bm{\bar{3}_{0}}}(c^2d^2)$&$(0,-1,i)^{T}$&$\multirow{1}*{$Z_{2}^{b}$}$&$\mathbb{1}_{3}$ &$(1,0,-1)^{T}$\\	\hline
	$\multirow{1}*{$Z_{2}^{bcd}$}$&$\rho_{\bm{\bar{3}_{0}}}(d^2)$ &$(1,0,i)^{T}$&$\multirow{1}*{$Z_{2}^{bc^2d^2}$}$&$\mathbb{1}_{3}$&$(1,0,1)^{T}$\\
					\hline
					$\multirow{1}*{$Z_{2}^{bc^3d^3}$}$&$\rho_{\bm{\bar{3}_{0}}}(d^2)$ &$(1,0,-i)^{T}$&$\multirow{1}*{$Z_{3}^{a}$}$&$\mathbb{1}_{3}$ &$(1,1,1)^{T}$\\
					\hline
					$\multirow{1}*{$Z_{3}^{ac}$}$&$\rho_{\bm{\bar{3}_{0}}}(d^2)$ &$(1,-i,-i)^{T}$&$\multirow{1}*{$Z_{3}^{ac^2}$}$&$\mathbb{1}_{3}$ &$(1,-1,-1)^{T}$\\
					\hline
					$\multirow{1}*{$Z_{3}^{ac^3}$}$&$\rho_{\bm{\bar{3}_{0}}}(d^2)$ &$(1,i,i)^{T}$&$\multirow{1}*{$Z_{3}^{ad}$}$&$\rho_{\bm{\bar{3}_{0}}}(bc^2d^2)$ &$(1,i,1)^{T}$\\
					\hline
					$\multirow{1}*{$Z_{3}^{ad^2}$}$&$\mathbb{1}_{3}$ &$(1,-1,1)^{T}$&$\multirow{1}*{$Z_{3}^{ad^3}$}$&$\rho_{\bm{\bar{3}_{0}}}(bc^{2}d^{2})$ &$(1,-i,1)^{T}$\\
					\hline
					$\multirow{1}*{$Z_{3}^{acd}$}$&$\rho_{\bm{\bar{3}_{0}}}(bc^{3}d)$ &$(1,1,-i)^{T}$&$\multirow{1}*{$Z_{3}^{acd^2}$}$&$\rho_{\bm{\bar{3}_{0}}}(d^2)$ &$(1,i,-i )^{T}$\\
					\hline
					$\multirow{1}*{$Z_{3}^{acd^3}$}$&$\rho_{\bm{\bar{3}_{0}}}(bc^3d)$ &$(1,-1,-i )^{T}$&$\multirow{1}*{$Z_{3}^{ac^2d}$}$&$\rho_{\bm{\bar{3}_{0}}}(b)$ &$(1,-i,-1)^{T}$\\
					\hline
					$\multirow{1}*{$Z_{3}^{ac^2d^2}$}$&$\mathbb{1}_{3}$ &$(1,1,-1 )^{T}$&$\multirow{1}*{$Z_{3}^{ac^2d^3}$}$&$\rho_{\bm{\bar{3}_{0}}}(b)$ &$(1,i,-1)^{T}$\\
					\hline
					$\multirow{1}*{$Z_{3}^{ac^3d}$}$&$\rho_{\bm{\bar{3}_{0}}}(bcd^3)$ &$(1,-1,i)^{T}$&$\multirow{1}*{$Z_{3}^{ac^3d^2}$}$&$\rho_{\bm{\bar{3}_{0}}}(d^2)$ &$(1,-i,i)^{T}$\\
					\hline
	$\multirow{1}*{$Z_{3}^{ac^3d^3}$}$&$\rho_{\bm{\bar{3}_{0}}}(bcd^3)$ &$(1,1,i)^{T}$&& &\\
					\hline

 \multicolumn{6}{c}{ } \\[-0.18in] \hline

\multicolumn{6}{|c|}{Invariant vacuum alignment in the representation $\bm{\bar{3}_{1}}$} \\ 					\hline 
$G_{\text{sol}}$ & $X^{\text{sol}}_{\bm{\bar{3}_{1}}}$ & $\langle\phi_{\text{sol}}\rangle/v_{\phi_s}$ & $G_{\text{sol}}$ & $X^{\text{sol}}_{\bm{\bar{3}_{1}}}$ & $\langle\phi_{\text{sol}}\rangle/v_{\phi_s}$ \\
					\hline
$\multirow{4}*{$Z_{2}^{ab}$}$&$\mathbb{1}_{3}$ &$(1,1,x)^{T}$&$\multirow{4}*{$Z_{2}^{abc}$}$&$\rho_{\bm{\bar{3}_{1}}}(d^2)$ &$(-1,i,i x)^{T}$\\
					\cline{2-3}\cline{5-6}
					~
					&$\rho_{\bm{\bar{3}_{1}}}(cd^2)$&$(i-1,i-1,i x)^{T}$&	~
					&$\rho_{\bm{\bar{3}_{1}}}(c)$&$((i-1) x,(1+i) x,1)^{T}$\\
					\cline{2-3}\cline{5-6}
					~
					&$\rho_{\bm{\bar{3}_{1}}}(c^2)$&$(i x,i x,1)^{T}$&~
					&$\rho_{\bm{\bar{3}_{1}}}(c^2d^2)$&$(i,1,i x)^{T}$ \\
					\cline{2-3}\cline{5-6}
					~
					&$\rho_{\bm{\bar{3}_{1}}}(c^3d^2)$&$(1+i,1+i,i x)^{T}$ &~
					&$\rho_{\bm{\bar{3}_{1}}}(c^3)$&$((-1-i) x,(i-1) x,1)^{T}$\\
		\hline
		$\multirow{4}*{$Z_{2}^{abc^2}$}$&$\mathbb{1}_{3}$ &$(1,-1,x)^{T}$&$\multirow{4}*{$Z_{2}^{abc^3}$}$&$\rho_{\bm{\bar{3}_{1}}}(d^2)$ &$(1,i,i x)^{T}$\\
		\cline{2-3}\cline{5-6}
		~
		&$\rho_{\bm{\bar{3}_{1}}}(cd^2)$&$(1-i,i-1,i x)^{T}$&~
		&$\rho_{\bm{\bar{3}_{1}}}(c)$&$((1-i) x,(1+i) x,1)^{T}$\\
		\cline{2-3}\cline{5-6}
		~
		&$\rho_{\bm{\bar{3}_{1}}}(c^2)$&$(-i x,i x,1)^{T}$&~
		&$\rho_{\bm{\bar{3}_{1}}}(c^2d^2)$&$(i,-1,i x)^{T}$\\
		\cline{2-3}\cline{5-6}
		~
		&$\rho_{\bm{\bar{3}_{1}}}(c^3d^2)$&$(-1-i,1+i,i x)^{T}$&~
		&$\rho_{\bm{\bar{3}_{1}}}(c^3)$&$((1+i) x,(i-1) x,1)^{T}$\\	
		\hline
		$\multirow{4}*{$Z_{2}^{a^2b}$}$&$\mathbb{1}_{3}$ &$(1,x,x)^{T}$&$\multirow{4}*{$Z_{2}^{a^2bd}$}$&$\rho_{\bm{\bar{3}_{1}}}(d)$ &$(1,(i-1) x,(1+i) x)^{T}$\\
		\cline{2-3}\cline{5-6}
		~
		&$\rho_{\bm{\bar{3}_{1}}}(d^2)$&$(1,i x,i x)^{T}$&~
		&$\rho_{\bm{\bar{3}_{1}}}(d^3)$&$(1,(-1-i) x,(i-1) x)^{T}$\\
		\cline{2-3}\cline{5-6}
		~
		&$\rho_{\bm{\bar{3}_{1}}}(c^2d)$&$(i,(1+i) x,(1+i) x)^{T}$&~
		&$\rho_{\bm{\bar{3}_{1}}}(c^2)$&$(i,i x,x)^{T}$\\
		\cline{2-3}\cline{5-6}
		~
		&$\rho_{\bm{\bar{3}_{1}}}(c^2d^3)$&$(i,(i-1) x,(i-1) x)^{T}$&~
		&$\rho_{\bm{\bar{3}_{1}}}(c^2d^2)$&$(i,-x,i x)^{T}$\\	
		\hline
		$\multirow{4}*{$Z_{2}^{a^2bd^2}$}$&$\mathbb{1}_{3}$ &$(1,x,-x )^{T}$&$\multirow{4}*{$Z_{2}^{a^2bd^3}$}$&$\rho_{\bm{\bar{3}_{1}}}(d)$ &$(1,(1-i) x,(1+i) x)^{T}$\\
		\cline{2-3}\cline{5-6}
		~
		&$\rho_{\bm{\bar{3}_{1}}}(d^2)$&$(1,-i x,i x)^{T}$&~
		&$\rho_{\bm{\bar{3}_{1}}}(d^3)$&$(1,(1+i) x,(i-1) x)^{T}$\\
		\cline{2-3}\cline{5-6}
		~
		&$\rho_{\bm{\bar{3}_{1}}}(c^2d)$&$(i,(-1-i) x,(1+i) x)^{T}$&~
		&$\rho_{\bm{\bar{3}_{1}}}(c^2)$&$(i,i x,-x)^{T}$\\
		\cline{2-3}\cline{5-6}
		~
		&$\rho_{\bm{\bar{3}_{1}}}(c^2d^3)$&$(i,(1-i) x,(i-1) x)^{T}$&~
		&$\rho_{\bm{\bar{3}_{1}}}(c^2d^2)$&$(i,x,i x)^{T}$\\	
		\hline
		$\multirow{4}*{$Z_{2}^{b}$}$&$\mathbb{1}_{3}$ &$(1,x,1)^{T}$&$\multirow{4}*{$Z_{2}^{bcd}$}$&$\rho_{\bm{\bar{3}_{1}}}(d^2)$ &$(-x,i,i x)^{T}$\\
		\cline{2-3}\cline{5-6}
		~
		&$\rho_{\bm{\bar{3}_{1}}}(cd^3)$&$((i-1) x,i,(i-1) x)^{T}$&~
		&$\rho_{\bm{\bar{3}_{1}}}(cd)$&$((i-1) x,1,(1+i) x)^{T}$\\
		\cline{2-3}\cline{5-6}
		~
		&$\rho_{\bm{\bar{3}_{1}}}(c^2d^2)$&$(i x,1,i x)^{T}$&~
		&$\rho_{\bm{\bar{3}_{1}}}(c^2)$&$(i,i x,1)^{T}$\\
		\cline{2-3}\cline{5-6}
		~
		&$\rho_{\bm{\bar{3}_{1}}}(c^3d)$&$((1+i) x,i,(1+i) x)^{T}$&~
		&$\rho_{\bm{\bar{3}_{1}}}(c^3d^3)$&$((-1-i) x,1,(i-1) x)^{T}$\\
		\hline
		$\multirow{4}*{$Z_{2}^{bc^2d^2}$}$&$\mathbb{1}_{3}$&$(1,x,-1)^{T}$&$\multirow{4}*{$Z_{2}^{bc^3d^3}$}$&$\rho_{\bm{\bar{3}_{1}}}(d^2)$ &$(x,i,i x)^{T}$\\
		\cline{2-3}\cline{5-6}
		~
		&$\rho_{\bm{\bar{3}_{1}}}(cd^3)$&$((1-i) x,i,(i-1) x)^{T}$&~
		&$\rho_{\bm{\bar{3}_{1}}}(cd)$&$((1-i) x,1,(1+i) x)^{T}$\\
		\cline{2-3}\cline{5-6}
		~
		&$\rho_{\bm{\bar{3}_{1}}}(c^2d^2)$&$(-i x,1,i x)^{T}$&~
		&$\rho_{\bm{\bar{3}_{1}}}(c^2)$&$(i,i x,-1)^{T}$\\
		\cline{2-3}\cline{5-6}
		~
		&$\rho_{\bm{\bar{3}_{1}}}(c^3d)$&$((-1-i) x,i,(1+i) x)^{T}$&	~
		&$\rho_{\bm{\bar{3}_{1}}}(c^3d^3)$&$((1+i) x,1,(i-1) x)^{T}$\\
					\hline
\specialrule{0em}{1.0pt}{1.0pt}
\midrule
	\end{longtable}
\end{center}

In the tri-direct CP approach, the original symmetry  $\Delta(96)\rtimes H_{CP}$ breaks spontaneously into $G_{\text{atm}}\rtimes H^{\text{atm}}_{CP}$ (atmospheric sector) and $G_{\text{sol}}\rtimes H^{\text{sol}}_{CP}$ (solar sector) through flavons $\phi_{\text{atm}}$ and $\phi_{\text{sol}}$, where $G_{\text{atm}}$ and $G_{\text{sol}}$ are Abelian subgroups of $\Delta(96)$. These flavons adopt vacuum alignments  which are invariant under the action of the corresponding residual symmetries. Here, the flavon $\phi_{\text{atm}}$ transforms as triplet representation $\bm{\bar{3}_{0}}$, and the flavon $\phi_{\text{sol}}$ is assigned to triplet representation $\bm{\bar{3}_{0}}$ or $\bm{\bar{3}_{1}}$. The most general VEVs of the flavons $\phi_{\text{atm}}$ and $\phi_{\text{sol}}$ which preserve the possible residual symmetries are summarized in table~\ref{tab:invariant_VEVs}.

\section{\label{sec:Diag_mnup}Diagonalization of the neutrino mass matrix $m^\prime_{\nu}$ }

This appendix outlines the diagonalization procedure for the modified neutrino mass matrix $m^\prime_{\nu}$. As specified in Eq.~\eqref{eq:mnup}, the matrix adopts the following explicit structure:
\begin{equation}
	m^\prime_{\nu}=\left\{\begin{array}{ccc}
	\begin{pmatrix}
	0 & ~ 0 &~ 0 \\
	0 &~ |y|e^{i\phi_{y}}  ~&~  |z|e^{i\phi_{z}} \\
	0 &~ |z|e^{i\phi_{z}}  ~&~  |w|e^{i\phi_{w}}
	\end{pmatrix}&\mathrm{for}&\mathrm{NO\,,}\\\\\begin{pmatrix}
	|y|e^{i\phi_{y}} &|z|e^{i\phi_{z}} &~ 0 \\
	|z|e^{i\phi_{z}} &~ |w|e^{i\phi_{w}}  ~&~  0\\
	0 &0  ~&~  0
	\end{pmatrix}&\mathrm{for}&\mathrm{IO\,.}
	\end{array}\right.
\end{equation}
where the parameters $|y|$, $|z|$, $|w|$, $\phi_{y}$, $\phi_{z}$ and $\phi_{w}$ can be obtained from Eq.~\eqref{eq:yzw}. The precise diagonalization of this matrix is achieved through a unitary transformation $U_{\nu2}$ in Eq.~\eqref{eq:Unu2},
\begin{equation}\label{eq:diag_2nd}
U_{\nu2}^Tm_\nu^{\prime}U_{\nu2}=\left\{\begin{array}{ccc}\mathrm{diag}(0,m_2,m_3)&\mathrm{for}&\mathrm{NO}\,,\\\mathrm{diag}(m_1,m_2,0)&\mathrm{for}&\mathrm{IO}\,.\end{array}\right.
\end{equation}
with the physical neutrino masses derived as
\begin{equation}\label{eq:nu_masses}
		m^2_i=\frac{1}{2}\left[|y|^2+|w|^2+2|z|^2-\frac{|w|^2-|y|^2}{\cos2\theta}\right], \quad
		m^2_j=\frac{1}{2}\left[|y|^2+|w|^2+2|z|^2+\frac{|w|^2-|y|^2}{\cos2\theta}\right]\,.
\end{equation}
with $m_{i}$= $m_{2}$, $m_{j}$= $m_{3}$ for the NO case and $m_{i}$= $m_{1}$, $m_{j}$= $m_{2}$ for the IO case.
The rotational parameter $\theta$, the phase parameters $\psi$, $\rho$ and $\sigma$ are determined by the following angular relations:
\begin{eqnarray}
\nonumber&&\sin2\theta=\frac{2|z|\sqrt{|y|^2+|w|^2+2|y||w|\cos(\phi_{y}+\phi_{w}-2\phi_{z})}}
{\sqrt{(|w|^2-|y|^2)^2+4|z|^2\left[|y|^2+|w|^2+2|y||w|\cos(\phi_{y}+\phi_{w}-2\phi_{z})\right]}},\\
\nonumber &&\cos2\theta=\frac{|w|^2-|y|^2}{\sqrt{(|w|^2-|y|^2)^2+4|z|^2
\left[|y|^2+|w|^2+2|y||w|\cos(\phi_{y}+\phi_{w}-2\phi_{z})\right]}}\,, \\
\nonumber&&\sin\psi=\frac{-|y|\sin(\phi_{y}-\phi_{z})+|w|\sin(\phi_{w}-\phi_{z})}
{\sqrt{|y|^2+|w|^2+2|y||w|\cos(\phi_{y}+\phi_{w}-2\phi_{z})}}\,, \\
\nonumber && \cos\psi=\frac{|y|\cos(\phi_{y}-\phi_{z})+|w|\cos(\phi_{w}-\phi_{z})}{\sqrt{|y|^2+|w|^2+2|y||w|\cos(\phi_{y}+\phi_{w}-2\phi_{z})}}\,,\\
\nonumber&&\sin\rho=-\frac{(m^2_2-|z|^2)\sin\phi_{z}+|y||w|\sin(\phi_{y}+\phi_{w}-\phi_{z})}
{m_2\sqrt{|y|^2+|w|^2+2|y||w|\cos(\phi_{y}+\phi_{w}-2\phi_{z})}}\,,\\
\nonumber && \cos\rho=\frac{(m^2_2-|z|^2)\cos\phi_{z}+|y||w|\cos(\phi_{y}+\phi_{w}-\phi_{z})}
{m_2\sqrt{|y|^2+|w|^2+2|y||w|\cos(\phi_{y}+\phi_{w}-2\phi_{z})}}\,,\\
\nonumber &&\sin\sigma=-\frac{(m^2_3-|z|^2)\sin\phi_{z}+|y||w|\sin(\phi_{y}+\phi_{w}-\phi_{z})}
{m_3\sqrt{|y|^2+|w|^2+2|y||w|\cos(\phi_{y}+\phi_{w}-2\phi_{z})}}\,,\\
\label{eq:prs_U2} && \cos\sigma=\frac{(m^2_3-|z|^2)\cos\phi_{z}+|y||w|\cos(\phi_{y}+\phi_{w}-\phi_{z})}
{m_3\sqrt{|y|^2+|w|^2+2|y||w|\cos(\phi_{y}+\phi_{w}-2\phi_{z})}}\,.
\end{eqnarray}

\end{appendix}

\providecommand{\href}[2]{#2}\begingroup\raggedright\endgroup


\begin{thebibliography}{10}

\bibitem{King:2014nza}
S.~F. King, A.~Merle, S.~Morisi, Y.~Shimizu, and M.~Tanimoto, ``{Neutrino Mass
  and Mixing: from Theory to Experiment},''
  \href{http://dx.doi.org/10.1088/1367-2630/16/4/045018}{{\em New J. Phys.}
  {\bfseries 16} (2014) 045018},
  \href{http://arxiv.org/abs/1402.4271}{{\ttfamily arXiv:1402.4271 [hep-ph]}}.

\bibitem{King:2015aea}
S.~F. King, ``{Models of Neutrino Mass, Mixing and CP Violation},''
  \href{http://dx.doi.org/10.1088/0954-3899/42/12/123001}{{\em J. Phys. G}
  {\bfseries 42} (2015) 123001},
  \href{http://arxiv.org/abs/1510.02091}{{\ttfamily arXiv:1510.02091
  [hep-ph]}}.

\bibitem{King:2017guk}
S.~F. King, ``{Unified Models of Neutrinos, Flavour and CP Violation},''
  \href{http://dx.doi.org/10.1016/j.ppnp.2017.01.003}{{\em Prog. Part. Nucl.
  Phys.} {\bfseries 94} (2017) 217--256},
  \href{http://arxiv.org/abs/1701.04413}{{\ttfamily arXiv:1701.04413
  [hep-ph]}}.

\bibitem{Feruglio:2019ybq}
F.~Feruglio and A.~Romanino, ``{Lepton flavor symmetries},''
  \href{http://dx.doi.org/10.1103/RevModPhys.93.015007}{{\em Rev. Mod. Phys.}
  {\bfseries 93} no.~1, (2021) 015007},
  \href{http://arxiv.org/abs/1912.06028}{{\ttfamily arXiv:1912.06028
  [hep-ph]}}.

\bibitem{Xing:2020ijf}
Z.-z. Xing, ``{Flavor structures of charged fermions and massive neutrinos},''
  \href{http://dx.doi.org/10.1016/j.physrep.2020.02.001}{{\em Phys. Rept.}
  {\bfseries 854} (2020) 1--147},
  \href{http://arxiv.org/abs/1909.09610}{{\ttfamily arXiv:1909.09610
  [hep-ph]}}.

\bibitem{Almumin:2022rml}
Y.~Almumin, M.-C. Chen, M.~Cheng, V.~Knapp-Perez, Y.~Li, A.~Mondol,
  S.~Ramos-Sanchez, M.~Ratz, and S.~Shukla, ``{Neutrino Flavor Model Building
  and the Origins of Flavor and CP Violation},''
  \href{http://dx.doi.org/10.3390/universe9120512}{{\em Universe} {\bfseries 9}
  no.~12, (2023) 512}, \href{http://arxiv.org/abs/2204.08668}{{\ttfamily
  arXiv:2204.08668 [hep-ph]}}.

\bibitem{Altarelli:2010gt}
G.~Altarelli and F.~Feruglio, ``{Discrete Flavor Symmetries and Models of
  Neutrino Mixing},'' \href{http://dx.doi.org/10.1103/RevModPhys.82.2701}{{\em
  Rev. Mod. Phys.} {\bfseries 82} (2010) 2701--2729},
  \href{http://arxiv.org/abs/1002.0211}{{\ttfamily arXiv:1002.0211 [hep-ph]}}.

\bibitem{Ishimori:2010au}
H.~Ishimori, T.~Kobayashi, H.~Ohki, Y.~Shimizu, H.~Okada, and M.~Tanimoto,
  ``{Non-Abelian Discrete Symmetries in Particle Physics},''
  \href{http://dx.doi.org/10.1143/PTPS.183.1}{{\em Prog. Theor. Phys. Suppl.}
  {\bfseries 183} (2010) 1--163},
  \href{http://arxiv.org/abs/1003.3552}{{\ttfamily arXiv:1003.3552 [hep-th]}}.

\bibitem{King:2013eh}
S.~F. King and C.~Luhn, ``{Neutrino Mass and Mixing with Discrete Symmetry},''
  \href{http://dx.doi.org/10.1088/0034-4885/76/5/056201}{{\em Rept. Prog.
  Phys.} {\bfseries 76} (2013) 056201},
  \href{http://arxiv.org/abs/1301.1340}{{\ttfamily arXiv:1301.1340 [hep-ph]}}.

\bibitem{Petcov:2017ggy}
S.~T. Petcov, ``{Discrete Flavour Symmetries, Neutrino Mixing and Leptonic CP
  Violation},'' \href{http://dx.doi.org/10.1140/epjc/s10052-018-6158-5}{{\em
  Eur. Phys. J. C} {\bfseries 78} no.~9, (2018) 709},
  \href{http://arxiv.org/abs/1711.10806}{{\ttfamily arXiv:1711.10806
  [hep-ph]}}.

\bibitem{Ding:2024ozt}
G.-J. Ding and J.~W.~F. Valle, ``{The symmetry approach to quark and lepton
  masses and mixing},''
  \href{http://dx.doi.org/10.1016/j.physrep.2024.12.005}{{\em Phys. Rept.}
  {\bfseries 1109} (2025) 1--105},
  \href{http://arxiv.org/abs/2402.16963}{{\ttfamily arXiv:2402.16963
  [hep-ph]}}.

\bibitem{Feruglio:2012cw}
F.~Feruglio, C.~Hagedorn, and R.~Ziegler, ``{Lepton Mixing Parameters from
  Discrete and CP Symmetries},''
  \href{http://dx.doi.org/10.1007/JHEP07(2013)027}{{\em JHEP} {\bfseries 07}
  (2013) 027}, \href{http://arxiv.org/abs/1211.5560}{{\ttfamily arXiv:1211.5560
  [hep-ph]}}.

\bibitem{Holthausen:2012dk}
M.~Holthausen, M.~Lindner, and M.~A. Schmidt, ``{CP and Discrete Flavour
  Symmetries},'' \href{http://dx.doi.org/10.1007/JHEP04(2013)122}{{\em JHEP}
  {\bfseries 04} (2013) 122}, \href{http://arxiv.org/abs/1211.6953}{{\ttfamily
  arXiv:1211.6953 [hep-ph]}}.

\bibitem{Ding:2013hpa}
G.-J. Ding, S.~F. King, C.~Luhn, and A.~J. Stuart, ``{Spontaneous CP violation
  from vacuum alignment in $S_4$ models of leptons},''
  \href{http://dx.doi.org/10.1007/JHEP05(2013)084}{{\em JHEP} {\bfseries 05}
  (2013) 084}, \href{http://arxiv.org/abs/1303.6180}{{\ttfamily arXiv:1303.6180
  [hep-ph]}}.

\bibitem{Ding:2013bpa}
G.-J. Ding, S.~F. King, and A.~J. Stuart, ``{Generalised CP and $A_4$ Family
  Symmetry},'' \href{http://dx.doi.org/10.1007/JHEP12(2013)006}{{\em JHEP}
  {\bfseries 12} (2013) 006}, \href{http://arxiv.org/abs/1307.4212}{{\ttfamily
  arXiv:1307.4212 [hep-ph]}}.

\bibitem{Li:2013jya}
C.-C. Li and G.-J. Ding, ``{Generalised CP and trimaximal $TM_1$ lepton mixing
  in $S_4$ family symmetry},''
  \href{http://dx.doi.org/10.1016/j.nuclphysb.2014.02.002}{{\em Nucl. Phys. B}
  {\bfseries 881} (2014) 206--232},
  \href{http://arxiv.org/abs/1312.4401}{{\ttfamily arXiv:1312.4401 [hep-ph]}}.

\bibitem{Ding:2013nsa}
G.-J. Ding and Y.-L. Zhou, ``{Predicting lepton flavor mixing from $\Delta$(48)
  and generalized $CP$ symmetries},''
  \href{http://dx.doi.org/10.1088/1674-1137/39/2/021001}{{\em Chin. Phys. C}
  {\bfseries 39} no.~2, (2015) 021001},
  \href{http://arxiv.org/abs/1312.5222}{{\ttfamily arXiv:1312.5222 [hep-ph]}}.

\bibitem{Ding:2014ssa}
G.-J. Ding and S.~F. King, ``{Generalized $CP$ and $\Delta(96)$ family
  symmetry},'' \href{http://dx.doi.org/10.1103/PhysRevD.89.093020}{{\em Phys.
  Rev. D} {\bfseries 89} no.~9, (2014) 093020},
  \href{http://arxiv.org/abs/1403.5846}{{\ttfamily arXiv:1403.5846 [hep-ph]}}.

\bibitem{Ding:2014hva}
G.-J. Ding and Y.-L. Zhou, ``{Lepton mixing parameters from $\Delta(48)$ family
  symmetry and generalised CP},''
  \href{http://dx.doi.org/10.1007/JHEP06(2014)023}{{\em JHEP} {\bfseries 06}
  (2014) 023}, \href{http://arxiv.org/abs/1404.0592}{{\ttfamily arXiv:1404.0592
  [hep-ph]}}.

\bibitem{Li:2014eia}
C.-C. Li and G.-J. Ding, ``{Deviation from bimaximal mixing and leptonic CP
  phases in S$_{4}$ family symmetry and generalized CP},''
  \href{http://dx.doi.org/10.1007/JHEP08(2015)017}{{\em JHEP} {\bfseries 08}
  (2015) 017}, \href{http://arxiv.org/abs/1408.0785}{{\ttfamily arXiv:1408.0785
  [hep-ph]}}.

\bibitem{Ding:2014ora}
G.-J. Ding, S.~F. King, and T.~Neder, ``{Generalised CP and $\Delta(6n^2)$
  family symmetry in semi-direct models of leptons},''
  \href{http://dx.doi.org/10.1007/JHEP12(2014)007}{{\em JHEP} {\bfseries 12}
  (2014) 007}, \href{http://arxiv.org/abs/1409.8005}{{\ttfamily arXiv:1409.8005
  [hep-ph]}}.

\bibitem{Chen:2014wxa}
P.~Chen, C.-C. Li, and G.-J. Ding, ``{Lepton Flavor Mixing and CP Symmetry},''
  \href{http://dx.doi.org/10.1103/PhysRevD.91.033003}{{\em Phys. Rev. D}
  {\bfseries 91} (2015) 033003},
  \href{http://arxiv.org/abs/1412.8352}{{\ttfamily arXiv:1412.8352 [hep-ph]}}.

\bibitem{Everett:2015oka}
L.~L. Everett, T.~Garon, and A.~J. Stuart, ``{A Bottom-Up Approach to Lepton
  Flavor and CP Symmetries},''
  \href{http://dx.doi.org/10.1007/JHEP04(2015)069}{{\em JHEP} {\bfseries 04}
  (2015) 069}, \href{http://arxiv.org/abs/1501.04336}{{\ttfamily
  arXiv:1501.04336 [hep-ph]}}.

\bibitem{Branco:2015hea}
G.~C. Branco, I.~de~Medeiros~Varzielas, and S.~F. King, ``{Invariant approach
  to CP in family symmetry models},''
  \href{http://dx.doi.org/10.1103/PhysRevD.92.036007}{{\em Phys. Rev. D}
  {\bfseries 92} no.~3, (2015) 036007},
  \href{http://arxiv.org/abs/1502.03105}{{\ttfamily arXiv:1502.03105
  [hep-ph]}}.

\bibitem{Li:2015jxa}
C.-C. Li and G.-J. Ding, ``{Lepton Mixing in $A_5$ Family Symmetry and
  Generalized CP},'' \href{http://dx.doi.org/10.1007/JHEP05(2015)100}{{\em
  JHEP} {\bfseries 05} (2015) 100},
  \href{http://arxiv.org/abs/1503.03711}{{\ttfamily arXiv:1503.03711
  [hep-ph]}}.

\bibitem{DiIura:2015kfa}
A.~Di~Iura, C.~Hagedorn, and D.~Meloni, ``{Lepton mixing from the interplay of
  the alternating group A$_{5}$ and CP},''
  \href{http://dx.doi.org/10.1007/JHEP08(2015)037}{{\em JHEP} {\bfseries 08}
  (2015) 037}, \href{http://arxiv.org/abs/1503.04140}{{\ttfamily
  arXiv:1503.04140 [hep-ph]}}.

\bibitem{Ballett:2015wia}
P.~Ballett, S.~Pascoli, and J.~Turner, ``{Mixing angle and phase correlations
  from A5 with generalized CP and their prospects for discovery},''
  \href{http://dx.doi.org/10.1103/PhysRevD.92.093008}{{\em Phys. Rev. D}
  {\bfseries 92} no.~9, (2015) 093008},
  \href{http://arxiv.org/abs/1503.07543}{{\ttfamily arXiv:1503.07543
  [hep-ph]}}.

\bibitem{Branco:2015gna}
G.~C. Branco, I.~de~Medeiros~Varzielas, and S.~F. King, ``{Invariant approach
  to $\mathcal {CP}$ in unbroken $\Delta(27)$},''
  \href{http://dx.doi.org/10.1016/j.nuclphysb.2015.07.024}{{\em Nucl. Phys. B}
  {\bfseries 899} (2015) 14--36},
  \href{http://arxiv.org/abs/1505.06165}{{\ttfamily arXiv:1505.06165
  [hep-ph]}}.

\bibitem{Chen:2015nha}
P.~Chen, C.-Y. Yao, and G.-J. Ding, ``{Neutrino Mixing from CP Symmetry},''
  \href{http://dx.doi.org/10.1103/PhysRevD.92.073002}{{\em Phys. Rev. D}
  {\bfseries 92} no.~7, (2015) 073002},
  \href{http://arxiv.org/abs/1507.03419}{{\ttfamily arXiv:1507.03419
  [hep-ph]}}.

\bibitem{Ding:2015rwa}
G.-J. Ding and S.~F. King, ``{Generalized CP and $\Delta (3n^2)$ Family
  Symmetry for Semi-Direct Predictions of the PMNS Matrix},''
  \href{http://dx.doi.org/10.1103/PhysRevD.93.025013}{{\em Phys. Rev. D}
  {\bfseries 93} (2016) 025013},
  \href{http://arxiv.org/abs/1510.03188}{{\ttfamily arXiv:1510.03188
  [hep-ph]}}.

\bibitem{Chen:2015siy}
P.~Chen, G.-J. Ding, F.~Gonzalez-Canales, and J.~W.~F. Valle, ``{Generalized
  $\mu-\tau$ reflection symmetry and leptonic CP violation},''
  \href{http://dx.doi.org/10.1016/j.physletb.2015.12.069}{{\em Phys. Lett. B}
  {\bfseries 753} (2016) 644--652},
  \href{http://arxiv.org/abs/1512.01551}{{\ttfamily arXiv:1512.01551
  [hep-ph]}}.

\bibitem{Li:2016ppt}
C.-C. Li, C.-Y. Yao, and G.-J. Ding, ``{Lepton Mixing Predictions from Infinite
  Group Series $D^{(1)}_{9n, 3n}$ with Generalized CP},''
  \href{http://dx.doi.org/10.1007/JHEP05(2016)007}{{\em JHEP} {\bfseries 05}
  (2016) 007}, \href{http://arxiv.org/abs/1601.06393}{{\ttfamily
  arXiv:1601.06393 [hep-ph]}}.

\bibitem{Chen:2016ptr}
P.~Chen, G.-J. Ding, and S.~F. King, ``{Leptogenesis and residual CP
  symmetry},'' \href{http://dx.doi.org/10.1007/JHEP03(2016)206}{{\em JHEP}
  {\bfseries 03} (2016) 206}, \href{http://arxiv.org/abs/1602.03873}{{\ttfamily
  arXiv:1602.03873 [hep-ph]}}.

\bibitem{Yao:2016zev}
C.-Y. Yao and G.-J. Ding, ``{CP Symmetry and Lepton Mixing from a Scan of
  Finite Discrete Groups},''
  \href{http://dx.doi.org/10.1103/PhysRevD.94.073006}{{\em Phys. Rev. D}
  {\bfseries 94} no.~7, (2016) 073006},
  \href{http://arxiv.org/abs/1606.05610}{{\ttfamily arXiv:1606.05610
  [hep-ph]}}.

\bibitem{Li:2016nap}
C.-C. Li, J.-N. Lu, and G.-J. Ding, ``{$A_4$ and CP symmetry and a model with
  maximal CP violation},''
  \href{http://dx.doi.org/10.1016/j.nuclphysb.2016.09.005}{{\em Nucl. Phys. B}
  {\bfseries 913} (2016) 110--131},
  \href{http://arxiv.org/abs/1608.01860}{{\ttfamily arXiv:1608.01860
  [hep-ph]}}.

\bibitem{Lu:2016jit}
J.-N. Lu and G.-J. Ding, ``{Alternative Schemes of Predicting Lepton Mixing
  Parameters from Discrete Flavor and CP Symmetry},''
  \href{http://dx.doi.org/10.1103/PhysRevD.95.015012}{{\em Phys. Rev. D}
  {\bfseries 95} no.~1, (2017) 015012},
  \href{http://arxiv.org/abs/1610.05682}{{\ttfamily arXiv:1610.05682
  [hep-ph]}}.

\bibitem{Everett:2016jsk}
L.~L. Everett and A.~J. Stuart, ``{Lepton Sector Phases and Their Roles in
  Flavor and Generalized CP Symmetries},''
  \href{http://dx.doi.org/10.1103/PhysRevD.96.035030}{{\em Phys. Rev. D}
  {\bfseries 96} no.~3, (2017) 035030},
  \href{http://arxiv.org/abs/1611.03020}{{\ttfamily arXiv:1611.03020
  [hep-ph]}}.

\bibitem{Li:2017zmk}
C.-C. Li and G.-J. Ding, ``{Implications of residual CP symmetry for
  leptogenesis in a model with two right-handed neutrinos},''
  \href{http://dx.doi.org/10.1103/PhysRevD.96.075005}{{\em Phys. Rev. D}
  {\bfseries 96} no.~7, (2017) 075005},
  \href{http://arxiv.org/abs/1701.08508}{{\ttfamily arXiv:1701.08508
  [hep-ph]}}.

\bibitem{Li:2017abz}
C.-C. Li, J.-N. Lu, and G.-J. Ding, ``{Toward a unified interpretation of quark
  and lepton mixing from flavor and CP symmetries},''
  \href{http://dx.doi.org/10.1007/JHEP02(2018)038}{{\em JHEP} {\bfseries 02}
  (2018) 038}, \href{http://arxiv.org/abs/1706.04576}{{\ttfamily
  arXiv:1706.04576 [hep-ph]}}.

\bibitem{Lu:2018oxc}
J.-N. Lu and G.-J. Ding, ``{Quark and lepton mixing patterns from a common
  discrete flavor symmetry with a generalized CP symmetry},''
  \href{http://dx.doi.org/10.1103/PhysRevD.98.055011}{{\em Phys. Rev. D}
  {\bfseries 98} no.~5, (2018) 055011},
  \href{http://arxiv.org/abs/1806.02301}{{\ttfamily arXiv:1806.02301
  [hep-ph]}}.

\bibitem{Lu:2019gqp}
J.-N. Lu and G.-J. Ding, ``{Dihedral flavor group as the key to understand
  quark and lepton flavor mixing},''
  \href{http://dx.doi.org/10.1007/JHEP03(2019)056}{{\em JHEP} {\bfseries 03}
  (2019) 056}, \href{http://arxiv.org/abs/1901.07414}{{\ttfamily
  arXiv:1901.07414 [hep-ph]}}.

\bibitem{Chen:2018lsv}
P.~Chen, S.~Centelles~Chuli\'a, G.-J. Ding, R.~Srivastava, and J.~W.~F. Valle,
  ``{Neutrino Predictions from Generalized CP Symmetries of Charged Leptons},''
  \href{http://dx.doi.org/10.1007/JHEP07(2018)077}{{\em JHEP} {\bfseries 07}
  (2018) 077}, \href{http://arxiv.org/abs/1802.04275}{{\ttfamily
  arXiv:1802.04275 [hep-ph]}}.

\bibitem{Hagedorn:2016lva}
C.~Hagedorn and E.~Molinaro, ``{Flavor and CP symmetries for leptogenesis and 0
  \ensuremath{\nu}\ensuremath{\beta}\ensuremath{\beta} decay},''
  \href{http://dx.doi.org/10.1016/j.nuclphysb.2017.03.015}{{\em Nucl. Phys. B}
  {\bfseries 919} (2017) 404--469},
  \href{http://arxiv.org/abs/1602.04206}{{\ttfamily arXiv:1602.04206
  [hep-ph]}}.

\bibitem{Delgadillo:2018tza}
L.~A. Delgadillo, L.~L. Everett, R.~Ramos, and A.~J. Stuart, ``{Predictions for
  the Dirac CP-Violating Phase from Sum Rules},''
  \href{http://dx.doi.org/10.1103/PhysRevD.97.095001}{{\em Phys. Rev. D}
  {\bfseries 97} no.~9, (2018) 095001},
  \href{http://arxiv.org/abs/1801.06377}{{\ttfamily arXiv:1801.06377
  [hep-ph]}}.

\bibitem{Minkowski:1977sc}
P.~Minkowski, ``{$\mu \to e\gamma$ at a Rate of One Out of $10^{9}$ Muon
  Decays?},'' \href{http://dx.doi.org/10.1016/0370-2693(77)90435-X}{{\em Phys.
  Lett. B} {\bfseries 67} (1977) 421--428}.

\bibitem{Mohapatra:1979ia}
R.~N. Mohapatra and G.~Senjanovic, ``{Neutrino Mass and Spontaneous Parity
  Nonconservation},'' \href{http://dx.doi.org/10.1103/PhysRevLett.44.912}{{\em
  Phys. Rev. Lett.} {\bfseries 44} (1980) 912}.

\bibitem{Schechter:1980gr}
J.~Schechter and J.~W.~F. Valle, ``{Neutrino Masses in SU(2) x U(1)
  Theories},'' \href{http://dx.doi.org/10.1103/PhysRevD.22.2227}{{\em Phys.
  Rev. D} {\bfseries 22} (1980) 2227}.

\bibitem{King:1998jw}
S.~F. King, ``{Atmospheric and solar neutrinos with a heavy singlet},''
  \href{http://dx.doi.org/10.1016/S0370-2693(98)01055-7}{{\em Phys. Lett. B}
  {\bfseries 439} (1998) 350--356},
  \href{http://arxiv.org/abs/hep-ph/9806440}{{\ttfamily arXiv:hep-ph/9806440}}.

\bibitem{King:1999cm}
S.~F. King, ``{Atmospheric and solar neutrinos from single right-handed
  neutrino dominance and U(1) family symmetry},''
  \href{http://dx.doi.org/10.1016/S0550-3213(99)00542-8}{{\em Nucl. Phys. B}
  {\bfseries 562} (1999) 57--77},
  \href{http://arxiv.org/abs/hep-ph/9904210}{{\ttfamily arXiv:hep-ph/9904210}}.

\bibitem{Esteban:2024eli}
I.~Esteban, M.~C. Gonzalez-Garcia, M.~Maltoni, I.~Martinez-Soler, J.~a.~P.
  Pinheiro, and T.~Schwetz, ``{NuFit-6.0: updated global analysis of
  three-flavor neutrino oscillations},''
  \href{http://dx.doi.org/10.1007/JHEP12(2024)216}{{\em JHEP} {\bfseries 12}
  (2024) 216}, \href{http://arxiv.org/abs/2410.05380}{{\ttfamily
  arXiv:2410.05380 [hep-ph]}}.

\bibitem{King:1999mb}
S.~F. King, ``{Large mixing angle MSW and atmospheric neutrinos from single
  right-handed neutrino dominance and U(1) family symmetry},''
  \href{http://dx.doi.org/10.1016/S0550-3213(00)00109-7}{{\em Nucl. Phys. B}
  {\bfseries 576} (2000) 85--105},
  \href{http://arxiv.org/abs/hep-ph/9912492}{{\ttfamily arXiv:hep-ph/9912492}}.

\bibitem{Frampton:2002qc}
P.~H. Frampton, S.~L. Glashow, and T.~Yanagida, ``{Cosmological sign of
  neutrino CP violation},''
  \href{http://dx.doi.org/10.1016/S0370-2693(02)02853-8}{{\em Phys. Lett. B}
  {\bfseries 548} (2002) 119--121},
  \href{http://arxiv.org/abs/hep-ph/0208157}{{\ttfamily arXiv:hep-ph/0208157}}.

\bibitem{King:2002nf}
S.~F. King, ``{Constructing the large mixing angle MNS matrix in seesaw models
  with right-handed neutrino dominance},''
  \href{http://dx.doi.org/10.1088/1126-6708/2002/09/011}{{\em JHEP} {\bfseries
  09} (2002) 011}, \href{http://arxiv.org/abs/hep-ph/0204360}{{\ttfamily
  arXiv:hep-ph/0204360}}.

\bibitem{Guo:2006qa}
W.-l. Guo, Z.-z. Xing, and S.~Zhou, ``{Neutrino Masses, Lepton Flavor Mixing
  and Leptogenesis in the Minimal Seesaw Model},''
  \href{http://dx.doi.org/10.1142/S0218301307004898}{{\em Int. J. Mod. Phys. E}
  {\bfseries 16} (2007) 1--50},
  \href{http://arxiv.org/abs/hep-ph/0612033}{{\ttfamily arXiv:hep-ph/0612033}}.

\bibitem{Harigaya:2012bw}
K.~Harigaya, M.~Ibe, and T.~T. Yanagida, ``{Seesaw Mechanism with Occam's
  Razor},'' \href{http://dx.doi.org/10.1103/PhysRevD.86.013002}{{\em Phys. Rev.
  D} {\bfseries 86} (2012) 013002},
  \href{http://arxiv.org/abs/1205.2198}{{\ttfamily arXiv:1205.2198 [hep-ph]}}.

\bibitem{Zhang:2015tea}
J.~Zhang and S.~Zhou, ``{A Further Study of the Frampton-Glashow-Yanagida Model
  for Neutrino Masses, Flavor Mixing and Baryon Number Asymmetry},''
  \href{http://dx.doi.org/10.1007/JHEP09(2015)065}{{\em JHEP} {\bfseries 09}
  (2015) 065}, \href{http://arxiv.org/abs/1505.04858}{{\ttfamily
  arXiv:1505.04858 [hep-ph]}}.

\bibitem{King:2005bj}
S.~F. King, ``{Predicting neutrino parameters from SO(3) family symmetry and
  quark-lepton unification},''
  \href{http://dx.doi.org/10.1088/1126-6708/2005/08/105}{{\em JHEP} {\bfseries
  08} (2005) 105}, \href{http://arxiv.org/abs/hep-ph/0506297}{{\ttfamily
  arXiv:hep-ph/0506297}}.

\bibitem{Antusch:2011ic}
S.~Antusch, S.~F. King, C.~Luhn, and M.~Spinrath, ``{Trimaximal mixing with
  predicted $\theta_{13}$ from a new type of constrained sequential
  dominance},'' \href{http://dx.doi.org/10.1016/j.nuclphysb.2011.11.009}{{\em
  Nucl. Phys. B} {\bfseries 856} (2012) 328--341},
  \href{http://arxiv.org/abs/1108.4278}{{\ttfamily arXiv:1108.4278 [hep-ph]}}.

\bibitem{King:2013iva}
S.~F. King, ``{Minimal predictive see-saw model with normal neutrino mass
  hierarchy},'' \href{http://dx.doi.org/10.1007/JHEP07(2013)137}{{\em JHEP}
  {\bfseries 07} (2013) 137}, \href{http://arxiv.org/abs/1304.6264}{{\ttfamily
  arXiv:1304.6264 [hep-ph]}}.

\bibitem{King:2015dvf}
S.~F. King, ``{Littlest Seesaw},''
  \href{http://dx.doi.org/10.1007/JHEP02(2016)085}{{\em JHEP} {\bfseries 02}
  (2016) 085}, \href{http://arxiv.org/abs/1512.07531}{{\ttfamily
  arXiv:1512.07531 [hep-ph]}}.

\bibitem{King:2016yvg}
S.~F. King and C.~Luhn, ``{Littlest Seesaw model from S$_{4} \times$ U(1)},''
  \href{http://dx.doi.org/10.1007/JHEP09(2016)023}{{\em JHEP} {\bfseries 09}
  (2016) 023}, \href{http://arxiv.org/abs/1607.05276}{{\ttfamily
  arXiv:1607.05276 [hep-ph]}}.

\bibitem{Ballett:2016yod}
P.~Ballett, S.~F. King, S.~Pascoli, N.~W. Prouse, and T.~Wang, ``{Precision
  neutrino experiments vs the Littlest Seesaw},''
  \href{http://dx.doi.org/10.1007/JHEP03(2017)110}{{\em JHEP} {\bfseries 03}
  (2017) 110}, \href{http://arxiv.org/abs/1612.01999}{{\ttfamily
  arXiv:1612.01999 [hep-ph]}}.

\bibitem{King:2018fqh}
S.~F. King, S.~Molina~Sedgwick, and S.~J. Rowley, ``{Fitting high-energy
  Littlest Seesaw parameters using low-energy neutrino data and
  leptogenesis},'' \href{http://dx.doi.org/10.1007/JHEP10(2018)184}{{\em JHEP}
  {\bfseries 10} (2018) 184}, \href{http://arxiv.org/abs/1808.01005}{{\ttfamily
  arXiv:1808.01005 [hep-ph]}}.

\bibitem{King:2013xba}
S.~F. King, ``{Minimal see-saw model predicting best fit lepton mixing
  angles},'' \href{http://dx.doi.org/10.1016/j.physletb.2013.06.013}{{\em Phys.
  Lett. B} {\bfseries 724} (2013) 92--98},
  \href{http://arxiv.org/abs/1305.4846}{{\ttfamily arXiv:1305.4846 [hep-ph]}}.

\bibitem{King:2013hoa}
S.~F. King, ``{A model of quark and lepton mixing},''
  \href{http://dx.doi.org/10.1007/JHEP01(2014)119}{{\em JHEP} {\bfseries 01}
  (2014) 119}, \href{http://arxiv.org/abs/1311.3295}{{\ttfamily arXiv:1311.3295
  [hep-ph]}}.

\bibitem{Bjorkeroth:2014vha}
F.~Bj\"orkeroth and S.~F. King, ``{Testing constrained sequential dominance
  models of neutrinos},''
  \href{http://dx.doi.org/10.1088/0954-3899/42/12/125002}{{\em J. Phys. G}
  {\bfseries 42} no.~12, (2015) 125002},
  \href{http://arxiv.org/abs/1412.6996}{{\ttfamily arXiv:1412.6996 [hep-ph]}}.

\bibitem{Chen:2019oey}
P.-T. Chen, G.-J. Ding, S.~F. King, and C.-C. Li, ``{A New Littlest Seesaw
  Model},'' \href{http://dx.doi.org/10.1088/1361-6471/ab7e8d}{{\em J. Phys. G}
  {\bfseries 47} no.~6, (2020) 065001},
  \href{http://arxiv.org/abs/1906.11414}{{\ttfamily arXiv:1906.11414
  [hep-ph]}}.

\bibitem{Feruglio:2017spp}
F.~Feruglio, {\em {Are neutrino masses modular forms?}},
  \href{http://dx.doi.org/10.1142/9789813238053_0012}{pp.~227--266}.
\newblock 2019.
\newblock \href{http://arxiv.org/abs/1706.08749}{{\ttfamily arXiv:1706.08749
  [hep-ph]}}.

\bibitem{Kobayashi:2023zzc}
T.~Kobayashi and M.~Tanimoto, ``{Modular flavor symmetric models},''
  \href{http://dx.doi.org/10.1142/S0217751X24410124}{{\em Int. J. Mod. Phys. A}
  {\bfseries 39} no.~09n10, (2024) 2441012},
  \href{http://arxiv.org/abs/2307.03384}{{\ttfamily arXiv:2307.03384
  [hep-ph]}}.

\bibitem{Ding:2023htn}
G.-J. Ding and S.~F. King, ``{Neutrino mass and mixing with modular
  symmetry},'' \href{http://dx.doi.org/10.1088/1361-6633/ad52a3}{{\em Rept.
  Prog. Phys.} {\bfseries 87} no.~8, (2024) 084201},
  \href{http://arxiv.org/abs/2311.09282}{{\ttfamily arXiv:2311.09282
  [hep-ph]}}.

\bibitem{Ding:2019gof}
G.-J. Ding, S.~F. King, X.-G. Liu, and J.-N. Lu, ``{Modular S$_{4}$ and A$_{4}$
  symmetries and their fixed points: new predictive examples of lepton
  mixing},'' \href{http://dx.doi.org/10.1007/JHEP12(2019)030}{{\em JHEP}
  {\bfseries 12} (2019) 030}, \href{http://arxiv.org/abs/1910.03460}{{\ttfamily
  arXiv:1910.03460 [hep-ph]}}.

\bibitem{Ding:2021zbg}
G.-J. Ding, S.~F. King, and C.-Y. Yao, ``{Modular $S_4\times SU(5)$ GUT},''
  \href{http://dx.doi.org/10.1103/PhysRevD.104.055034}{{\em Phys. Rev. D}
  {\bfseries 104} no.~5, (2021) 055034},
  \href{http://arxiv.org/abs/2103.16311}{{\ttfamily arXiv:2103.16311
  [hep-ph]}}.

\bibitem{deMedeirosVarzielas:2022fbw}
I.~de~Medeiros~Varzielas, S.~F. King, and M.~Levy, ``{Littlest modular
  seesaw},'' \href{http://dx.doi.org/10.1007/JHEP02(2023)143}{{\em JHEP}
  {\bfseries 02} (2023) 143}, \href{http://arxiv.org/abs/2211.00654}{{\ttfamily
  arXiv:2211.00654 [hep-ph]}}.

\bibitem{deAnda:2023udh}
F.~J. de~Anda and S.~F. King, ``{Modular flavour symmetry and orbifolds},''
  \href{http://dx.doi.org/10.1007/JHEP06(2023)122}{{\em JHEP} {\bfseries 06}
  (2023) 122}, \href{http://arxiv.org/abs/2304.05958}{{\ttfamily
  arXiv:2304.05958 [hep-ph]}}.

\bibitem{deMedeirosVarzielas:2023ujt}
I.~de~Medeiros~Varzielas, S.~F. King, and M.~Levy, ``{A modular SU (5) littlest
  seesaw},'' \href{http://dx.doi.org/10.1007/JHEP05(2024)203}{{\em JHEP}
  {\bfseries 05} (2024) 203}, \href{http://arxiv.org/abs/2309.15901}{{\ttfamily
  arXiv:2309.15901 [hep-ph]}}.

\bibitem{Ding:2018fyz}
G.-J. Ding, S.~F. King, and C.-C. Li, ``{Tri-Direct CP in the Littlest Seesaw
  Playground},'' \href{http://dx.doi.org/10.1007/JHEP12(2018)003}{{\em JHEP}
  {\bfseries 12} (2018) 003}, \href{http://arxiv.org/abs/1807.07538}{{\ttfamily
  arXiv:1807.07538 [hep-ph]}}.

\bibitem{Ding:2018tuj}
G.-J. Ding, S.~F. King, and C.-C. Li, ``{Lepton mixing predictions from $S_4$
  in the tridirect CP approach to two right-handed neutrino models},''
  \href{http://dx.doi.org/10.1103/PhysRevD.99.075035}{{\em Phys. Rev. D}
  {\bfseries 99} no.~7, (2019) 075035},
  \href{http://arxiv.org/abs/1811.12340}{{\ttfamily arXiv:1811.12340
  [hep-ph]}}.

\bibitem{Ding:2017hdv}
G.-J. Ding, S.~F. King, and C.-C. Li, ``{Golden Littlest Seesaw},''
  \href{http://dx.doi.org/10.1016/j.nuclphysb.2017.10.019}{{\em Nucl. Phys. B}
  {\bfseries 925} (2017) 470--499},
  \href{http://arxiv.org/abs/1705.05307}{{\ttfamily arXiv:1705.05307
  [hep-ph]}}.

\bibitem{deAdelhartToorop:2011nfg}
R.~de~Adelhart~Toorop, F.~Feruglio, and C.~Hagedorn, ``{Discrete Flavour
  Symmetries in Light of T2K},''
  \href{http://dx.doi.org/10.1016/j.physletb.2011.08.013}{{\em Phys. Lett. B}
  {\bfseries 703} (2011) 447--451},
  \href{http://arxiv.org/abs/1107.3486}{{\ttfamily arXiv:1107.3486 [hep-ph]}}.

\bibitem{deAdelhartToorop:2011re}
R.~de~Adelhart~Toorop, F.~Feruglio, and C.~Hagedorn, ``{Finite Modular Groups
  and Lepton Mixing},''
  \href{http://dx.doi.org/10.1016/j.nuclphysb.2012.01.017}{{\em Nucl. Phys. B}
  {\bfseries 858} (2012) 437--467},
  \href{http://arxiv.org/abs/1112.1340}{{\ttfamily arXiv:1112.1340 [hep-ph]}}.

\bibitem{Ding:2012xx}
G.-J. Ding, ``{TFH Mixing Patterns, Large $\theta_{13}$ and $\Delta(96)$ Flavor
  Symmetry},'' \href{http://dx.doi.org/10.1016/j.nuclphysb.2012.04.002}{{\em
  Nucl. Phys. B} {\bfseries 862} (2012) 1--42},
  \href{http://arxiv.org/abs/1201.3279}{{\ttfamily arXiv:1201.3279 [hep-ph]}}.

\bibitem{King:2009mk}
S.~F. King and C.~Luhn, ``{A New family symmetry for SO(10) GUTs},''
  \href{http://dx.doi.org/10.1016/j.nuclphysb.2009.05.020}{{\em Nucl. Phys. B}
  {\bfseries 820} (2009) 269--289},
  \href{http://arxiv.org/abs/0905.1686}{{\ttfamily arXiv:0905.1686 [hep-ph]}}.

\bibitem{Chen:2014tpa}
M.-C. Chen, M.~Fallbacher, K.~T. Mahanthappa, M.~Ratz, and A.~Trautner, ``{CP
  Violation from Finite Groups},''
  \href{http://dx.doi.org/10.1016/j.nuclphysb.2014.03.023}{{\em Nucl. Phys. B}
  {\bfseries 883} (2014) 267--305},
  \href{http://arxiv.org/abs/1402.0507}{{\ttfamily arXiv:1402.0507 [hep-ph]}}.

\bibitem{Girardi:2013sza}
I.~Girardi, A.~Meroni, S.~T. Petcov, and M.~Spinrath, ``{Generalised
  geometrical CP violation in a T' lepton flavour model},''
  \href{http://dx.doi.org/10.1007/JHEP02(2014)050}{{\em JHEP} {\bfseries 02}
  (2014) 050}, \href{http://arxiv.org/abs/1312.1966}{{\ttfamily arXiv:1312.1966
  [hep-ph]}}.

\bibitem{ParticleDataGroup:2024cfk}
{\bfseries Particle Data Group} Collaboration, S.~Navas {\em et~al.}, ``{Review
  of particle physics},''
  \href{http://dx.doi.org/10.1103/PhysRevD.110.030001}{{\em Phys. Rev. D}
  {\bfseries 110} no.~3, (2024) 030001}.

\bibitem{Jarlskog:1985ht}
C.~Jarlskog, ``{Commutator of the Quark Mass Matrices in the Standard
  Electroweak Model and a Measure of Maximal CP Nonconservation},''
  \href{http://dx.doi.org/10.1103/PhysRevLett.55.1039}{{\em Phys. Rev. Lett.}
  {\bfseries 55} (1985) 1039}.

\bibitem{Branco:1986gr}
G.~C. Branco, L.~Lavoura, and M.~N. Rebelo, ``{Majorana Neutrinos and {CP}
  Violation in the Leptonic Sector},''
  \href{http://dx.doi.org/10.1016/0370-2693(86)90307-2}{{\em Phys. Lett. B}
  {\bfseries 180} (1986) 264--268}.

\bibitem{Nieves:1987pp}
J.~F. Nieves and P.~B. Pal, ``{Minimal Rephasing Invariant {CP} Violating
  Parameters With Dirac and Majorana Fermions},''
  \href{http://dx.doi.org/10.1103/PhysRevD.36.315}{{\em Phys. Rev. D}
  {\bfseries 36} (1987) 315}.

\bibitem{Nieves:2001fc}
J.~F. Nieves and P.~B. Pal, ``{Rephasing invariant CP violating parameters with
  Majorana neutrinos},''
  \href{http://dx.doi.org/10.1103/PhysRevD.64.076005}{{\em Phys. Rev. D}
  {\bfseries 64} (2001) 076005},
  \href{http://arxiv.org/abs/hep-ph/0105305}{{\ttfamily arXiv:hep-ph/0105305}}.

\bibitem{Jenkins:2007ip}
E.~E. Jenkins and A.~V. Manohar, ``{Rephasing Invariants of Quark and Lepton
  Mixing Matrices},''
  \href{http://dx.doi.org/10.1016/j.nuclphysb.2007.09.031}{{\em Nucl. Phys. B}
  {\bfseries 792} (2008) 187--205},
  \href{http://arxiv.org/abs/0706.4313}{{\ttfamily arXiv:0706.4313 [hep-ph]}}.

\bibitem{Branco:2011zb}
G.~C. Branco, R.~G. Felipe, and F.~R. Joaquim, ``{Leptonic CP Violation},''
  \href{http://dx.doi.org/10.1103/RevModPhys.84.515}{{\em Rev. Mod. Phys.}
  {\bfseries 84} (2012) 515--565},
  \href{http://arxiv.org/abs/1111.5332}{{\ttfamily arXiv:1111.5332 [hep-ph]}}.

\bibitem{Antusch:2005gp}
S.~Antusch, J.~Kersten, M.~Lindner, M.~Ratz, and M.~A. Schmidt, ``{Running
  neutrino mass parameters in see-saw scenarios},''
  \href{http://dx.doi.org/10.1088/1126-6708/2005/03/024}{{\em JHEP} {\bfseries
  03} (2005) 024},
\href{http://arxiv.org/abs/hep-ph/0501272}{{\ttfamily arXiv:hep-ph/0501272
  [hep-ph]}}.


\bibitem{King:2016yef}
S.~F. King, J.~Zhang, and S.~Zhou, ``{Renormalisation Group Corrections to the
  Littlest Seesaw Model and Maximal Atmospheric Mixing},''
  \href{http://dx.doi.org/10.1007/JHEP12(2016)023}{{\em JHEP} {\bfseries 12}
  (2016) 023},
\href{http://arxiv.org/abs/1609.09402}{{\ttfamily arXiv:1609.09402 [hep-ph]}}.


\bibitem{Geib:2017bsw}
T.~Geib and S.~F. King, ``{Comprehensive renormalization group analysis of the
  littlest seesaw model},''
  \href{http://dx.doi.org/10.1103/PhysRevD.97.075010}{{\em Phys. Rev.}
  {\bfseries D97} no.~7, (2018) 075010},
\href{http://arxiv.org/abs/1709.07425}{{\ttfamily arXiv:1709.07425 [hep-ph]}}.


\bibitem{JUNO:2022mxj}
{\bfseries JUNO} Collaboration, A.~Abusleme {\em et~al.}, ``{Sub-percent
  precision measurement of neutrino oscillation parameters with JUNO},''
  \href{http://dx.doi.org/10.1088/1674-1137/ac8bc9}{{\em Chin. Phys. C}
  {\bfseries 46} no.~12, (2022) 123001},
  \href{http://arxiv.org/abs/2204.13249}{{\ttfamily arXiv:2204.13249
  [hep-ex]}}.

\bibitem{DUNE:2020ypp}
{\bfseries DUNE} Collaboration, B.~Abi {\em et~al.}, ``{Deep Underground
  Neutrino Experiment (DUNE), Far Detector Technical Design Report, Volume II:
  DUNE Physics},'' \href{http://arxiv.org/abs/2002.03005}{{\ttfamily
  arXiv:2002.03005 [hep-ex]}}.

\bibitem{Hyper-Kamiokande:2018ofw}
{\bfseries Hyper-Kamiokande} Collaboration, K.~Abe {\em et~al.},
  ``{Hyper-Kamiokande Design Report},''
  \href{http://arxiv.org/abs/1805.04163}{{\ttfamily arXiv:1805.04163
  [physics.ins-det]}}.

\bibitem{Planck:2018vyg}
{\bfseries Planck} Collaboration, N.~Aghanim {\em et~al.}, ``{Planck 2018
  results. VI. Cosmological parameters},''
  \href{http://dx.doi.org/10.1051/0004-6361/201833910}{{\em Astron. Astrophys.}
  {\bfseries 641} (2020) A6}, \href{http://arxiv.org/abs/1807.06209}{{\ttfamily
  arXiv:1807.06209 [astro-ph.CO]}}. [Erratum: Astron.Astrophys. 652, C4
  (2021)].

\bibitem{Euclid:2024imf}
{\bfseries Euclid} Collaboration, M.~Archidiacono {\em et~al.}, ``{Euclid
  preparation - LIV. Sensitivity to neutrino parameters},''
  \href{http://dx.doi.org/10.1051/0004-6361/202450859}{{\em Astron. Astrophys.}
  {\bfseries 693} (2025) A58},
  \href{http://arxiv.org/abs/2405.06047}{{\ttfamily arXiv:2405.06047
  [astro-ph.CO]}}.

\bibitem{KamLAND-Zen:2024eml}
{\bfseries KamLAND-Zen} Collaboration, S.~Abe {\em et~al.}, ``{Search for
  Majorana Neutrinos with the Complete KamLAND-Zen Dataset},''
  \href{http://arxiv.org/abs/2406.11438}{{\ttfamily arXiv:2406.11438
  [hep-ex]}}.

\bibitem{LEGEND:2021bnm}
{\bfseries LEGEND} Collaboration, N.~Abgrall {\em et~al.}, ``{The Large
  Enriched Germanium Experiment for Neutrinoless $\beta\beta$ Decay}:
  {LEGEND-1000 Preconceptual Design Report},''
  \href{http://arxiv.org/abs/2107.11462}{{\ttfamily arXiv:2107.11462
  [physics.ins-det]}}.

\bibitem{nEXO:2021ujk}
{\bfseries nEXO} Collaboration, G.~Adhikari {\em et~al.}, ``{nEXO: neutrinoless
  double beta decay search beyond 10$^{28}$ year half-life sensitivity},''
  \href{http://dx.doi.org/10.1088/1361-6471/ac3631}{{\em J. Phys. G} {\bfseries
  49} no.~1, (2022) 015104}, \href{http://arxiv.org/abs/2106.16243}{{\ttfamily
  arXiv:2106.16243 [nucl-ex]}}.

\bibitem{Escobar:2008vc}
J.~A. Escobar and C.~Luhn, ``{The Flavor Group Delta(6n**2)},''
  \href{http://dx.doi.org/10.1063/1.3046563}{{\em J. Math. Phys.} {\bfseries
  50} (2009) 013524}, \href{http://arxiv.org/abs/0809.0639}{{\ttfamily
  arXiv:0809.0639 [hep-th]}}.





\end{thebibliography}
\end{document}